\documentclass[]{aa}  

\usepackage{xcolor}
\usepackage{multirow}
\usepackage{graphicx}
\usepackage{amsmath,amssymb}
\usepackage{rotating, lscape, longtable}
\usepackage{txfonts}
\usepackage{hyperref}
\begin{document}

   \title{X-Shooting ULLYSES: Massive Stars at Low Metallicity}

  \subtitle{XI. Pipeline-determined Physical Properties of Magellanic Cloud OB Stars}

   \titlerunning{XShootU XI: Pipeline Properties of OB stars}

   \authorrunning{Bestenlehner et al.}

   \author{J.M. Bestenlehner{\inst{\ref{inst:uos1}, \ref{inst:uos2} }
          \and
          Paul A. Crowther\inst{\ref{inst:uos1}}\thanks{Corresponding author; (paul.crowther@sheffield.ac.uk)}
         \and
          C. Hawcroft\inst{\ref{inst:stsci}}
          \and
          H. Sana\inst{\ref{inst:kul}}
          \and
          F. Tramper\inst{\ref{inst:cab}}
           \and
          J.S. Vink\inst{\ref{inst:aop}}
          \and
          S.A. Brands\inst{\ref{inst:ams}} 
          \and \newline
          A.A.C. Sander\inst{\ref{inst:ari}}
           \and
          the XShootU collaboration}
}
   \institute{
      {Astrophysics Research Cluster, School of Mathematical and Physical Sciences, University of Sheffield, Hicks Building, Hounsfield Road, Sheffield S3 7RH, United Kingdom}\label{inst:uos1}
      \and
      {School of Chemical, Materials and Biological Engineering, University of Sheffield, Sir Robert Hadfield Building, Mappin Street, Sheffield S1 3JD, United Kingdom}\label{inst:uos2}
        \and
       {Space Telescope Science Institute, 3700 San Martin Drive, Baltimore, MD 21218, USA}\label{inst:stsci}
        \and
       {Institute of Astronomy, KU Leuven, Celestijnenlaan 200D, B-3001 Leuven, Belgium}\label{inst:kul}
       \and
        {Departamento de Astrof\'{\i}sica, Centro de Astrobiolog\'{\i}a, (CSIC-INTA), Ctra. Torrej\'on a Ajalvir, km 4,  28850 Torrej\'{o}n de Ardoz, Madrid, Spain}\label{inst:cab}
       \and
       {Armagh Observatory and Planetarium, College Hill, BT61 9DG Armagh, Northern Ireland}\label{inst:aop}
       \and
       {Astronomical Institute Anton Pannekoek, Amsterdam University, Science Park 904, 1098 XH Amsterdam, The Netherlands}\label{inst:ams}
       \and
       {Astronomisches Rechen-Institut, Zentrum f\"{u}r Astronomie der Universit\"{a}t Heidelberg, M\"{o}nchhofstr. 12-14, 69120 Heidelberg, Germany}\label{inst:ari}
}

   \date{Received 4 Oct 2024; accepted 25 Jan 2025}

 
  \abstract
%
    {%
    The proximity and low dust extinction of the Magellanic Clouds provides an ideal environment for metal-poor massive stars to be studied in detail. The {\it Hubble Space Telescope} ULLYSES initiative has provided exquisite ultraviolet spectroscopy of a large sample of OB stars in the Magellanic Clouds, and its legacy value has been enhanced through the acquisition of optical VLT/XShooter spectroscopy (XShootU). 
    }
    {We aim to determine the spectral types and physical properties of 122 LMC and 103 SMC OB stars observed via XShootU. Physical parameters are required for these to serve as templates in spectral libraries of metal poor massive stars. We also aim to identify double-lined binaries and OeBe stars for which analysis requires non-standard treatment.}  
    {We apply a pipeline designed to analyse large spectroscopic samples of hot luminous stars to XShootU spectroscopic datasets, together with grids of synthetic model spectra computed with the non-LTE atmospheric code {\sc fastwind} at LMC and SMC metallicities. }
    {We determine physical and wind properties of 97 LMC and 77 SMC massive stars, ranging from O2 to B9 subtypes, representing the majority of the XShootU OB sample (OeBe and candidate SB2 systems are excluded). Results are broadly in agreement with previous optical spectroscopic studies, with evolutionary masses spanning $12-117 M_{\odot}$ in the LMC and $11-74 M_{\odot}$ in the SMC. We determine a revised $T_{\rm eff}$-spectral type calibration for Magellanic Cloud stars, identify stars with peculiar radial velocities, and compare wind properties of high luminosity O stars with dense winds, revealing $\sim$0.27 dex higher wind momenta of LMC stars with respect to SMC counterparts. Incorporating the recent empirical metallicity dependence of $Z^{0.22}$ for wind velocities, this suggests a mass-loss dependence of $Z^{0.5}$ for luminous O stars. Studies incorporating ultraviolet mass-loss diagnostics are required for OB stars with weak winds and/or low luminosities.}
    {}%

   
   \maketitle

\section{Introduction}
\label{intro}

Massive stars, in spite of their rarity, are responsible for energetic transients \citep{Smartt2015, Levan+2016}, multi-messenger astrophysics \citep{Abbott+2016} plus direct (stellar) and indirect (nebular) diagnostics in high redshift galaxies observed with {\it James Webb Space Telescope} \citep{Arellano-Cordova+2022, Cameron+2023, Curti+2024}. Many of these phenomena are favoured at low metallicity, so there is a renewed interest in the properties and evolution of metal-poor massive stars.

Locally, the proximity of the Magellanic Clouds provides our best view of individual metal-poor massive stars, with present day metallicities 1/2 Z$_{\odot}$ (LMC) and 1/5 {Z$_{\odot}$ (SMC) \citep{RussellDopita1990}. Extremely metal-poor massive stars have been identified in the Local Group \citep{Garcia+2019, Telford+2021}, but these are exceptionally rare owing to the modest star-formation rates of their host galaxies \citep{Crowther2019}.

The rest-frame ultraviolet and optical spectroscopic appearance of young stellar populations are dominated by massive stars ($\geq 8 M_{\odot}$), owing to their high bolometric luminosities. However, interpretation of observations is hindered by uncertainties in the evolution of massive stars as a result of rotational mixing, stellar winds and binarity, especially the dependence of these quantities on metallicity \citep{Langer2012}.

Evolutionary calculations for main-sequence OB stars generally adopt theoretical mass-loss prescriptions \citep{Vink+2001} which likely {\it underestimate} rates for very massive stars \citep{Bestenlehner+2014, Bestenlehner+2020} and {\it overestimate} rates for B supergiants \citep{BerniniPeron+2024}. Rotational velocities peak at fairly modest rates, but extend to high velocities \citep{Ramirez-Agudelo+2013, Dufton+2013}, likely due to binary interaction \citep{deMink+2014}. The close binary fraction of massive stars is high in the Milky Way and LMC \citep{Sana+2012, Sana+2013}, with SMC results awaiting analysis of the BLOeM survey \citep{Shenar+2024}.

Contemporary population synthesis models typically employ synthetic ultraviolet and optical spectra of OB stars \citep{BruzualCharlot2003, Leitherer10, PYPOPSTAR}. Historically, empirical libraries of OB stars have been limited to solar metallicity \citep{Howarth89, XSH}. Spectroscopic optical surveys such as VFTS \citep{Evans+2011} have targeted the Magellanic Clouds, but these lack absolute flux calibration, and only modest ultraviolet samples have been acquired \citep{Walborn+1995, Bouret+2003, Crowther+2016}. 

The recent {\it Hubble Space Telescope} Director's Discretionary programme ULLYSES \citep{ULLYSES} Roman-Duval et al. (submitted) and associated VLT Large Programme XShootU \citep{Vink+2023} provide high quality UV {\it and} optical spectroscopy of OB stars in the Magellanic Clouds. This initative permits an improved understanding of stellar winds \citep{Backs+2024} and CNO abundances \citep{Martins+2024} at different metallicities, plus template OB stars at sub-solar composition \citep{Crowther2024}.

It is necessary to determine the physical properties of ULLYSES/XShootU stars in order to incorporate these observations into population synthesis codes. This is the primary goal of the present study, exploiting a new pipeline for the efficient analysis of large samples of optical OB spectra \citep{Bestenlehner2024}, and is intended to complement detailed studies of more limited sub-samples \citep[e.g.][]{BerniniPeron+2024, Backs+2024}. Physical
parameters of Magellanic Cloud OB stars are also useful in the context of the Sloan Digital Sky Survey V (SDSS-V) Local Volume Mapper which is characterizing nebular emission in the Magellanic Clouds and southern Milky Way \citep{LVM}.

We present observational datasets in Section~\ref{obs}, including  contemporary spectral classification of XShootU targets, briefly describe the pipeline used to analyse ULLYSES OB stars in Section~\ref{analysis}. We present our derived physical parameters in Section~\ref{properties}, allowing us to compare the properties of XShootU OB stars between the Magellanic Clouds. The inclusion of H$\alpha$ allows  wind properties of stars with dense winds to be quantified in Section~\ref{wind-properties}, incorporating wind velocities courtesy of ULLYSES spectroscopy \citep[e.g.][]{Hawcroft+2023}, so extends the previous empirical study of \citet{Mokiem+2007}. Finally, 
comparisons with literature results are made in Section~\ref{literature} and brief conclusions are drawn in Section~\ref{conclusions}.


\section{Spectroscopic datasets and classification}\label{obs}

\subsection{Optical spectroscopy}\label{s:os}

Our primary datasets are VLT/Xshooter spectroscopy obtained with the ESO Large Programme XShootU \citep{Vink+2023}, which provides complete spectral coverage of 114 LMC and 112 SMC stars between Oct 2020 and Jan 2022. Spectral coverage involved 3100--5500\AA\ (UBV arm), 5500--10200\AA\ (VIS arm) at resolving powers of R = 6,700 (45 km\,s$^{-1}$) and 11,400 (26 km\,s$^{-1}$), respectively, using the 0.8 and 0.7 arcsec slits. 
Version 2 of the early data release (eDR2) was used for our spectroscopic analysis. Full
details of the bespoke data reduction process are provided in \citet{Sana+2024}, including slit-loss correction, 
co-addition, flux calibration, telluric correction and normalisation. Sky regions were adjusted to ensure nearby companions were excluded from background regions.

The ULLYSES Magellanic Cloud target selection is described in Roman-Duval et al. (submitted). In brief, representative OB stars in both Clouds were identified spanning spectral type and luminosity class (O and early B), plus a modest selection of B supergiants at all spectral types. Targets were selected on the basis of literature (heterogeneous) classifications, with an attempt made to exclude known binaries, with a few notable exceptions (e.g. HD~5980, LMC X-4). 

Consequently, in contrast to other large spectroscopic surveys of the OB stars in the Magellanic Clouds, such as VFTS \citep{Evans+2011} or BLOeM \citep{Shenar+2024}, ULLYSES/XShootU is not intended to provide a representative subset of massive stars in the LMC or SMC. Targets were selected to minimise binarity, and
preferentially select UV bright targets along low extinction sight-lines. 

Nevertheless, several examples of spectroscopic or eclipsing binaries were known from the pre-XShootU literature \citep{Vink+2023}. The majority of XShootU targets were observed via two back-to-back exposures, but multiple observations of selected targets were obtained \citep[][see their Appendix C]{Sana+2024} which permit (a subset of) close binary systems to be identified. XShootU omitted several ULLYSES targets for which archival VLT/Xshooter spectroscopy was available. These are excluded from our spectroscopic analysis since the tailored data reduction process of \citet{Sana+2024} was not followed.

In addition, supplementary observations of selected narrow-lined ULLYSES stars (21 in LMC, 27 in SMC) have been obtained with the Magellan 6.5m Clay dual-echelle spectrograph MIKE between Dec 2021 and Dec 2022, which provides spectral coverage from 3350--5000\AA\ (blue arm) and 4900-9500\AA\ (red arm) at resolving powers of R = 35,000 and 40,000 ($\sim$8 km\,s$^{-1}$), respectively, using the 0.7 arcsec slit. A standard echelle data reduction process was followed. Examples of representative spectra are presented in fig.~9 of \citet{Crowther2022}.

The Magellan/MIKE dataset was not used for the spectroscopic pipeline, but its higher resolution with respect to VLT/Xshooter greatly helped to identify double-lined binaries. By way of example, BI272 was classified as O7\,II by \citet{Conti+1986}, with O9\,II favoured from inspection of XShootU datasets, despite unusually strong Si\,{\sc iii} $\lambda\lambda$4553--4575 features. Close inspection of the MIKE spectrum indicated broad and narrow components in He\,{\sc i} lines, plus broad He\,{\sc ii} lines, indicating a O7\,V primary and B2 secondary component, the latter responsible for Si\,{\sc iii} and Mg\,{\sc ii} $\lambda$4481. Brands et al. (submitted) attributed their inability to obtain a satisfactory fit to BI272 to binarity.

\subsection{UV spectroscopy}

Although the primary goal of the present study is to determine the physical parameters of XShootU targets, we also consider wind properties in Section~\ref{wind-properties}. In order to convert wind densities (largely obtained from H$\alpha$) to mass-loss rates and wind momenta, we require wind velocities, which are primarily obtained from far ultraviolet P Cygni profiles of N\,{\sc v} $\lambda\lambda$1238--42, Si\,{\sc iv} $\lambda\lambda$1393--1402 and C\,{\sc iv} $\lambda\lambda$1548--51.

\begin{figure}
\centering
  \includegraphics[width=0.6\columnwidth,angle=-90,bb=46 45 540 757]{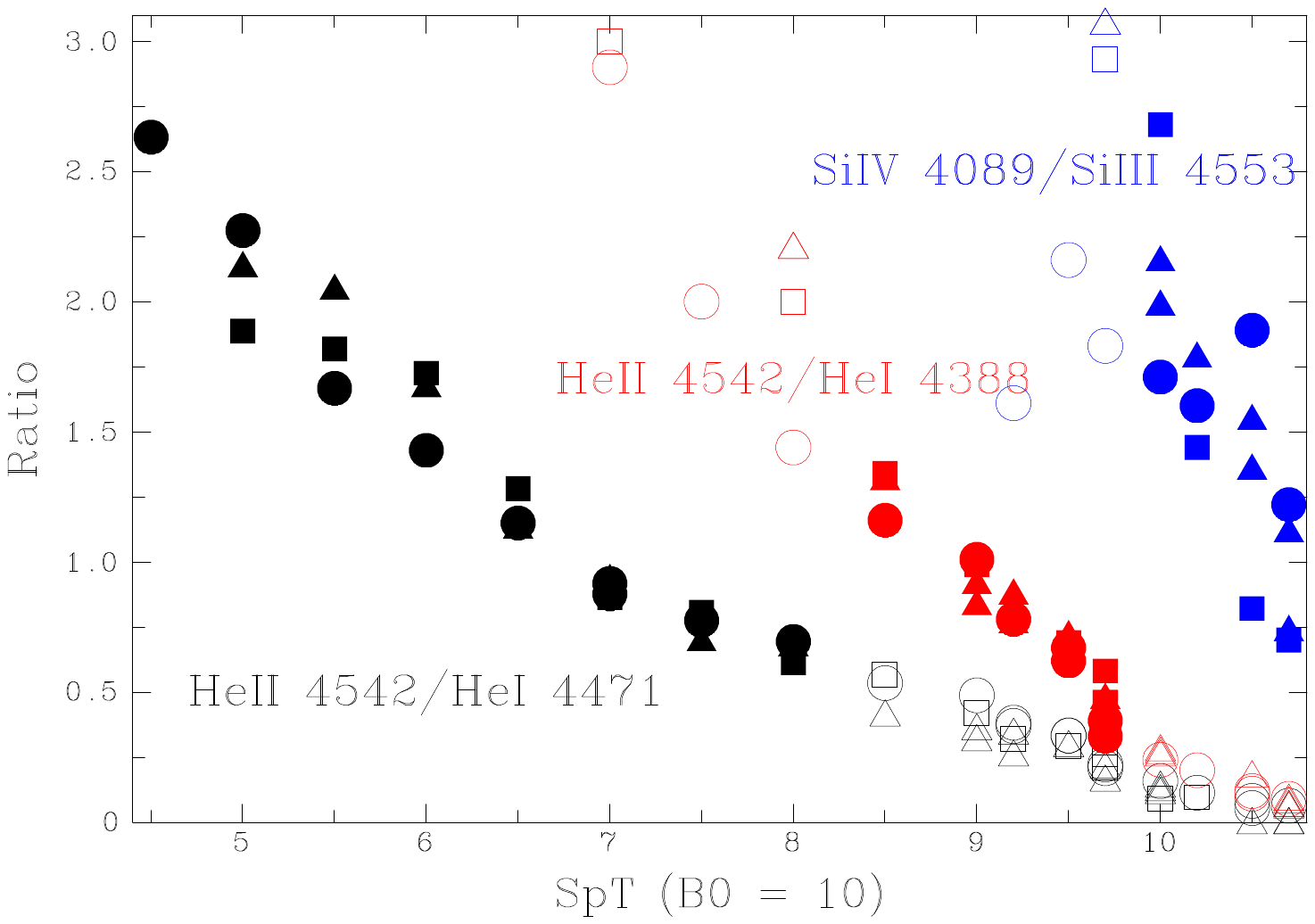}
  \includegraphics[width=0.6\columnwidth,angle=-90,bb=46 45 540 757]{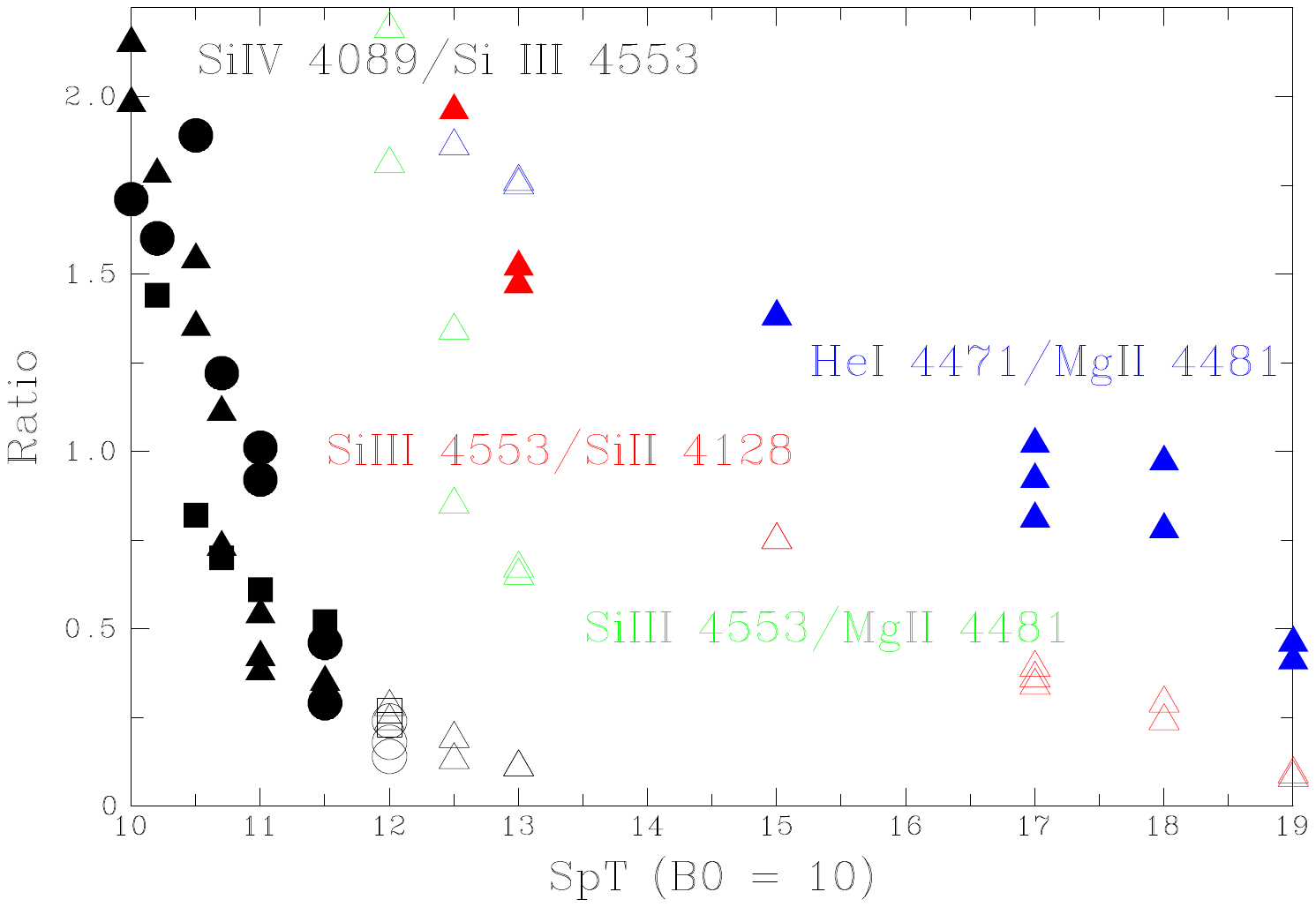}
  \caption{Upper panel: Overview of He\,{\sc i-ii} and Si\,{\sc iii-iv} classification diagnostics for Galactic O and early B templates \citep{Sota+2011, Negueruela+2024}, circles for luminosity classes V-IV, squares for III-II and triangles for I, filled for primary diagnostics as applied to LMC stars. Lower panel: As above, for Si\,{\sc ii-iv} and He\,{\sc i}/Mg\,{\sc ii} classification diagnostics for Galactic B stars (restricted to supergiants for B2.5+), and excluding B4, B6 subtypes for LMC stars \citep{Fitzpatrick1991}.
  We also show the ratio of Si\,{\sc iii} $\lambda$4553 to Mg\,{\sc ii} $\lambda$4481 for Galactic B2-3 supergiants (green) since this is the qualitative diagnostic for SMC mid-B supergiants \citep{Lennon1997}.}
  \label{OB}
\end{figure}

ULLYSES\footnote{https://ullyses.stsci.edu/} \citep{ULLYSES} provides far-UV spectroscopy for nearly all XShootU targets, courtesy of a STScI Director's Discretionary Award using the COS or STIS instruments aboard {\it Hubble Space Telescope}. COS G130M/1291 + G160M/1611 spectroscopy achieved a wavelength coverage of $\lambda\lambda$1132--1433\AA\ at $R$ = 14,000, while STIS E140M spectroscopy achieve a wavelength coverage of $\lambda\lambda$1143-1710\AA\ at $R$ = 46,000. Further details of ULLYSES are provided by Roman-Duval et al. (submitted). Four targets initially selected for observation were ultimately dropped from the ULLYSES programme, namely AzV 255 (SMC), Sk --67$^{\circ}$ 51 (LMC), BI 128 (LMC) and [ST92] 5-52 (LMC), though were retained for XShootU.

Wind velocities for a subset of XShootU sample have previously been determined by Sobolev with Exact Integral (SEI) fitting \citep{Hawcroft+2023}. For the remainder, we accessed DR7 datasets from ULLYSES in order to either fit (SEI method) or directly measure wind velocities, following the method set out by \citet{Prinja+1990}, \citet{PrinjaCrowther1998} and \citet{Crowther+2016}.

\begin{figure*}
\centering
  \includegraphics[width=10cm,angle=-90,bb=47 66 532 757]{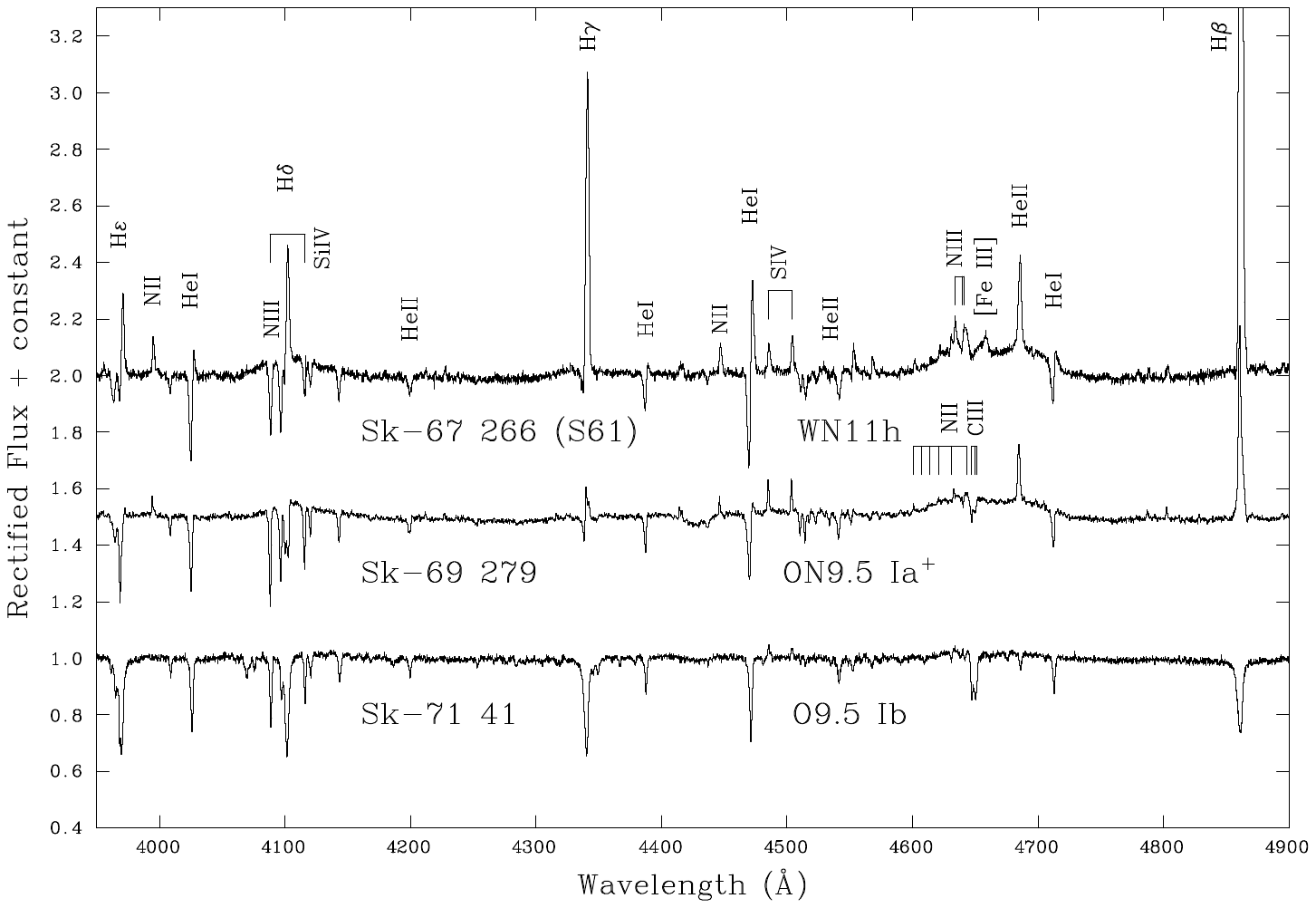}
  \caption{Sequence of late O supergiant (Sk --71$^{\circ}$ 41, O9.5~Ib, this study), late O hypergiant (Sk --69$^{\circ}$ 279, ON9.5 Ia$^{+}$,  this study) and very late WN star \citep[Sk --67$^{\circ}$ 266, WN11h,][]{CrowtherSmith1997} in the LMC, illustrating the transition from absorption line dominated stars to emission line dominated (e.g. He\,{\sc i} $\lambda$4471, He\,{\sc ii} $\lambda$4686). Note that Balmer lines include nebular contributions for Sk --67$^{\circ}$ 266 (S61).}
  \label{SK67D266}
\end{figure*}

\subsection{Classification}\label{class}

We have established LMC and SMC reference OB stars from comparison with Galactic O star templates from \citet{MaizApellaniz+2016} and Galactic B star templates from \citet{Negueruela+2024}. These stars are drawn from the present datasets, supplemented by archival VLT/FLAMES \citep{Evans+2006, Dufton+2019} and BLOeM \citep{Shenar+2024} datasets. Reference stars are ideally sharp-lined, permitting rotational broadening to be applied for comparison with fast rotators.

Montages of XShootU spectroscopy of OB dwarfs, giants and supergiants for each galaxy are available in online material (see Sect.~\ref{data_availability}). Owing to the absence of SMC B5 and B9 supergiants amongst the XShootU sample, we have included spectroscopy of these subtypes from the BLOeM survey \citep{Shenar+2024} which have been classified using an identical approach. 

\subsubsection{O stars}

Spectral types of O stars generally involved the relative strengths of He\,{\sc ii} $\lambda$4541 to He\,{\sc i} $\lambda$4471, or He\,{\sc i} $\lambda$4388, following \citet{Sota+2011} and \citet{MaizApellaniz+2016}. For very early O-type stars we use the scheme of \citet{Walborn+2002}, which is based on N\,{\sc iii} $\lambda\lambda$4634-41 and N\,{\sc iv} $\lambda$4058 diagnostics. 
In general one would expect extremely weak He\,{\sc i} signatures for such stars, but He\,{\sc i} is relatively strong in a few instances (AzV 14, AzV 435, W61 28--5), which is suggestive of binarity. 

Luminosity classes generally follow the morphology of He\,{\sc ii} $\lambda$4686 following \citet{Walborn1971, Walborn1972}, updated for O8.5--9.7 subtypes by \citet{Sota+2011} which uses the ratio of He\,{\sc ii} $\lambda$4686 to He\,{\sc i} $\lambda$4713. 

\subsubsection{B stars}

 For B0--0.7 subtypes, our primary diagnostic is the ratio of Si\,{\sc iv} $\lambda$4088 to Si\,{\sc iii} $\lambda$4553 \citep{Negueruela+2024}. This
is metallicity dependent \citep{Walborn1983}, so the high quality of XShootU datasets permits the use of helium diagnostics too \citep{Sota+2011}. In Fig.~\ref{OB} we present
line ratios of He and Si diagnostics for O4 to early B stars (upper panel) highlighting primary ratios for different spectral types. In practice, classifications based on Si or He are in close agreement. For example, the B0\,Ia standard Sk --68$^{\circ}$ 52 from \citet{Fitzpatrick1991} remains as B0\,Ia from both Si and He diagnostics, while his B0.5\,Ia standard Sk --68$^{\circ}$ 41 is reclassified as B0.7\,Ia from both diagnostics.

Spectral types of LMC B1+ supergiants is based on Galactic templates adapted from \citet{Fitzpatrick1991} to ensure consistency in Si\,{\sc ii-iv} line strengths with Galactic standards of \citet{Negueruela+2024}, together with Mg\,{\sc ii} $\lambda$4481 to He\,{\sc i} $\lambda$4471 for B5+ stars, as shown in the lower panel of Fig.~\ref{OB}. The B1.5\,Ia standard Sk --67$^{\circ}$ 14 from \citet{Fitzpatrick1991} is reclassified  B1\,Ia from the ratio of Si\,{\sc iv} $\lambda$4088 to Si\,{\sc iii} $\lambda$4553, while his B3\,Ia standard Sk --67$^{\circ}$ 78 remains unchanged.

The high quality of XShootU observations could permit the addition of a new B4\,Ia LMC standard (Sk --70$^{\circ}$ 50), on the basis of the ratio of Si\,{\sc iii} $\lambda$4553 to Si\,{\sc ii} $\lambda$4128 and Si\,{\sc iii} $\lambda$4553 to Mg\,{\sc ii} $\lambda$4481, although we retain B3\,Ia \citep{Fitzpatrick1991} following the discussion relating to the B4 subtype by \citet{Negueruela+2024}.

For SMC B1+ supergiants we largely follow the scheme of \citet{Lennon1997} owing to the weakness of metallic lines. Figure~\ref{OB} demonstrates that the Si\,{\sc iii} $\lambda$4553 to Mg\,{\sc ii} $\lambda$4481 is a robust discriminator of B2--3 supergiants.
Classification of late B supergiants qualitatively follows the LMC scheme, except that no B7 subtypes are defined for the SMC. The high quality of our XShootU datasets leads to reclassification of several B supergiants with respect to \citet{Lennon1997}, including AzV 96 (B1.5 to B1 since Si\,{\sc iv} $\lambda$4116 is clearly detected), AzV 22 (B5  to B3  since Si\,{\sc iii} $\lambda$4553 is clearly detected). B-type luminosity classes in both galaxies are based on comparison with H$\gamma$ morphologies of contemporary Galactic templates \citep{Negueruela+2024}.

\subsubsection{OBC/OB/OBN classifications}

Regarding OC/O/ON and BC/B/BN nomenclature, we follow past convention \citep{WalbornFitzpatrick1990} in assigning ON for unusually strong N\,{\sc v} $\lambda\lambda$4603--20 absorption for early O stars (e.g. VFTS 506), unusually strong N\,{\sc iii} $\lambda$4520 for late O stars (e.g. Sk --68$^{\circ}$ 135). We reassign AzV~215 from BN0\,Ia \citep{Lennon1997} to B0\,Ia  since its nitrogen lines are comparable in strength to normal B0 supergiants (e.g. AzV  235).
Conversely, we assign OC to early (e.g. Sk --68$^{\circ}$ 133) and mid (e.g. AzV 69) O stars for which C\,{\sc iii} $\lambda\lambda$4647--51 emission exceeds N\,{\sc iii} $\lambda$4634, and early B stars for which N\,{\sc ii} $\lambda$4447 and $\lambda\lambda$4601--43 are unusually weak (e.g. Sk --68$^{\circ}$ 26).

\subsubsection{OB hypergiants versus Wolf-Rayet stars}

The emphasis of the present study is the analysis of OB stars, with Wolf-Rayet stars excluded owing to our grid selection. Classification diagnostics for O2--4 stars versus WN5--7 stars includes H$\beta$ \citep{CrowtherWalborn2011}, while late O stars are discriminated from WN9--11 stars on the basis of He\,{\sc ii} $\lambda$4686 and He\,{\sc i} $\lambda$5876 (or $\lambda$4471) emission  \citep{CrowtherSmith1997}. 

By way of example, Sk --68$^{\circ}$ 135 is a late O hypergiant \citep[ON9.7 Ia$^{+}$,][]{Walborn1977} despite strong H$\beta$ emission. Sk --69$^{\circ}$ 279  has previously been classified as O9.2 Iaf \citep{Gvaramadze+2018} but is reclassified as a late O hypergiant (ON9.5 Ia$^{+}$) owing to its morphological similarity to Sk --68$^{\circ}$ 135, despite even stronger H$\beta$ emission. Figure~\ref{SK67D266} compares blue VLT/Xshooter spectra of Sk --69$^{\circ}$ 279 to a normal late O supergiant (Sk --71$^{\circ}$ 41), and a very late WN star (Sk --67$^{\circ}$ 266, S61) showing the transition from absorption line dominated to emission line dominated stars.

\subsection{XShootU sample}

Fig.~\ref{XShootU_SpT} provides a histogram of spectral types and luminosity classes of LMC and SMC ULLYSES OB stars, based on XShootU spectral classifications, including several stars with archival VLT/Xshooter datasets. As noted above, ULLYSES targets were selected to provide complete coverage of spectral type and luminosity class for O and early B stars in both Clouds, plus full spectral type coverage of B supergiants. In reality, early O stars are exceptionally rare in the SMC plus some historical classifications required adjustment, such that all SMC ULLYSES targets later than B3 are B8 supergiants. In addition, several literature late O or early B dwarfs are newly classified as giants or binaries or both (e.g. N206-FS 170, B1 III: + B). This is not wholly unexpected since HST orbit requirements were biased towards far-UV bright targets (Roman-Duval et al. submitted).

\begin{figure}
\centering
  \includegraphics[width=1\columnwidth]{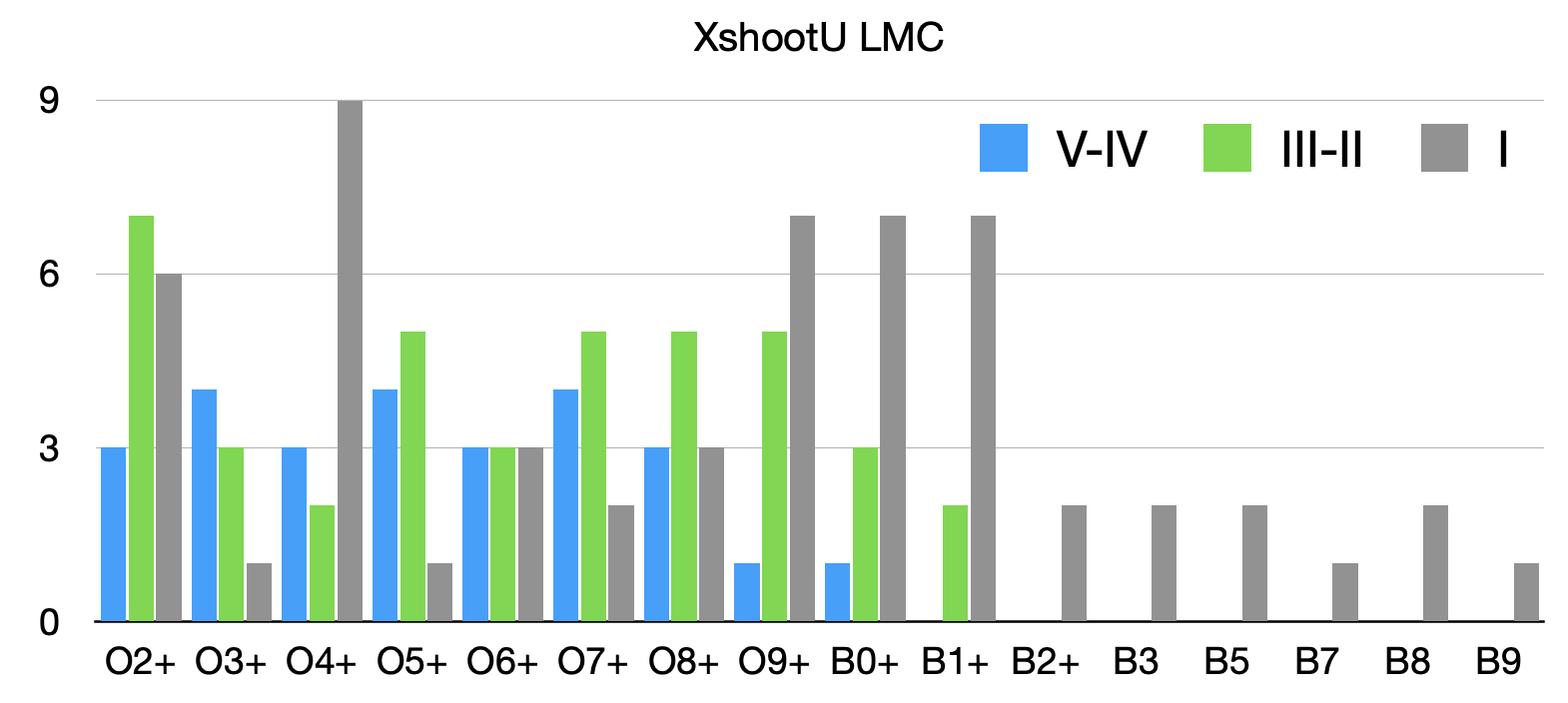}
  \includegraphics[width=1\columnwidth]{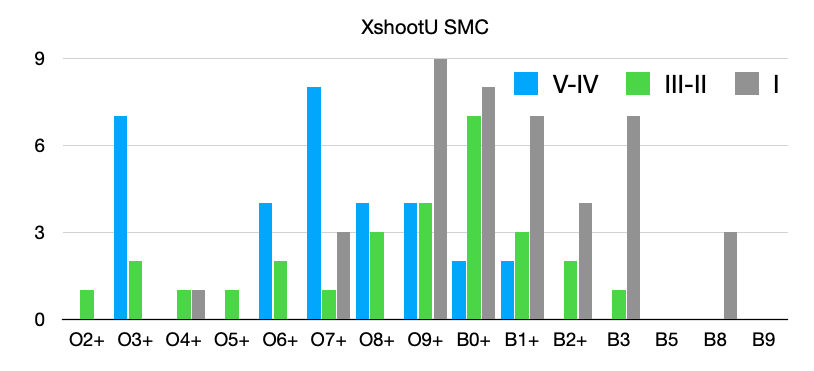}
  \caption{Spectral types and luminosity classes of LMC (upper panel) and SMC (lower panel) ULLYSES OB stars based on XShootU spectral classifications, including archival VLT/Xshooter stars (blue: V-IV, green III-II, grey: I).}
  \label{XShootU_SpT}
\end{figure}

\begin{figure}
\centering
  \includegraphics[width=0.8\columnwidth,bb=42 17  533 776]{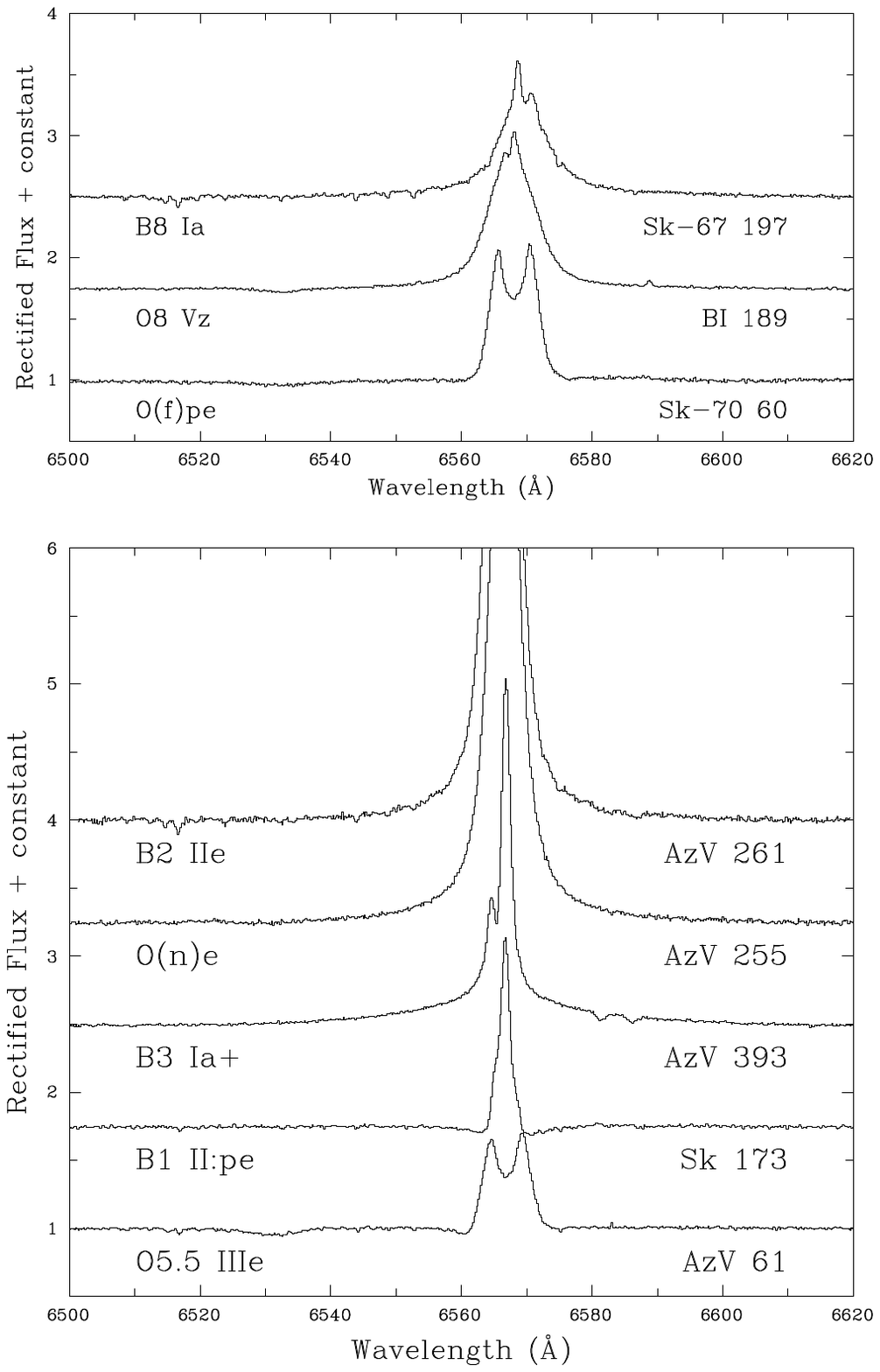}
  \caption{H$\alpha$ profiles of OeBe stars in the SMC (lower panel) and LMC (upper panel), excluded from our pipeline analysis.}
  \label{OBe}
\end{figure}

For the current spectroscopic analysis all observations were coadded. Since we exclude SB2's from our sample this should not impact our analysis. We have also excluded LMC X-4 (SB1) for which spectral variability was identified (our classification is based on the first epoch of observation). Wolf-Rayet stars were excluded from the analysis, together with OB stars whose spectra include problematic features for the pipeline (e.g. strong nebulosity). Several OeBe stars were also identified and excluded from analysis. H$\alpha$ profiles of OeBe stars are shown in Fig.~\ref{OBe}. Spectral types of excluded stars are provided in online material (see Sect.~\ref{data_availability}. Literature results for the majority of these are provided in \citet{Vink+2023}.

\section{Pipeline}\label{analysis}

For our spectroscopic analysis pipeline we employ grids of synthetic model spectra computed with v10.6 of
the non-LTE atmosphere code {\sc fastwind} \citep{Puls+2005, Rivero-Gonzalez+2012} including  H, He, C, N, O, Si and Mg as explicit elements. Separate grids were calculated for LMC and SMC metallicities, assuming $0.5Z_{\odot}$ and $0.2Z_{\odot}$ respectively with initial, semi-processed and fully processed CNO composition. Grids covered the following parameter space $\log T_{\rm eff}$ (K) over [4.0, 4.775] in 0.025 dex steps, corresponding to 10kK $\leq T_{\rm eff} \leq$ 60kK, $\log g$ (cm\,s$^{-2}$) over [1.5, 4.5] in 0.2 dex steps, wind strength $\log Q$ over [--11.4, --15.0] in 0.3 dex steps and Helium abundances in mass-fraction $Y$ over [0.15, 0.55] in 0.05 steps. 
Convergence difficulties were experienced at the lowest temperatures ($T_{\rm eff} \leq$ 12kK) impacting on fits to B9 supergiants.

The wind--strength parameter $Q = \dot{M} (R_{\ast} v_{\infty})^{-3/2}$, with units $M_{\odot}$\,yr$^{-1}$, $R_{\odot}$ and km\,s$^{-1}$, so mass-loss rates can be determined from $Q, R_{\ast}$ and $v_{\infty}$. In addition, we varied nitrogen abundances for high temperature grids from $\log T_{\rm eff}$/K = 4.6 to 4.775, ranging from initial N-abundance \citep{Vink+2023} to fully CNO processed (4 values for LMC, 3 values for SMC) as the ionisation balance between nitrogen ions becomes the main temperature diagnostic.

A smooth wind with volume filling factor $f_{\rm v} =1 $ and $\beta =1$ velocity law was assumed and micro-turbulent velocity was set to $\varv_{\rm mic}=10$ km\,s$^{-1}$. For the terminal wind velocities, $v_{\infty}$, in our grid we adopted the empirical calibration of Magellanic Cloud OB stars by \citet{Hawcroft+2023}, namely
\begin{equation}
    \varv_{\infty} = \left[ (92 \pm 3) T_{\rm eff}/{\rm kK)} - (1040 \pm 100) \right] Z/Z_{
    \odot}^{(0.22\pm0.03)} 
\end{equation}\label{Hawcroft}
in km\,s$^{-1}$ for LMC ($T_{\rm eff} \geq$ 15 kK, $Z = 0.5 Z_{\odot}$) and SMC ($T_{\rm eff} \geq$ 20 kK, $Z = 0.2 Z_{\odot}$) OB stars. These results involved Sobolev with Exact Integral (SEI) fits to ULLYSES \citep{ULLYSES} observations supplemented with literature results, and was extrapolated to lower temperature with a minimum $\varv_{\infty}$ = 250 km\,s$^{-1}$. Individual wind velocities are used when determining mass-loss rates from wind densities, $Q$, as discussed in Section~\ref{vely}.

A complete description of the pipeline\footnote{https://github.com/jbestenlehner/mdi\_analysis\_pipeline} is provided in \citet{Bestenlehner2024}. In brief, we used the spectral range $\lambda\lambda3800 - 7200$\AA\ including the observational error spectrum \citep{Sana+2024} by utilising a $\chi^2$ minimisation Ansatz: 
\begin{equation}
    \chi^2 = (\vec{d} - \mathrm{R}\vec{s})^{\mathrm T}\mathrm{N}^{-1}(\vec{d} - \mathrm{R}\vec{s})
\end{equation}
with $\vec{d}$ the observed and $\vec{s}$ the synthetic spectra, $\mathrm{R}$ the instrumental responds matrix and observational, diagonal error matrix $\mathrm{N}$. As model uncertainties should be budgeted into the parameter determination we de-idealised the model spectrum $\vec{s}$ according to \citet{Bestenlehner2024}. 

Our sample is fairly heterogeneous, ranging from early O dwarfs to late B supergiants, and extends to Of/WN stars, but excludes WR stars, which are outside the intended parameter space of {\sc fastwind}. Therefore, the model-error is averaged over the entire parameter space of our sample. This impacted the overall performance of the pipeline. For example, targets with stronger winds ($\log Q > -12.5$) are affected by stellar wind assumptions, which would be included into the model-error. However, weak-wind targets would not provide any information on the accuracy of the wind assumptions made in the stellar atmosphere model and potentially averaging out those contributions. Therefore, a meaningful model-error should be ideally based on a sample of similar objects \citep[c.f. the discussion in][]{Bestenlehner2024}.


Typical macro-turbulent velocities for OB stars are in the range between a few km\,s$^{-1}$ to $\sim$30 km\,s$^{-1}$. The velocity resolution of the UBV and VIS arm are 45 and 26 km\,s$^{-1}$, respectively (Section~\ref{s:os}). Therefore, we convolved our synthetic grid with a fixed $\varv_{\rm mac} = 20$ km\,s$^{-1}$ and assumed any additional broadening is due to rotation, with projected rotational velocities of $\varv_{\rm e} \sin i = [0, 10, 20, 35, 50, 75, 100, 150, 200, 250, 300, 350, 400, 450, 500]$ km\,s$^{-1}$.

\begin{figure}
\centering
  \includegraphics[width=0.8\columnwidth]{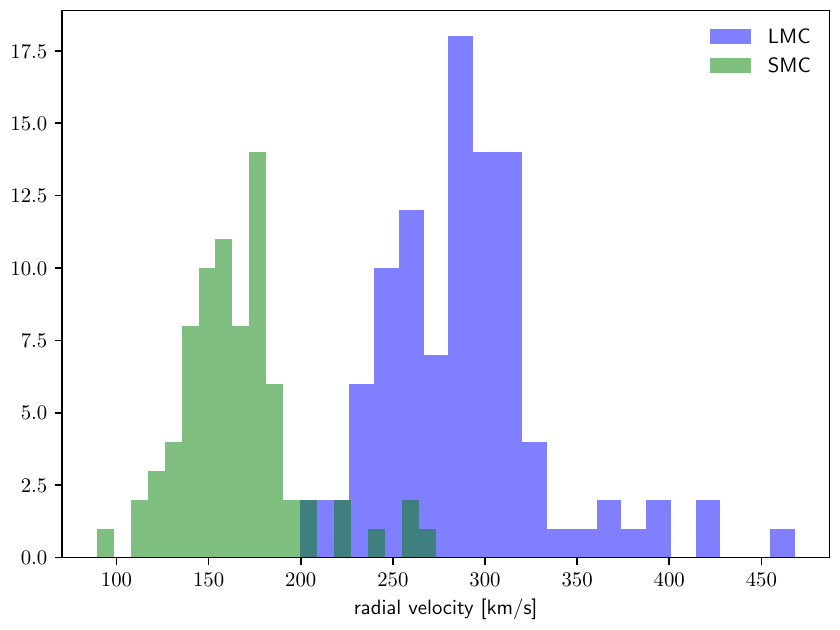}
  \caption{Histogram of radial velocities (in km\,s$^{-1}$) of LMC (blue) and SMC (green) OB stars, from
  cross-correlation with synthetic template spectra.}
  \label{RV}
\end{figure}

Observations are cross-correlated with synthetic template spectra to determine their radial velocity (RV) and then corrected for the RV shift before sampled on the wavelength grid of the synthetic spectra. Figure~\ref{RV} shows a bimodal distribution of RVs corresponding to the LMC and SMC, with average values of $\varv_{\rm rad}$ = 290$\pm$46 km\,s$^{-1}$, and $\varv_{\rm rad}$ = 166$\pm$33 km\,s$^{-1}$, respectively. Outliers are potential multiple systems or runaway stars, and include Sk --67$^{\circ}$ 22 (LMC) with $\varv_{\rm rad}$ = 468 km\,s$^{-1}$ and NGC 346 ELS 26 (SMC) with 227 km\,s$^{-1}$.



Hydrogen lines are the most prominent lines in OB stars and dominate the $\chi^2$. As our sample includes O and B stars, He lines can be as weak as metal lines. Therefore, we chose the following wavelength sampling: First, we initialize a wavelength array with 0.1\AA\ spacing around the spectral lines in our {\sc fastwind} LINES-list. Second, we increased the number of wavelength points by a factor of 5 beyond $\pm5$\AA\ of the central wavelength of the Balmer lines, because $\log g$ is based on the pressure-broadened wings. Third, we increased the number of wavelength points by a factor of 25 within $\pm 1$\AA\ of the central wavelength of the He and metal lines. 

The stellar atmosphere grid is none-rectilinear as models did not converge or failed to compute due to unphysical parameter space (e.g. Eddington limit). Before determining the uncertainties we fill the gaps in the probability distribution function (PDF) with zero-values, so that the PDF becomes a 4D$-T_{\rm eff}-\log g - log Q - Y$ rectilinear grid. The 4D grid was then interpolated to artificially increase the grid resolution using the multidimensional interpolation function {\it scipy.interpolate.interpn} with cubic-spline method to obtain more accurate parameters and less gridded uncertainties. 

We used the following standard deviations in 4D; $1\sigma$: 0.0902, $2\sigma$: 0.5940 and $3\sigma$: 0.9389, following \citet{Wang+2015}. CNO abundances and $\varv_{\rm e} \sin i$ were not included as they mainly improve the fit to the nitrogen lines and the line broadening, but also a 6D grid interpolation becomes computationally very expensive. We estimated N-abundances in the 2D$-T_{\rm eff}-N$ PDF for stars hotter than $\log T_{\rm eff}$/K $\geq 4.6$, because the ionisation balance of the nitrogen lines becomes the main temperature diagnostic, and projected rotational velocities in 1D-PDF, because line broadening is largely independent of the stellar parameters.

\begin{figure}
\centering
\includegraphics[width=\columnwidth]{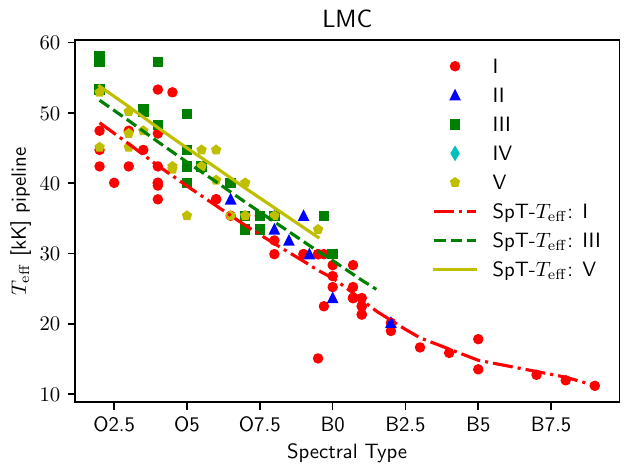} \includegraphics[width=\columnwidth]{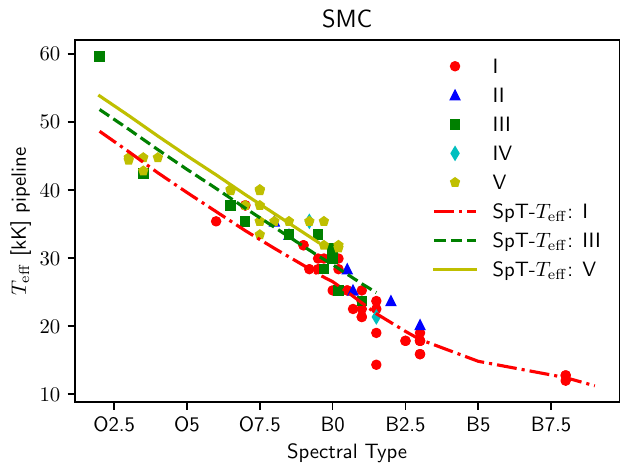}
  \caption{Pipeline temperatures (kK) versus XShootU spectral types for single+SB1 stars in the LMC (upper panel) and SMC (lower panel), colour coded by luminosity class: V (yellow pentagon), IV (cyan diamond), III (green square), II (blue triangle), I (red circle). We also include our calibration for dwarfs (solid yellow line), giants (dashed green line) and supergiants (dot-dashed red line).}
  \label{SpT_T}
\end{figure}

In order to determine bolometric luminosities we adopted distance moduli of 18.48 mag \citep{Pietrzynski+2019} and 18.98 mag \citep{Graczyk+2020} for the LMC and SMC, respectively, and incorporated optical and IR photometry listed in \citet{Vink+2023}. 

\section{Physical properties of Magellanic Cloud OB stars}\label{properties}

Online material (see Sect.~\ref{data_availability}) includes spectral fits for each star while electronic tables provide pipeline physical parameters, notes for which include details of problematic cases. No selected stars were excluded from our pipeline study, though parameters considered unreliable are flagged with parentheses. Parameters were not determined for ULLYSES targets with archival VLT/Xshooter spectroscopy since they were not processed using the XShootU pipeline \citep{Sana+2024}. These targets are included for completeness, since revised spectral classifications are provided. Wind properties are discussed in Sect.~\ref{wind-properties}, with parameters also provided in electronic form. Double-lined binaries (SB2), OeBe stars and Wolf-Rayet stars are excluded from our analysis, and are listed in online material.

Uncertainties presented here exclude systematic differences between different codes and approaches set out in \citet{Sander+2024}. Current results for stars in common with \citet[][models F4]{Sander+2024} are updated in some instances. Comparisons with previous
studies for a subset of our sample are deferred to Section~\ref{literature}.

\begin{table}[ht]
\caption{Spectral type-$T_{\rm eff}$ (kK) calibration of OB stars in the XShootU sample together with previous results for OB stars in the LMC \citep{Doran+2013} and SMC \citep{Dufton+2019}.}
\label{SpT_calib}
\begin{center}
    \begin{tabular}{lccccccc}
    \hline
    Spect. & \multicolumn{3}{c}{XShootU} & \multicolumn{2}{c}{LMC}
    & \multicolumn{2}{c}{SMC} \\
    Type   & V & III  & I     &  V   &  I &  V & I \\ 
    \hline\hline
    O2    & 53.8     & 51.8    & 48.6 &  54.0    &  46.0  & $\cdots$ & $\cdots$\\
    O3    & 50.9     & 48.9    & 45.6 &  48.0    &  42.0  & $\cdots$ & $\cdots$\\
    O4    & 47.9     &45.9     & 42.5 &  43.9    &  40.1  & $\cdots$ & $\cdots$\\
    O5    & 45.0     &43.0     & 39.6 &  41.9    &  38.3  & 45.2    & $\cdots$ \\
    O6    & 42.2     &40.2     & 36.8 &  39.9    &  36.4  & 43.0    & $\cdots$ \\
    O7    & 39.3     &37.3     & 34.0 &  37.9    &  34.5  & 40.7    & $\cdots$ \\
    O8    & 36.5     &34.5     & 31.4 &  35.9    &  32.6  & 38.5    & $\cdots$ \\
    O9    & 33.7     &31.7     & 28.9 &  33.9    &  30.7  & 36.3    & $\cdots$ \\
    O9.5  & 32.3     &30.3     & 27.6 &  32.9    &  29.8  & 35.1    & $\cdots$ \\
    B0    & $\cdots$ &29.0     & 26.5 &  31.4    &  28.6  & 32.0     & 27.2      \\
    B0.5  & $\cdots$ &27.6     & 25.0 &  29.1    &  25.4  & 29.6     & 24.3      \\
    B1    & $\cdots$ &26.2     & 23.4 & $\cdots$ &  22.2  & 27.3     & 22.3      \\
    B1.5  & $\cdots$ &24.9     & 21.8 & $\cdots$ &  20.6  & 26.1     & 20.6      \\
    B2    & $\cdots$ &$\cdots$  & 20.5 & $\cdots$ &  19.0  & 24.9     & 18.9      \\
    B2.5  & $\cdots$ &$\cdots$  & 19.2 & $\cdots$ & 17.4   & 23.9     & 17.2      \\
    B3    & $\cdots$ &$\cdots$  & 18.0 & $\cdots$ & 15.8   & 21.5     & 15.5      \\
    B5    & $\cdots$ &$\cdots$  & 14.8 & $\cdots$ & 14.2   & $\cdots$ & 13.8   \\
    B8    & $\cdots$ &$\cdots$  & 12.3 & $\cdots$ & 12.3   & $\cdots$ & $\cdots$ \\
    B9    & $\cdots$ &$\cdots$  & 11.2 & $\cdots$ &$\cdots$& $\cdots$ & $\cdots$ \\
    \hline
    \end{tabular}
    \end{center}
\end{table}

\subsection{Stellar temperatures}\label{temp}

Figure~\ref{SpT_T} compares spectral types to inferred effective temperatures for our sample. We are able to obtain satisfactory temperatures from early/mid O stars to mid/late B stars (comparison with literature results are provided in Section.~\ref{literature}). However the analysis is highly sensitive to the accuracy of the spectral normalisation process. For example, at the high temperature range of early O stars ($T_{\rm eff} \gtrsim 50$ kK) we observed a large scatter for the LMC stars. In general  the region between the N\,{\sc v} doublet ($\lambda\lambda4604 - 4620$) and He\,{\sc ii} $\lambda 4686$ is poorly normalised, which results in a lower weight for this wavelength range when minimising the $\chi^2$. 

For early O stars the temperature is derived based on the ionisation balance of N\,{\sc iv} to {N\,{\sc v}. The fairly weak line strength and the lower weight around N\,{\sc v} implies that the presence of N\,{\sc v} hardly contributes to the overall $\chi^2$.
Other outliers include late O hypergiants in the LMC (e.g. Fig.~\ref{SK67D266}) whose weak He\,{\sc ii} $\lambda$4686 is not reproduced, plus the B0.7 bright giant AzV 85 for which an unrealistically low temperature and low gravity is obtained. 

Table~\ref{SpT_calib} presents the temperature scale of Magellanic Cloud OB stars inferred from this analysis, separated into classes V, III and I, together with previous calibrations for the LMC \citep{Doran+2013} and SMC \citep{Dufton+2019}. These are overlaid on individual results in Fig.~\ref{SpT_T}. We do not attempt to provide separate LMC and SMC scales since pipeline results for early O stars are problematic. Note that previous calibrations usually originate from multiple studies of O and B stars, whereas the present study encompasses the full range of stars from early O to late B. Temperatures obtained are broadly consistent with previous calibrations, noting the modest number of late B supergiants included in ULLYSES/XShootU.

\begin{figure}
\centering
  \includegraphics[width=\columnwidth]{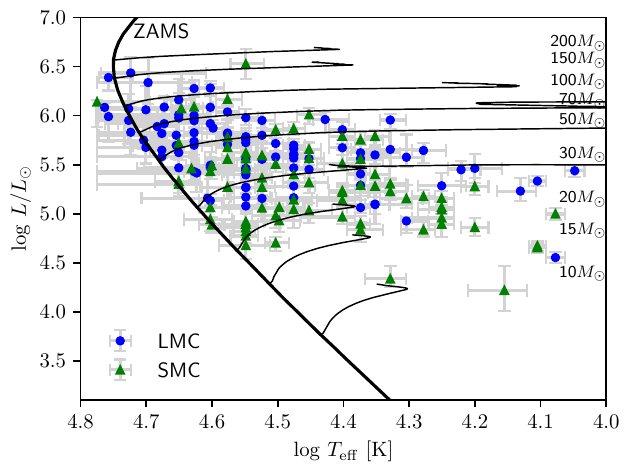}
  \caption{Hertzsprung-Russell diagram  of XshootU sample (LMC: blue circles, SMC: green triangles), together with LMC evolutionary tracks for 10--200 $M_{\odot}$ main-sequence stars from \citet{Brott+2011} and \citet{Kohler+2015}. Evolutionary masses of post-main sequence stars are determined from TAMS luminosities of SMC tracks from \citet{Schootemeijer+2019}.}
  \label{HRD}
\end{figure}

\subsection{Stellar luminosities}

Figure~\ref{HRD} presents pipeline results in a Hertzsprung-Russell (HR) diagram (LMC: blue, SMC: green), superimposed upon LMC metallicity evolutionary tracks from \citet{Brott+2011} and \citet{Kohler+2015}. This represents a more robust HR diagram than that presented in \citet{Vink+2023} since it was based, in part, on spectral type calibrations. 

Stellar luminosities plus individual reddening parameters $R_{5495}$ and $E(4405 - 5495)$ were obtained by fitting the photometric fluxes with the model spectral energy distribution (SED) employing the reddening law of \citet{MaizApellaniz+2014}. Online material includes SED fits for each star.

A histogram of interstellar extinctions is presented in Fig.~\ref{EBV} for LMC (blue) and SMC (green) OB stars. Unsurprisingly for a UV-selected sample, extinctions are low, with $E(B-V) \leq$ 0.25 mag for the majority of sources.

From inspection of photometry-SED fits we noticed a few inconsistencies. The $J$-Band photometry of Sk --67$^{\circ}$ 2 and Sk --68$^{\circ}$ 8 and the $K$-band of N11 ELS 32 do not follow the general trend of the other photometry and the flux calibrated XShootU spectrum. These were  removed from the fit. AzV\,85 requires a negative $E(B-V)$ to reproduce its SED, even though the trend of the flux calibrated spectrum and photometric points agree. In the case of AzV\,6 the optical photometry was inconsistent with the flux calibrated spectrum and the near-IR photometry was offset with respect to the optical data and flux calibrated spectra. This led to an unphysically high luminosity and poor SED fit.

While ULLYSES was designed to include representative OB stars from both  galaxies, the relatively modest star formation rate of the SMC resulted in a deficiency of (luminous) early O stars. This is evident in Fig.~\ref{XShootU_SpT}, and results in lower average luminosities. The mean luminosity among the O-star sample is $\log L/L_{\odot} = 5.74 \pm 0.31$ (LMC) versus $5.40 \pm 0.40$ (SMC), excluding AzV 6 (see above). For B stars $\log L/L_{\odot} = 5.40 \pm 0.31$ (LMC) versus $5.14 \pm 0.35$ (SMC). Too few Magellanic Cloud late B supergiants are included in the XShootU dataset to contribute to the Flux-weighted Gravity-Luminosity Relationship \citep[FGLR,][]{Urbaneja+2017}.

\begin{figure}
\centering
  \includegraphics[width=0.8\columnwidth]{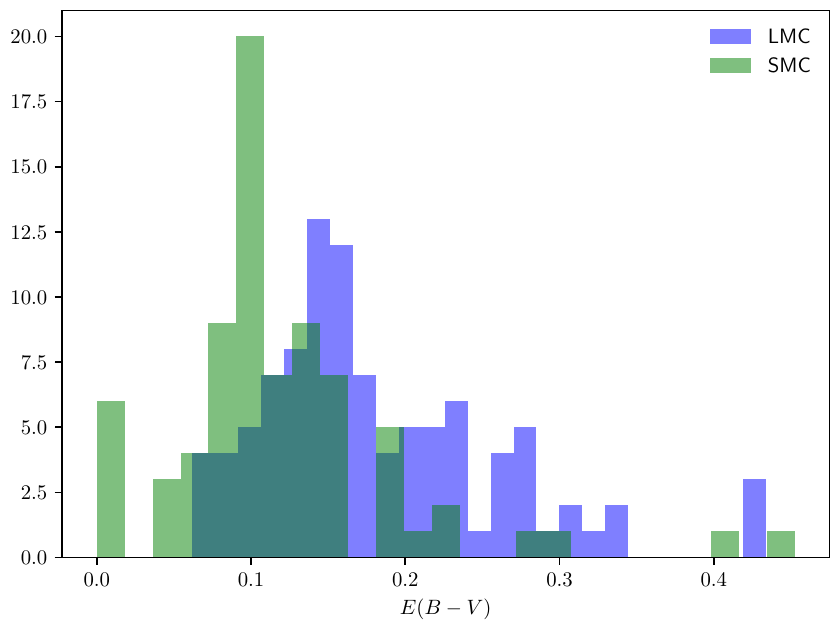}
  \caption{Histogram of $E(B-V)$ values for LMC (blue) and SMC (green) OB stars.}
  \label{EBV}
\end{figure}

\subsection{Stellar masses}\label{masses}

 \subsubsection{Evolutionary masses and ages}
 
 Evolutionary masses, $M_{\rm evol}$, presented in electronic data tables are obtained from contemporary evolutionary models. For main sequence (MS) stars we use {\sc bonnsai}\footnote{The BONNSAI web-service is available at www.astro.uni-bonn.de/stars/bonnsai} \citep{Schneider+2014}, coupled to evolutionary models from \citet{Brott+2011} for 5--60 $M_{\odot}$ at LMC and SMC metallicity plus \citet{Kohler+2015} for 60--500 $M_{\odot}$ at LMC metallicity}.
 
 Evolutionary masses use spectroscopic temperatures, luminosities and $v_{\rm e} \sin i$ as input observables. The only prior adopted was a Salpeter Initial Mass Function, with a uniform prior for initial rotational velocity. In a few instances no solution was possible for these parameters, in which case $v_{\rm e} \sin i$ was excluded (solution shown with ":" in electrond tables. 
It was necessary to resort to the LMC grid \citep[upper limit 500 $M_{\odot}$,][]{Kohler+2015} for a small subset of SMC stars close to, or above, 60 $M_{\odot}$ (e.g. AzV~232, O7\,Iaf$^{+}$). 
The use of higher metallicity models will overestimate their MS mass-loss, and in turn, overestimate their initial masses.
 
 Initial masses are very close to current evolutionary masses, with the exception of very massive stars (e.g. VFTS 482), initially fast rotators (e.g. VFTS 72), and luminous early-type B supergiants, although these are sensitive to adopted mass-loss prescriptions \citep{Vink+2001}. In a few instances current masses exceed initial masses using {\sc bonnsai}. This is because values cited are mode quantities, and occasionally current mass distributions differ from bell functions e.g. mean, median and mode $M_{\rm evol}$ for 2dFS 3694 (B0.7~III) are 16.7$^{+0.9}_{-1.4}, 16.6^{+1.0}_{-1.3}$ and $17.2^{+0.4}_{-2.0}~M_{\odot}$, respectively.

\begin{figure}
\centering
  \includegraphics[width=1\columnwidth]{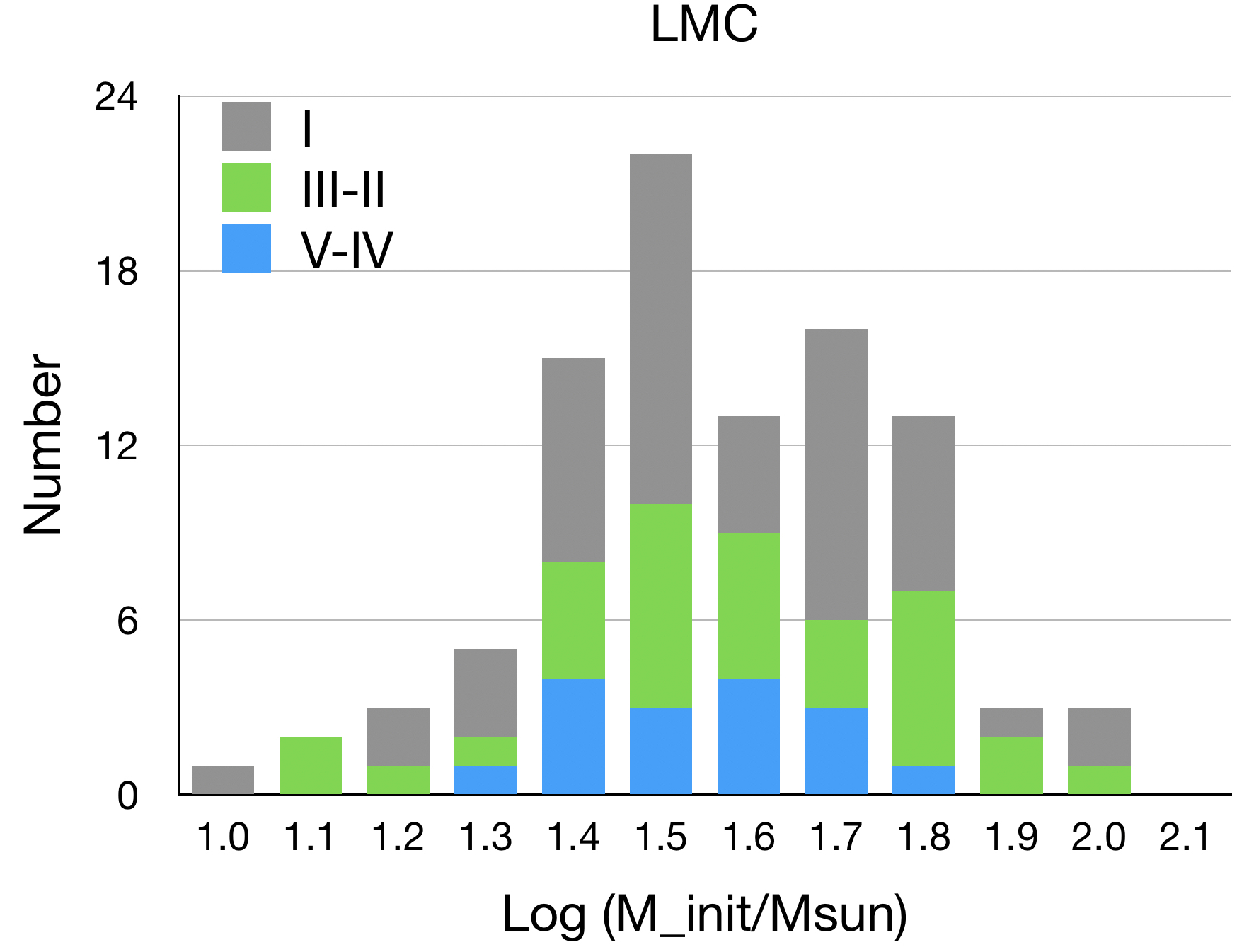}
  \includegraphics[width=1\columnwidth]{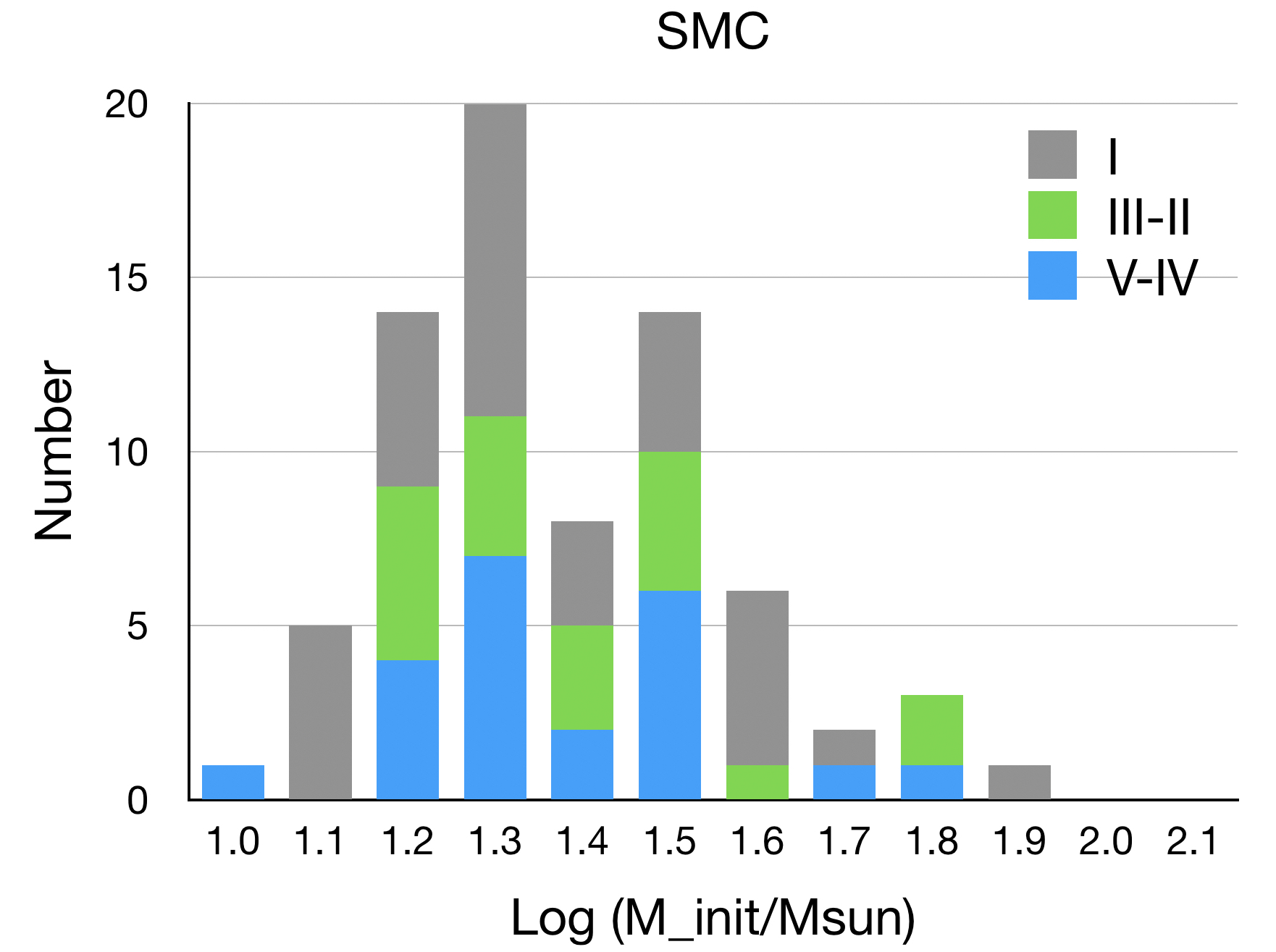}
  \caption{Distribution of initial masses, $M_{\rm ini}$ (in $M_{\odot}$) of LMC (upper panel) and SMC (lower panel) ULLYSES OB stars (blue: V-IV, green III-II, grey: I), based on {\sc bonnsai} results using \citet{Brott+2011} 
and \citet{Kohler+2015}  models for main sequence stars, or \citet{Schootemeijer+2019} SMC evolutionary models for post-main sequence late B supergiants. }
  \label{XShootU_mass}
\end{figure}
 
 \citet{Brott+2011} models incorporated into {\sc bonnsai} only extend to the terminal age main sequence (TAMS). 
For post-main sequence (post-MS) stars, the determination of masses is more problematic since evolutionary models exhibit considerably more variety than during the main sequence. 
The luminosity at the end of the contraction phase following the TAMS provides a credible mass estimate. 
TAMS masses can be approximated by initial masses (at most only a few percent of the initial mass is lost during the main sequence evolution for metal-poor stars with $M_{\rm init} \leq 30 M_{\odot}$).

We have estimate stellar masses of post-MS stars from comparison with SMC evolutionary models 
of \citet{Schootemeijer+2019} with $\alpha_{\rm SC} = 10$ and $\alpha_{\rm OV} = 0.33$, i.e. consistent semiconvection and overshooting parameters to \citet{Brott+2011} and \citet{Kohler+2015}, from which, 
 \begin{equation} 
 \log M_{\rm init} = 0.6288 - 0.1764 \log L_{\rm TAMS} + 0.0602 (\log L_{\rm TAMS})^2 \end{equation}
 with masses and luminosities in solar units, and TAMS lifetimes, $t$, in Myr
 \begin{equation}
 \log \tau_{\rm TAMS} = 4.7964 - 1.0105 \log L_{\rm TAMS} + 0.0523 (\log L_{\rm TAMS})^2
 \end{equation}
 These are flagged in electronic tables via []. By way of example, an initial mass (stellar age) of 20.4$^{+2.8}_{-2.6}~M_{\odot}$ (7.6$^{+1.3}_{-1.1}$ Myr) is obtained for Sk 179 (B3~II) from {\sc bonnsai}, versus 20.7$^{+3.8}_{-2.1}~M_{\odot}$ (9.5$^{+1.4}_{-1.5}$ Myr) from the TAMS luminosity based on \citet{Schootemeijer+2019} models. \citet{Brott+2011} models are known to predict unusually short main-sequence lifetimes (Marchant, priv. comm.). 
 
 The use of SMC metallicity models for seven LMC post-MS stars will (slightly) underestimate their cumulative mass-loss history, and in turn, introduce modest errors to initial mass calculations. In addition, current masses of post-MS stars in general may differ from the true mass since additional mass-loss may occur during the cool supergiant phase.

  The distribution of initial masses of XShootU OB stars in each galaxy is presented in Fig.~\ref{XShootU_mass}. In brief, the higher luminosities of LMC XShootU stars is reflected in 0.2 dex higher masses. Dwarfs in the SMC span 1.2 $\leq (\log M_{\rm init}/M_{\odot}) \leq 1.6$, with the exception of two O3\,V's, with supergiants primarily in the range 1.1 $\leq (\log M_{\rm init}/M_{\odot}) \leq 1.7$, with a similar mass distribution for giants. In contrast, LMC dwarfs typically range from  1.4 $\leq (\log M_{\rm init}/M_{\odot}) \leq 1.8$, supergiants span a broad mass range 1.2 $\leq (\log M_{\rm init}/M_{\odot}) \leq 2.1$, again with a similar distribution for giants.  Focusing on, for example, O6--8 dwarfs, average evolutionary masses are comparable for the SMC (27$\pm 6~M_{\odot}, N=10$) and LMC (31$\pm 10~M_{\odot}, N=6$).  \citet{Martins+2005} report stellar masses of 26$\pm$5 $M_{\odot}$ for O6--8 dwarfs, in their observational calibration of Galactic O stars.

Electronic data tables also include stellar ages, and range from 0$^{+1.5}_{-0}$ Myr (e.g. Sk --67$^{\circ}$ 211) to 19$^{+3}_{-1}$~Myr (e.g. Sk --67$^{\circ}$ 195). The range of evolutionary masses sampled is
10.8$\pm0.9~M_{\odot}$ for 2dFS 3947 (B1~IV) in the SMC to 117$^{+29}_{-39}~M_{\odot}$ for Sk --67$^{\circ}$ 211 (O2~III(f*)) in the LMC. The median mass of all XShootU O stars (B stars) in the SMC is 32.3 $M_{\odot}$ (19.6 $M_{\odot}$), versus 46.0~$M_{\odot}$ (27.3~$M_{\odot}$) in the LMC.

\subsubsection{Spectroscopic masses}
 
  Alternatively, spectroscopic masses, $M_{\rm spec}$, can be inferred from surface gravities and radii. Typical spectroscopic gravities presented in electronic data tables are $\log g$/(cm s$^{-2}$) = 3.7$\pm$0.4 for  O stars, $\log g$/(cm s$^{-2}$) = 2.7$\pm$0.6 for B stars, though exclude contributions from centrifugal forces owing to rotation. Corrected gravities, $g_{c}$, are obtained from \citet{Herrero+1992}
\[
g_{c} = g + (\varv_{\rm e} \sin i)^2 /R_{\ast}.\]
These are included in electronic data tables. In most instances corrections are modest, but can exceed 0.1 dex for rapid rotators e.g. $\log g_{c} - \log g$ = 0.23 dex for VFTS 190 with $\varv_{\rm e} \sin i \sim$ 300 km\,s$^{-1}$. 
%
%
Spectroscopic gravities are unrealistically high in some instances.  By way of example the O6\,Vz((f)) star
  N11 ELS 048 has an evolutionary mass of $M_{\rm evol} = 46.0^{+5.0}_{-4.2}~M_{\odot}$ but an unrealistically high spectroscopic gravity, implying an unphysical spectroscopic mass of $134^{+23}_{-33}~M_{\odot}$. 
Such cases
are indicated with parentheses in electronic data tables.
  
  Spectroscopic masses are compared to inferred evolutionary masses in Fig.~\ref{M_spec-M_evol}, separated into main sequence (MS) and post-MS populations, 
the latter selected for stars located beyond the terminal age main sequence (TAMS). There is a well established discrepancy between spectroscopic and evolutionary  masses \citep{Herrero+1992}, which is not so apparent from our analysis. Fits to evolutionary masses below 60~$M_{\odot}$ are shown in Fig.~\ref{M_spec-M_evol} for LMC (SMC) as dashed (dotted) lines, above which there are too few stars to draw conclusions. 
  For main sequence stars, evolutionary masses are often considered to be more realistic, in spite of challenges with evolutionary calculations.
 

\begin{figure}[htbp]
\centering
  \includegraphics[width=9 cm]{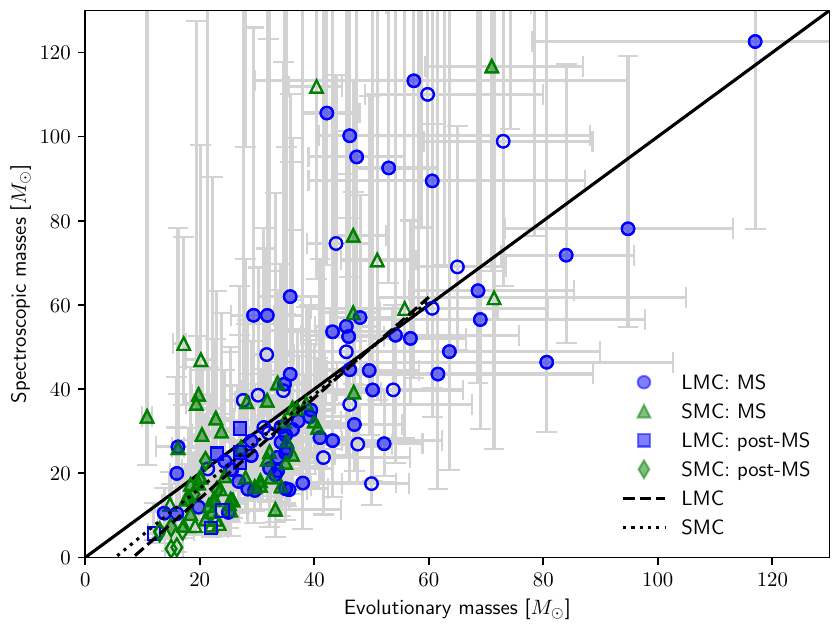}
  \caption{Evolutionary versus spectroscopic masses (in $M_{\odot}$) of LMC (blue) and SMC (green) OB stars, in which main sequence stars are circles/triangles and post-main sequence stars are squares/diamonds. Stars with unreliable surface gravities or uncertain stellar parameters are indicated with open symbols. The solid black line is a 1:1 relation, while dashed (dotted) black lines are fits to LMC (SMC) stars up to 60 $M_{\odot}$. }
  \label{M_spec-M_evol}
\end{figure}

Ultimately, careful analysis of eclipsing binaries may be necessary to address the mass discrepancy \citep{Tkachenko+2020}. Several XshootU targets are known eclipsing binaries (see Sect.~\ref{data_availability} for online material), plus some targets that are apparently single or SB1 binaries have been identified as double-lined SB2 binaries from spectroscopic monitoring (e.g. VFTS 482, Pollock et al. submitted). By way of example, the inferred evolutionary mass of VFTS 482 from our analysis is $101\pm26~M_{\odot}$, yet Pollock et al. (submitted) infer minimum dynamical masses of 105~$M_{\odot}$ and 80~$M_{\odot}$ for the O2.5\,If/WN6 primary and O3\,V-III secondary, respectively.

 Figure~\ref{sHRD} presents a spectroscopic HR (sHR) diagram \citep{LangerKudritzki2014} of the XShootU Magellanic Cloud OB stars, where $\mathcal{L} = T^4_{\rm eff}/g$, emphasising their extreme nature. Indeed some evolved B supergiants
 in both galaxies approach the Eddington limit, with surface gravities close to the low limit of the grid at $\log g$/(cm\,s$^{-2}$) = 1.5. Spectroscopic gravities of some late B supergiants are unreliable since they lead to unphysically low spectroscopic masses, potentially owing to assumptions in our pipeline analyses. By way of example $\log g_{c}$/(cm\,s$^{-2}$) = 1.53$^{+0.38}_{-0.03}$ for Sk --66$^{\circ}$ 50 (B7 Ia$^{+}$) implies $M_{\rm spec} = 11.1^{+14.2}_{-1.5}~M_{\odot}$, versus $M_{\rm evol} = 24^{+2}_{-2}~M_{\odot}$ from inspection of \citet{Schootemeijer+2019} tracks. 

\begin{figure}[htbp]
\centering
  \includegraphics[width=9 cm]{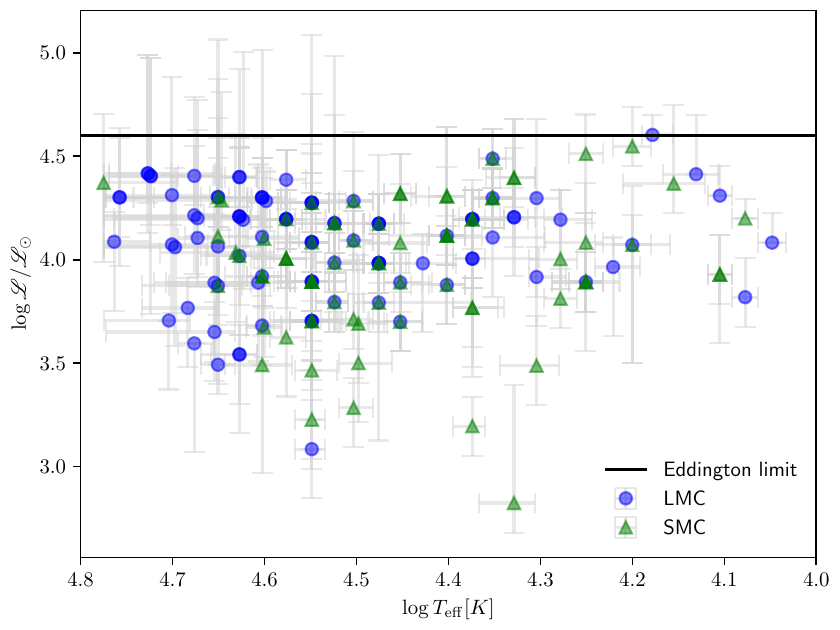}
  \caption{Pipeline results (LMC: blue circles, SMC: green triangles) presented in a spectroscopic HR diagram, with $T_{\rm eff}$ in K, together with the Eddington limit for $X$=73\%, at $\log \mathcal{L}/\mathcal{L}_{\odot} = 4.6$ (solid line).}
  \label{sHRD}
\end{figure}

\subsection{Ionizing photon rates}

Results presented in electronic form include
hydrogen ($\lambda \leq$ 912\AA, $Q_{\rm H}$) and helium ($\lambda \leq$ 504\AA, $Q_{\rm He I}$) ionizing photon rates (s$^{-1}$). As expected, the hottest stars (early O-type) dominate the Lyman  and especially the helium continuum ionizing production, as shown in Fig.~\ref{ion}. Very massive main sequence WN and classical Wolf-Rayet stars  can also play a major role \citep{Doran+2013, Crowther+2022}, although these were not included in this analysis, as discussed earlier. Ionizing properties of XShootU OB stars are relevant for studies of H\,{\sc ii} regions in the Magellanic Clouds, whose spectroscopic properties are being characterized by the SDSS-V Local Volume Mapper \citep{LVM}.

Ionized helium ($\lambda \leq$ 228\AA, $Q_{\rm He II}$) outputs are not included since these are modest for O stars and are strongly dependent on X-ray production. Indeed, relatively few XShootU targets have been detected in X-rays. The majority of these either host a compact object - examples include LMC X-4 \citep{Hutchings+1978} and SMC X-1 \citep{Reynolds+1993} - or are known colliding wind binaries \citep[HD~5980,][]{Naze+2002}. Several XShootU targets in the Tarantula Nebula have been detected in the deep T-ReX survey \citep{Crowther+2022, T-ReX}, but detection of the bulk of the OB population in the Magellanic Clouds awaits a high spatial resolution, high throughout X-ray mission.

\begin{figure}[htbp]
\centering
  \includegraphics[width=9 cm]{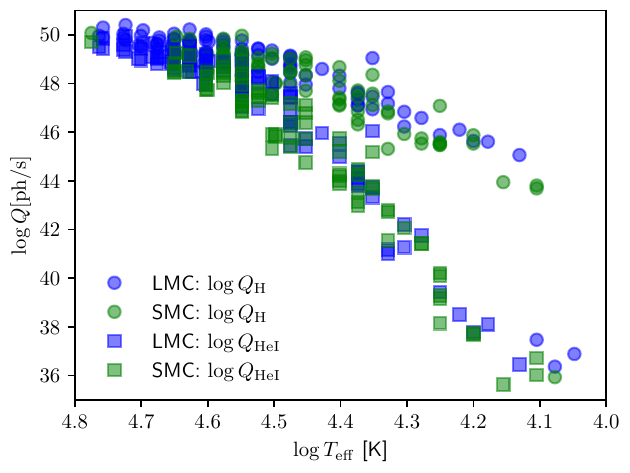}
  \caption{Hydrogen (circles) and neutral helium (squares) ionizing photon rates (s$^{-1}$) of LMC (blue) and SMC (green) OB stars, versus $T_{\rm eff}$ in K.}
  \label{ion}
\end{figure}

\subsection{Rotational velocities}

The distribution of rotational velocities for XShootU stars are presented in Fig.~\ref{vsini}. For the LMC (SMC) stars, average values are $\varv_{\rm e} \sin i = 121 \pm 61 (108\pm 75)$ km\,s$^{-1}$ for O stars and $64 \pm 41 (62 \pm 50)$ km\,s$^{-1}$ for B stars. 
There are relatively few fast rotators amongst the XShootU sample, with only 10\% of OB stars
 $\varv_{\rm e} \sin i \geq 200$ km\,s$^{-1}$, and only 1\% above $\varv_{\rm e} \sin i \geq 300$ km\,s$^{-1}$. Stars with the highest rotational velocities ($\sim300$ km\,s$^{-1}$) are flagged as such via `n' or `nn' spectral classifications, and are rarer than from unbiased surveys such as VFTS \citep{Evans+2011} or BLOeM \citep{Shenar+2024}. As anticipated, most XShootU B supergiants in both galaxies are slow rotators.  

\begin{figure}[htbp]
\centering
  \includegraphics[width=9 cm]{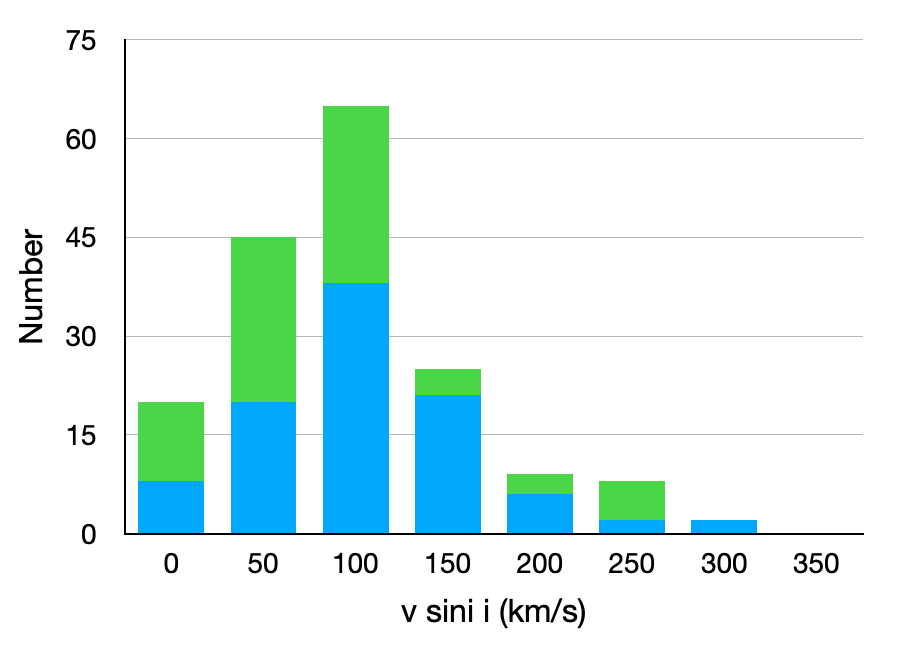}
  \caption{Histogram of projected rotational velocities $\varv_{\rm e} \sin  i$ (km\,s$^{-1}$) for LMC (blue) and SMC (green) OB stars, in 50 km\,s$^{-1}$ bins aside from 0 bin i.e. 50 refers to 50$\pm$25 km\,s$^{-1}$.  There are few rapid rotators amongst the XShootU sample.}
  \label{vsini}
\end{figure}

\subsection{Elemental abundances}

Helium is our primary focus regarding elemental abundances in XShootU OB stars.  The baseline He abundance from H\,{\sc ii} regions \citep{RussellDopita1990} is $Y \sim$ 25\%, whereas our grid permits lower helium mass fractions to avoid a truncated PDF. Such results should be viewed with caution. In contrast, high He mass fractions for a significant subset of OB supergiants are more credible. Indeed, the upper limit of $Y$ = 55\% from
our grid is obtained for some B supergiants, although such results are sensitive to the robustness of inferred $T_{\rm eff}, \log g$. 

Fig.\ref{Y} compares projected rotational velocities to $Y$, surface Helium mass fractions. It is apparent that there is no clear relationship between these properties, suggesting that processes other than rotation 
may dominate surface He enrichment. However, one should bear in mind that mixing
depends on several parameters, and the present sample is deliberately heterogeneous.





\begin{figure}[htbp]
\centering
  \includegraphics[width=9 cm]{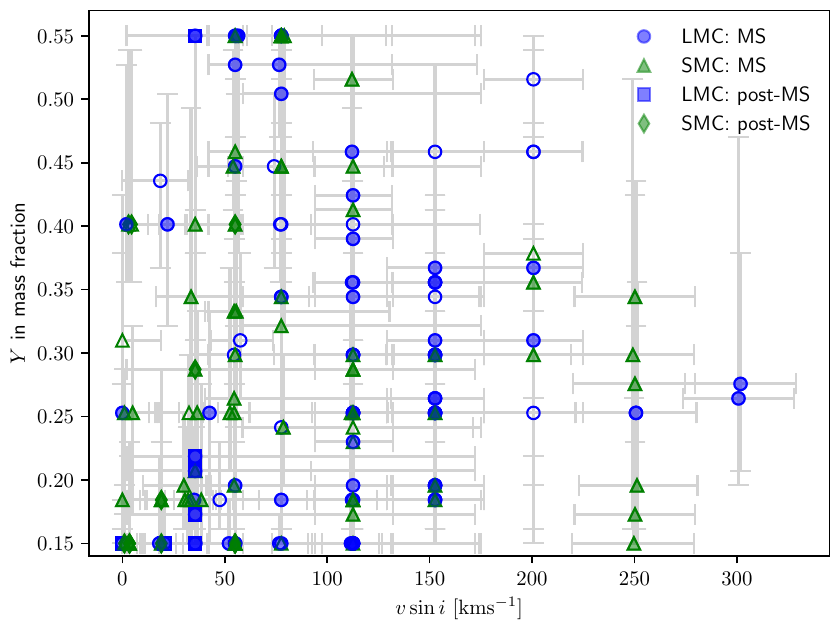}
  \caption{Projected $\varv_{e} \sin i$ (km\,s$^{-1})$ versus helium mass fractions for LMC (blue) and SMC (green) main sequence and post-main sequence stars.}
  \label{Y}
\end{figure}

\citet{Martins+2024} undertake a detailed study of CNO abundances in non-supergiant XShootU OB stars. We focus on estimating nitrogen in early O stars, whose baseline values are $\log$ (N/H) + 12 = 7.03 in the LMC and 6.66 in the SMC \citep{Vink+2023} by number.
For the LMC, nitrogen abundances range from baseline values to moderate enrichment in some (super)giants to fully processed nitrogen for some early O (super)giants and Of/WN stars. For the
SMC, modest enrichment is obtained for some early O dwarfs, with nitrogen fully processed in a few instances (e.g. NGC 346 MPG 355, ON2\,III(f$^{\ast}$)).

The overwhelming majority of main sequence evolutionary models \citep{Brott+2011} predict no surface He or N enrichment for the optimised {\sc bonnsai} parameters ($T_{\rm eff}, \log L/L_{\odot}, \varv_{\rm e} \sin i$). These are in clear tension with the preferred He and N mass fractions, indicating deficiencies in current evolutionary models for single stars or other processes at play (e.g. close binary evolution). 
By way of example, our analysis of the high mass LMC supergiant VFTS 545 (O2\,If$^{\ast}$/WN5, Mk~35) reveals considerable helium ($Y = 0.40^{+0.13}_{-0.05}$) and nitrogen ($\log$N/H + 12 = 8.7$\pm$0.2) enrichment, in spite of a negligible projected rotational velocity. Meanwhile the {\sc bonnsai} solution that is consistent with spectroscopic parameters ($T_{\rm eff}, \log L/L_{\odot}, \varv_{\rm e} \sin i$) predicts $Y = 0.25^{+0.05}_{-0.05}$ and $\log$N/H + 12 = 6.91$^{+0.10}_{-0.02}$. Similarly, the pipeline solution for the SMC B2 supergiant AzV 18 (Sk 13) favours the maximum helium enrichment ($Y = 0.55^{+0.00}_{-0.03}$) whereas the {\sc bonnsai} solution for spectroscopic parameters predicts no He enrichment $Y=0.25^{+0.01}_{-0.00}$.



\section{Wind properties of Magellanic Cloud OB stars}\label{wind-properties}

The primary focus of the present study is on the physical parameters
of Magellanic Cloud OB stars, rather than their wind properties for which ultraviolet (ULLYSES) datasets are crucial, especially for stars with relatively weak winds \citep{Vink+2023}. Nevertheless, our diagnostics include H$\alpha$, our grid spans a wide range of wind densities, such that estimates of mass-loss rates can be made, at least for those stars with relatively high wind densities $\log Q \geq -13.2$. 

\subsection{Wind velocities}\label{vely}

\begin{figure}
\centering
  \includegraphics[width=9cm]{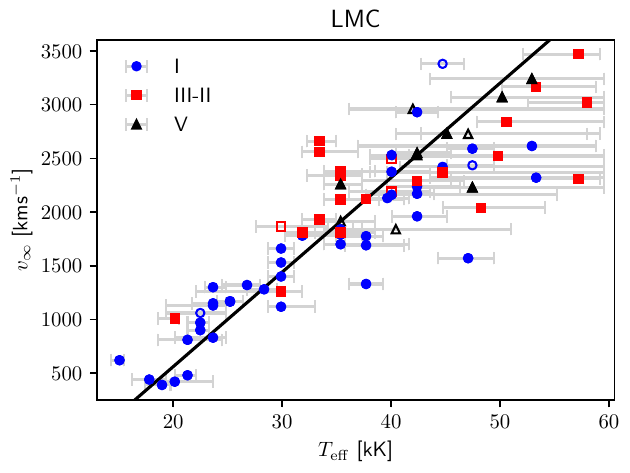}
  \includegraphics[width=9cm]{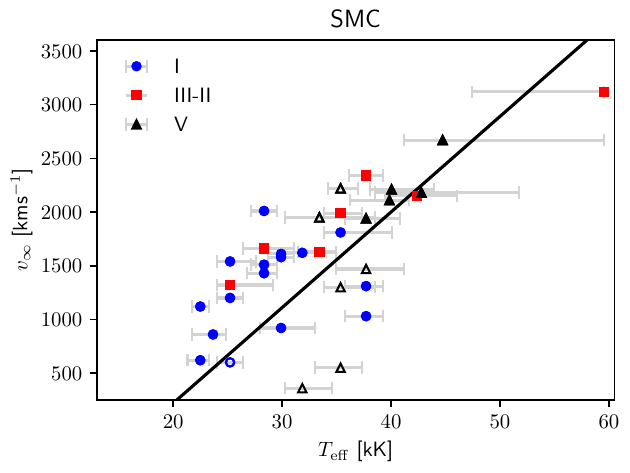}
  \caption{Upper panel: Comparison between SEI (quality i or ii filled symbols; quality iii open symbols) or $\varv_{\rm black}$ (open symbols)-derived wind velocities (km\,s$^{-1}$) and pipeline temperatures (kK) for LMC OB dwarfs (black circles), giants (red squares) and supergiants (blue triangles), together with the LMC $\varv_{\infty}-T_{\rm eff}$ calibration from \citet{Hawcroft+2023}. Lower panel: As above for SMC OB stars.}
  \label{v_inf}
\end{figure}


The metallicity-dependent $\varv_{\infty}$-$T_{\rm eff}$ calibration \citep{Hawcroft+2023} adopted for our grid-based pipeline analysis is based upon SEI-derived terminal wind velocities from ULLYSES observations of C\,{\sc iv} $\lambda\lambda$1548-51 \citep{Hawcroft+2023} plus literature-based temperatures. SEI-derived velocities from strong wind profiles (quality flags i or ii as defined by \citet{Hawcroft+2023}) should be fairly robust despite adopting a uniform Doppler shift of 146 km\,s$^{-1}$ (262 km\,s$^{-1}$) for SMC (LMC) stars. For stars lacking an SEI measurement, or for which SEI quality flags were poor (weak or negligible wind signature), we determine wind velocities from either $v_{\rm black}$ \citep{Prinja+1990} or $v_{\infty} \simeq 0.85 v_{\rm edge}$ \citep{PrinjaCrowther1998, Crowther+2016} using C\,{\sc iv} $\lambda$1548, N\,{\sc v} $\lambda$1238 or Si\,{\sc iv} $\lambda$1394, after applying individual radial velocities from our study. These are presented in electronic data tables (see Sect.~\ref{data_availability}). 

We reassess the $\varv_{\infty}$-$T_{\rm eff}$ calibration of \citet{Hawcroft+2023}, specifically the combined literature sample for each galaxy used in their Table~1, by substituting literature temperatures for ULLYSES stars for pipeline results. This is presented in Fig.~\ref{v_inf} for LMC (upper panel) and SMC (lower panel) OB stars (quality flags i-ii are filled symbols, open symbols otherwise). 

For LMC OB stars, the calibration provides reasonable wind velocities, though highlights the large temperature uncertainties of early O stars discussed in Sect.~\ref{temp}). For SMC OB stars, wind velocities are generally underpredicted, especially for cooler OB supergiants. \citet{Hawcroft+2023} included three B supergiants with reliable wind velocities for which $T_{\rm eff} <$ 25 kK, with a mean $\varv_{\infty}$ = 690 km\,s$^{-1}$. Two of 
these supergiants  are in common with our sample (AzV~78 is not an original ULLYSES target), to which we add AzV~242, resulting in a higher mean $\varv_{\infty}$ = 867 km\,s$^{-1}$ (Fig.~\ref{v_inf}), favouring a somewhat shallower gradient. \citet{Telford+2024} provide terminal wind velocities for O stars in other metal-poor galaxies.

For our subsequent analysis, including determination of mass-loss rates from wind densities $Q$, we either select individual values from \citet{Hawcroft+2023} or measure wind velocities from ULLYSES spectroscopy. Indeed, a subset of ULLYSES targets were observed after the study of \citet{Hawcroft+2023}, for which additional determinations are made, following an identical procedure.

\begin{figure}[htbp]
\centering
  \includegraphics[width=9 cm]{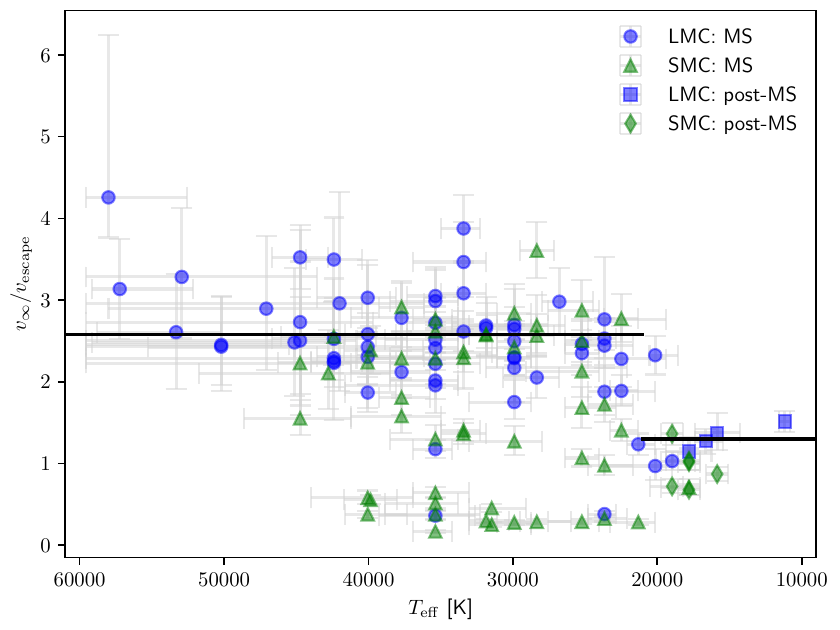}
  \caption{Temperature dependence of $v_{\infty}/v_{\rm esc}$ for LMC (SMC) main sequence stars shown as blue circles (green triangles) and post-main sequence stars shown as blue squares (green diamonds), together with the \citet{Lamers+1995} relation for Milky Way stars. $v_{\infty}/v_{\rm esc} \ll 1$ for a sizeable subset of (mostly) SMC OB stars.}
  \label{v_esc}
\end{figure}


In a few instances, significantly higher velocities are obtained from $v_{\rm edge}$ of N\,{\sc v} $\lambda$1238 versus an SEI fit to C\,{\sc iv} $\lambda$1548. 
These predominantly possess modest wind densities (e.g. AzV 377), so do not impact on wind momenta relations presented in Section.~\ref{mass-loss}. 
Wind velocities are generally similiar to previous determinations \citep[e.g.][]{Bouret+2013}, although in some cases previous work adopts standard $v_{\infty}-v_{\rm esc}$ scaling relations \citep{Lamers+1995}.

Figure~\ref{v_esc} presents the temperature dependence of $v_{\infty}/v_{\rm esc}$ for LMC (blue) and SMC (green) OB stars together with the \citet{Lamers+1995} relation for Milky Way stars. $v_{\rm esc}$ is determined from evolutionary masses, via
\[ v_{\rm esc} = (2 G M_{\rm evol} (1-\Gamma_{e})/R_{\ast})^{0.5},\]
where $\Gamma_{e}$ is the Eddington parameter \citep[e.g.][]{Groenewegen+1989}. Sk --69$^{\circ}$ 212 (O4.5\,If) is excluded from this diagram since $\Gamma_{e} \sim 1$. We favour evolutionary masses over spectroscopic masses owing to unrealistic values of the latter in some instances (recall Sect.~\ref{masses}). The \citet{Lamers+1995} relation is a reasonable approximation for LMC OB stars, but lower $v_{\infty}/v_{\rm esc}$ ratios are obtained for SMC stars, with a sizeable subset of late O and early B stars revealing $v_{\infty}/v_{\rm esc} \ll 1$. Wind velocities are difficult to measure for the majority of SMC stars, but there is no UV spectroscopic evidence of fast winds for many OB stars from the ULLYSES sample.

\begin{figure}[htbp]
\centering
  \includegraphics[width=9 cm]{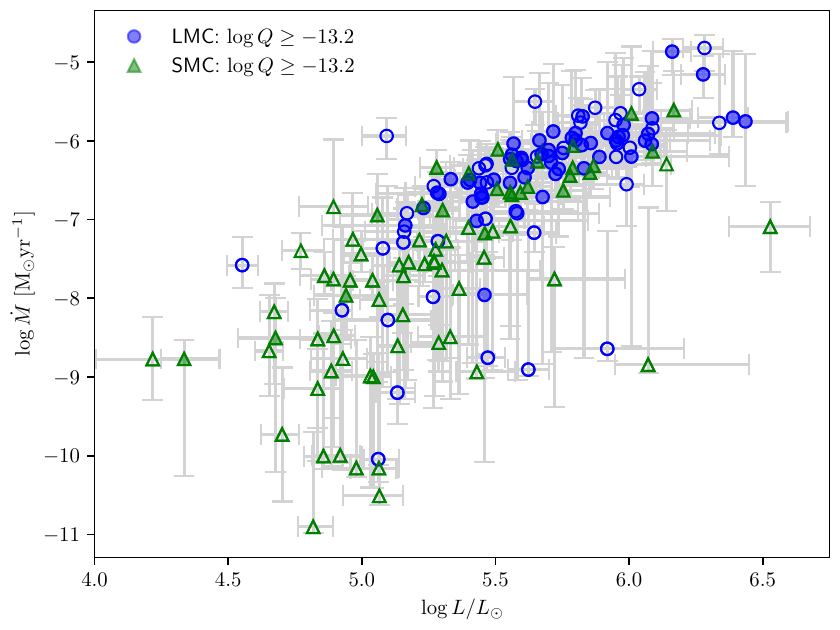}
  \caption{Inferred mass-loss rates of LMC (blue circles) and SMC (green triangles) OB stars versus luminosity, in which stars with weak winds ($\log Q \leq$  --13.2) or uncertain stellar parameters are indicated with open symbols.}
  \label{Mdot}
\end{figure}

\subsection{Mass-loss rates}\label{mass-loss}

Empirical wind velocities allow us to estimate mass-loss rates of OB stars from our pipeline wind densities. It is widely accepted that winds from hot luminous stars are clumped \citep{Owocki+1988,Hillier1991,Sundqvist+2011}. To estimate mass-loss rates we therefore adopt a clumping factor of 10} or volume filling factor of $f_{v}$ = 0.1 \citep{Crowther+2002, Evans+2006}, i.e. the derived rates are a factor of $\sqrt{10}$ lower than they would be with a smooth wind assumption. 

Figure~\ref{Mdot} presents inferred mass-loss rates of XShootU OB stars in the LMC and SMC. Filled symbols indicate stars with strong winds ($\log Q \geq -$13.2) while open symbols indicate stars with either weak winds or H$\alpha$ fits affected by nebular emission. Inferred mass-loss rates of stars with strong winds typically lie in the range --7 $\leq \log \dot{M}/(M_{\odot}$ yr$^{-1}) \leq$ --5.5. Overall OB stars in the LMC exhibit somewhat higher mass-loss rates than their SMC counterparts, many of which have low wind densities (open symbols).


\begin{table}[htbp]
\begin{center}
\caption{Comparison between fits to wind momenta of SMC and LMC OB stars according to \citet{Mokiem+2007} using optical spectroscopy with wind velocities largely estimated from $v_{\rm esc}$ scaling relations, plus
this study, for OB stars with dense winds ($\log Q \geq -13.2)$, plus empirical wind velocities.}\label{D_mom}
    \begin{tabular}{l@{\hspace{2mm}}l@{\hspace{2mm}}l@{\hspace{2mm}}l@{\hspace{2mm}}l}
        \hline
    $\log D_{0}$ & $x$ & $N$ & Sample & Ref \\ 
    \hline\hline
    \multicolumn{5}{c}{SMC} \\
    18.20$\pm$1.09 & 1.84$\pm$0.19 & 28 & O2--B1  & \citet{Mokiem+2007}\\
     14.78$\pm$1.33 & 2.38$\pm$0.24 & 26 & OB stars  & This study\\
    \multicolumn{5}{c}{LMC} \\
     17.88$\pm$0.91 & 1.96$\pm$0.16 &38 & O2--B0.7  & \citet{Mokiem+2007}\\
     15.29$\pm$1.14 & 2.32$\pm$0.20 & 58 & OB stars  & This study\\
    \hline
    \end{tabular}
    \end{center}
\end{table}

\begin{figure}[h!]
\centering
  \includegraphics[width=9 cm]{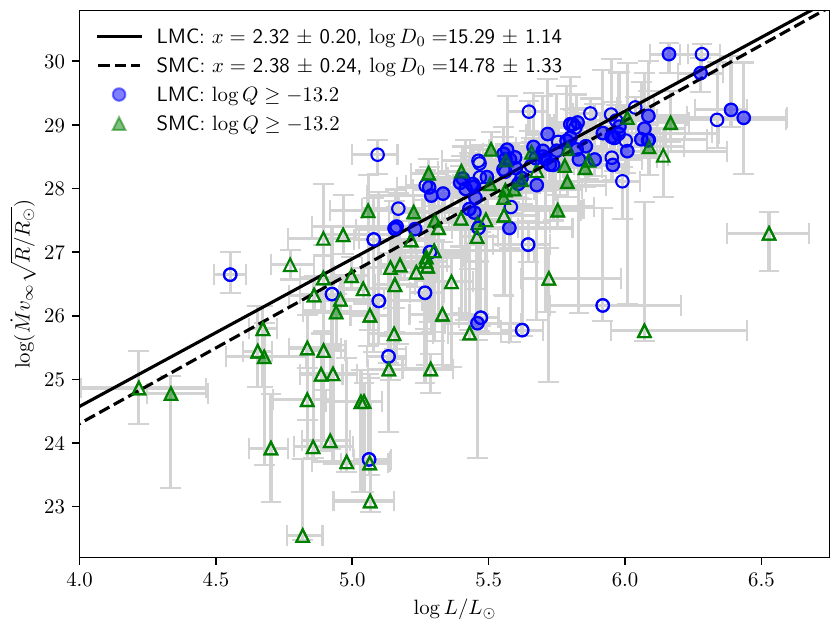}
  \includegraphics[width=9 cm]{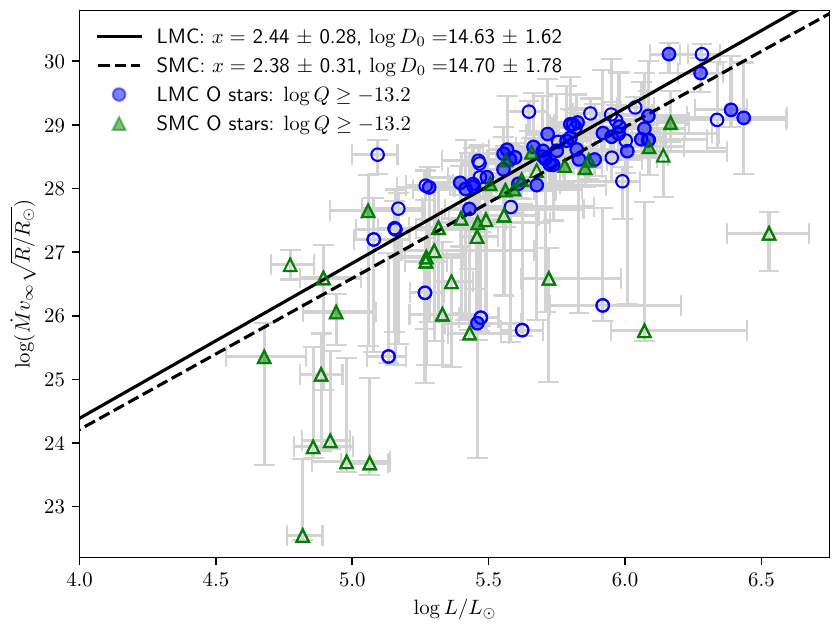}
  \includegraphics[width=9 cm]{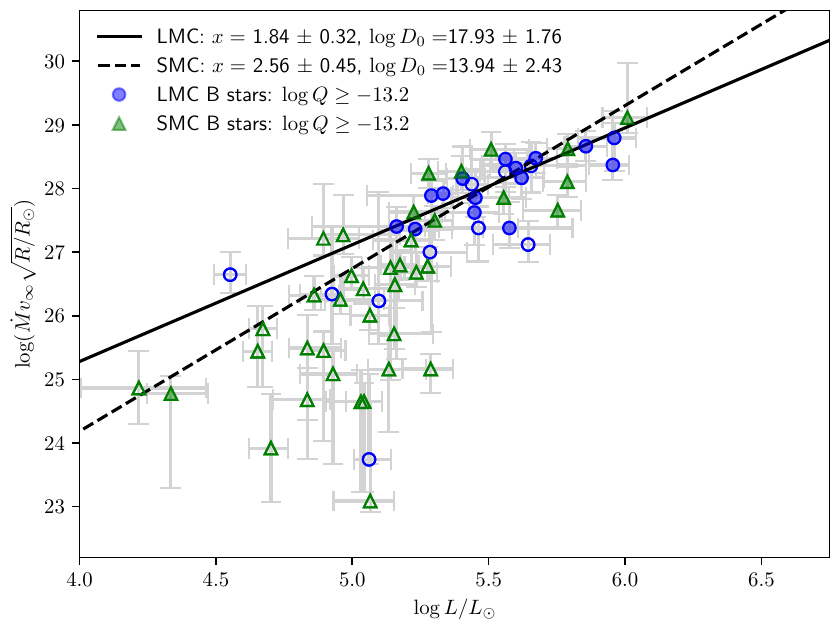}
  \caption{Upper panel: (Reduced) wind momentum versus luminosity results for LMC (blue circles) and SMC (green triangles) OB stars. Wind velocities are obtained from the calibration of \citet{Hawcroft+2023} and stars with weak winds ($\log Q \leq$  --13.2) or uncertain stellar parameters are indicated with open symbols. Middle panel: As above for O stars. Lower panel: As above for B stars.}
  \label{mom}
\end{figure}

Wind properties of OB stars in different environments are most directly compared via their (reduced) wind momenta, 
\[
D_{\rm mom} = \dot{M} v_{\infty} (R_{\ast}/R_{\odot})^{0.5},
\]
with units of g\,cm\,s$^{-2}$ \citep{Kudritzki+1999}. Figure~\ref{mom} shows a fit to wind momenta versus stellar luminosities for XShootU OB stars (upper panel), O stars (middle panel) and B stars (lower panel), excluding  stars whose parameters are not considered reliable (e.g. Sk --68$^{\circ}$ 135). A significant subset of the SMC sample possess relatively weak winds ($\log Q \leq$ --13.2), as indicated with open symbols in Fig.~\ref{mom}.

If we limit our sample to stars with wind densities of $\log Q \geq$ --13.2, plus individual wind velocities (presented in electronic data tables), we can parameterize their wind momenta as follows
\[
\log D_{\rm mom} = \log D_{0} + x \log (L/L_{\odot}).
\]
Fits are presented in Fig.~\ref{mom}. We find, on average, $\sim$0.27 dex higher wind momenta for O stars in the LMC with respect to the SMC for 5.5 $\leq \log (L/L_{\odot}) \leq 6.0$, as indicated in Fig.~\ref{mom}.

Wind velocities scale with $Z^{0.22\pm 0.03}$ according to \citet{Hawcroft+2023} suggesting $\dot{M} \propto Z^{0.5}$ for luminous O stars. Our results suggest a somewhat weaker metallicity dependence of mass-loss rates than those predicted by \citet{Vink+2001}, for which $\dot{M} \propto Z^{0.7}$ is widely used in evolutionary calculations \citep[see also][]{Bjorklund+2023}, though support updated Monte Carlo models \citep{VinkSander2021}.

Observationally, \citet{Mokiem+2007} have determined wind properties of Magellanic Cloud OB stars from analysis of visual spectroscopy, including H$\alpha$. They obtained $\dot{M} \propto Z^{0.83 \pm 0.16}$ by adopting a previous theoretical relation between wind velocity and metallicity $v_{\infty} \propto Z^{0.13}$ \citep{Leitherer+1992}. Use of $v_{\infty} \propto Z^{0.22}$ from \citet{Hawcroft+2023} would lead to a revised metallicity dependence of $\dot{M} \propto Z^{0.74}$, in close agreement with theory, albeit a stronger dependence than from our results. \citet{Mokiem+2007} included O stars with somewhat lower luminosities which are known to possess very weak winds at late O spectral types \citep{Marcolino+2009, Rickard+2022}, considered wind clumping solely for H$\alpha$ emission line stars, and crucially relied on {\it indirect} wind velocity indicators in most instances (e.g. $v_{\infty}/v_{\rm esc}$ relations). Table~\ref{D_mom} compares wind momenta coefficients from \citet{Mokiem+2007} with our results.


We separately consider B stars in the lower panel of Fig.~\ref{mom}, most of which are supergiants given our $\log Q \geq -13.2$ wind density criteria. 
However, considering our modest sample and reliance on H$\alpha$, we defer to quantitative studies which include UV diagnostics \citep{Bouret+2021, BerniniPeron+2024}, which are more suitable for a comprehensive comparison with theoretical predictions.

\section{Comparison to physical and wind properties from the literature}\label{literature}

Previous quantitative studies have included large samples of O stars \citep[e.g.][]{Sabin-Sanjulian+2017, Ramirez-Agudelo+2017} but our study is unprecedented in its breadth (early O to late B), access to UV spectroscopy from ULLYSES, and demonstrates the capabilities of the pipeline \citep{Bestenlehner2024} applied  to large grids of synthetic spectra computed with {\sc fastwind} \citep{Puls+2005, Rivero-Gonzalez+2012}.

Many of the XShootU sample stars were subject to earlier quantitative spectral analysis efforts using via optical or optical+UV spectroscopy. Since our analysis is restricted to optical diagnostics we compare our derived physical parameters to representative literature results in Table~\ref{lit}. Typically, previous studies  employed {\sc fastwind} with a Genetic Algorithm (GA) approach \citep{Mokiem+2006-SMC, Mokiem+2007-LMC} although some analyses involved fitting lines by-eye \citep{Trundle+2004, Massey+2009} or {\sc PoWR} \citep{Ramachandran+2018b}. The majority of Magellanic Cloud studies of OB stars with {\sc cmfgen} \citep{Hillier-Miller1998} involve fits to UV and optical diagnostics \citep[e.g.][]{Hillier+2003, Evans+2004-cmfgen}. \citet{Massey+2013} compare properties of OB stars obtained with {\sc cmfgen} and {\sc fastwind}. 

Overall temperatures, gravities and luminosities obtained from our pipeline agree with detailed studies within the uncertainties, at least for the targets listed. We consider our errors to be more realistic given the number of parameters involved. Our estimated mass-loss rates are however less reliable since H$\alpha$ is a poor diagnostic of stars with weak winds. Literature stellar parameters for a large subset of ULLYSES targets are provided in Table~B.2 of \citet{Vink+2023}.

Recently, a subset of ULLYSES/XShootU targets have been the subject of detailed UV and optical quantitative studies. LMC studies have included {\sc fastwind} GA analyses of OB supergiants by \citet{Verhamme+2024} and O stars by Brands et al. (submited), plus {\sc PoWR} analyses of O stars by \citet{GomezGonzalez+2024}, while SMC studies have included {\sc fastwind} GA analyses of O stars by \citet{Backs+2024} and {\sc cmfgen} studies of B supergiants by \citep{BerniniPeron+2024}.

\begin{figure}[h!]
\centering
  \includegraphics[width=9 cm]{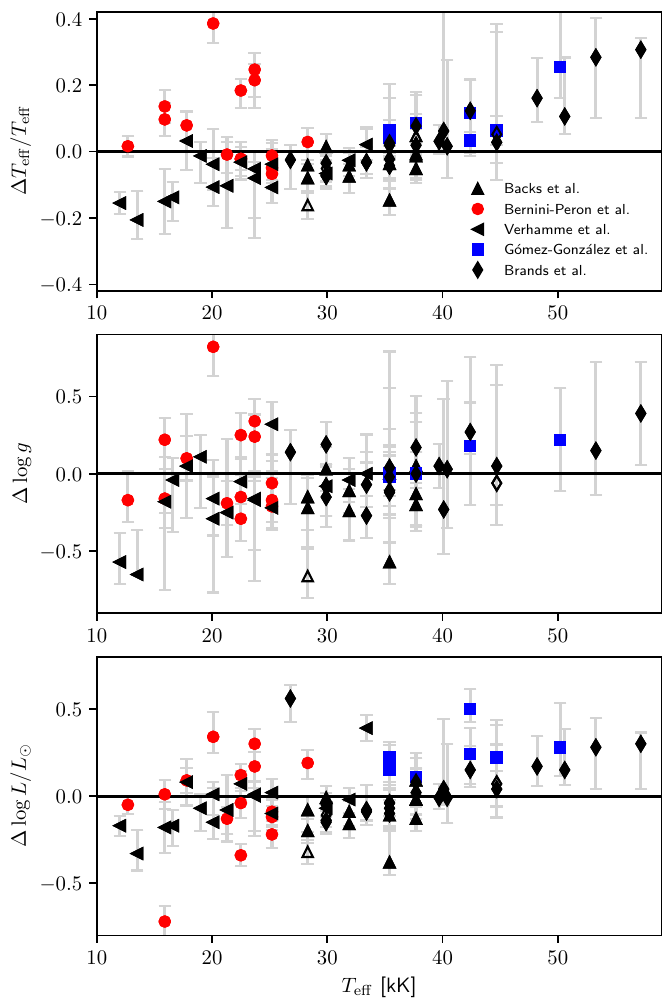}
  \caption{Upper panel: Comparison between pipeline-derived $T_{\rm eff}$ and $\Delta T_{\rm eff}/T_{\rm eff}$ from detailed XShootU analyses of OB stars using {\sc fastwind} (black), {\sc cmfgen} (red) and {\sc PoWR} (blue). Positive values indicate higher quantities from pipeline. Open symbols are suspected binaries according to \citet{Backs+2024} and Brands et al. (submitted). Middle and lower panels: As above for $\Delta \log g$ and $\Delta \log L/L_{\odot}$.} 
  \label{XShootU}
\end{figure}

The upper panel of Figure~\ref{XShootU} compares pipeline-derived temperatures $T_{\rm eff}$ with detailed results $\Delta T_{\rm eff} = [T_{\rm eff}$ (pipeline) - $T_{\rm eff}$ (study)]/$T_{\rm eff}$ (study). Satisfactory results are obtained for the majority of stars in common, albeit with some scatter. Nevertheless, there are several early O stars in the LMC for which significantly higher (and highly uncertain) temperatures are obtained with respect to Brands et al. (submitted). In addition, pipeline-derived temperatures of some SMC mid B supergiants are also significantly higher than \citet{BerniniPeron+2024}. Open symbols indicate suspected binaries according to \citet{Backs+2024} and Brands et al. (submitted).

The middle panel of Figure~\ref{XShootU} compares pipeline-derived surface gravities with other XShootU studies, uncorrected for centrifugal effects, where $\Delta \log g = \log g$ (pipeline) - $\log g$ (study). Overall agreement is generally good, albeit with large uncertainties in a number of instances. The lower panel of Figure~\ref{XShootU} compares pipeline-derived luminosities with other XShootU studies, where $\Delta \log L/L_{\odot} = \log L/L_{\odot}$ (pipeline) - $\log L/L_{\odot}$ (study). Once again, agreement is satisfactory, aside from a few outliers. Our results support the non-negligible scatter in results discussed in \citet{Sander+2024} for three XShootU O stars analysed using various techniques.

\begin{table*}[ht]
\caption{Comparison of physical and wind parameters of selected XShootU OB stars from this study with representative  literature results based on {\it optical} diagnostics. All analyses involve {\sc fastwind} \citep{Puls+2005, Rivero-Gonzalez+2012} with the exception of Sk --71$^{\circ}$ 41 which used {\sc PoWR} \citep{Grafener+2002, Sander+2015}. Mass-loss rates incorporate volume filling factors, $f_{v}$ to reflect differences in clumping factors. Wind velocities for our pipeline are taken from individual SEI fits of \citet{Hawcroft+2023} or measured via $\varv_{\rm black}$ or $\varv_{\rm edge}$ (see electronic data tables). 
}
\label{lit}
\begin{center}
    \begin{tabular}{l@{\hspace{2mm}}l@{\hspace{2mm}}l@{\hspace{2mm}}l@{\hspace{2mm}}l@{\hspace{2mm}}l@{\hspace{2mm}}c@{\hspace{2mm}}l@{\hspace{2mm}}c@{\hspace{2mm}}c@{\hspace{2mm}}l}
    \hline
    Star & Spect.& $T_{\rm eff}$ & $\log g$     & $\log L$   & $\log \dot{M}/\sqrt{f_{v}}$ & $\varv_{\infty}$ & $Y$  & $\varv_{\rm e} \sin i$ & Fitting & Ref\\
         & Type  &     kK        & cm\,s$^{-2}$ & $L_{\odot}$ & $M_{\odot}$\,yr$^{-1}$ & km\,s$^{-1}$ & $\cdots$ & km\,s$^{-1}$ & Tool \\ [2pt]
    \hline\hline
    \multicolumn{11}{c}{\bf SMC}\\
     AzV 435 & O3\,V((f$^{\ast}$))z+? & 46.0 & 3.90
 & 5.88  &$-6.7$ & 1500 & 0.28  & 110 & by-eye (N) & RPM12 \\ [2pt]
    $\cdots$ & $\cdots$ & $44.3^{+15.2}_{-3.5}$ & $3.69^{+0.52}_{-0.14}$ & $6.07^{+0.42}_{-0.14}$ & $-8.3^{+2.0}_{-0.0}$ & 1490: & $0.24^{+0.07}_{-0.07}$ & 113$^{+20}_{-19}$ & Pipeline    & This work \\ [4pt] 
    AzV 207 & O7\,V((f))z & 37.0 & 3.7
 & 5.34  & --7.0 & 2000: & 0.28 & $\cdots$ & by-eye (He) & MBK04 \\ [2pt]
    $\cdots$ & $\cdots$ & $37.7^{+3.5}_{-2.7}$ & $3.69^{+0.52}_{-0.24}$ & $5.27^{+0.16}_{-0.13}$ & $-7.0^{+0.5}_{-1.7}$ & 1470 & $0.25^{+0.14}_{-0.08}$ & 113$^{+20}_{-19}$ &Pipeline    & This work \\ [4pt] 
    AzV 232 & O7\,Iaf$^{+}$ & $34.1^{+0.6}_{-0.6}$ & $3.35^{+0.17}_{-0.12}$
 & $6.02^{+0.06}_{-0.06}$  &$-5.2^{+0.1}_{-0.1}$ & 1330 & $0.48^{+0.06}_{-0.03}$ & 74 & GA (He) & MKE06 \\ [2pt]
    $\cdots$ & $\cdots$ & $37.7^{+1.6}_{-2.0}$ & $3.69^{+0.24}_{-0.14}$ & $6.17^{+0.09}_{-0.10}$ & $-5.1^{+0.5}_{-0.3}$ & 1310 & $0.25^{+0.11}_{-0.05}$ & 54$^{+74}_{-18}$ & Pipeline    & This work \\ [4pt] 
    NGC346 ELS 25 & O9.2\,V(n) & $36.2^{+1.2}_{-0.8}$ & $4.07^{+0.24}_{-0.08}$
 & $4.90^{+0.08}_{-0.08}$  &$-9.5^{+0.4}_{-1.4}$ & 600: & $0.28^{+0.06}_{-0.04}$ & 138 & GA (He) & MKE06 \\ [2pt]
 $\cdots$ & $\cdots$ & $35.4^{+1.6}_{-2.3}$ & $4.12^{+0.14}_{-0.48}$ & $4.92^{+0.09}_{-0.12}$ & $-9.5^{+1.4}_{-0.0}$ &  610:    & $0.30^{+0.06}_{-0.13}$ & 153$^{+24}_{-24}$ & Pipeline    & This work \\ [4pt] 
 AzV 372 & O9.2\,Iab & $31.0^{+0.7}_{-1.2}$ & $3.19^{+0.16}_{-0.17}$
 & $5.83^{+0.09}_{-0.09}$  &$-5.7^{+0.1}_{-0.1}$ & 1550 & $0.30^{+0.07}_{-0.06}$ & 135 & GA (He) & MKE06 \\ [2pt]
 $\cdots$ & $\cdots$ & $28.3^{+1.2}_{-1.2}$ & $2.88^{+0.19}_{-0.14}$ & $5.56^{+0.09}_{-0.09}$ & $-5.7^{+0.2}_{-0.2}$ & 1510 & $0.29^{+0.06}_{-0.07}$ & 113$^{+20}_{-19}$ & Pipeline    & This work \\ [4pt] 
    NGC346 ELS 26 & B0\,III & $32.6^{+0.4}_{-1.2}$ & $3.76^{+0.05}_{-0.17}$
 & $4.93^{+0.09}_{-0.09}$  &$-7.3^{+0.3}_{-1.3}$ & (2210) & $0.30^{+0.05}_{-0.02}$ & 67 & GA (He) & MKE06 \\ [2pt]
 $\cdots$ & $\cdots$ & $31.5^{+1.6}_{-2.3}$ & $3.69^{+0.14}_{-0.29}$ & $4.93^{+0.10}_{-0.13}$ & $-8.3^{+0.4}_{-1.4}$ & 360  & $0.26^{+0.09}_{-0.08}$ & 55$^{+77}_{-14}$ & Pipeline    & This work \\ [4pt] 
    AzV 215 & B0\,Ia & $27.0^{+1.0}_{-1.0}$ & $2.90^{+0.10}_{-0.10}$
 & $5.63$  &$-5.9^{+0.1}_{-0.1}$ & 1400 & $\cdots$ & $\cdots$ & by-eye (Si, He) & TLP04 \\ [2pt]
 $\cdots$ & $\cdots$ & $25.2^{+2.3}_{-1.2}$ & $2.69^{+0.33}_{-0.14}$ & $5.51^{+0.16}_{-0.09}$ & $-5.6^{+0.2}_{-0.2}$ & 1540 & $0.55^{+0.00}_{-0.23}$ & 79$^{+19}_{-18}$ & Pipeline    & This work \\ [4pt] 
    AzV 22 & B3\,Ia & $14.5^{+1.5}_{-1.5}$ & $1.90^{+0.15}_{-0.15}$
 & $5.04$  &$-6.6^{+0.1}_{-0.1}$ & 280 & $\cdots$ & $\cdots$ & by-eye (Si) & TLP04 \\ [2pt]
 $\cdots$ & $\cdots$ & $15.9^{+0.8}_{-0.8}$ & $2.12^{+0.14}_{-0.57}$ & $5.28^{+0.10}_{-0.10}$ & $-6.9^{+0.3}_{-0.2}$ & 300   & $0.15^{+0.08}_{-0.00}$ & 4$^{+7}_{-4}$ & Pipeline    & This work \\ [4pt] 
    \multicolumn{11}{c}{\bf LMC}\\ [2pt]
     Sk --67$^{\circ}$ 22 & O2\,If$^{\ast}$/WN5 & 46.0 & 3.70
 & 5.80  &$-4.82$ & 2650 &  0.54 & 200 & by-eye (N) & RPM12 \\ [2pt]
    $\cdots$ & $\cdots$ & $47.5^{+2.0}_{-2.3}$ & $3.69^{+0.14}_{-0.14}$ & $5.65^{+0.09}_{-0.10}$ & $-5.0^{+0.1}_{-0.3}$ & 2590   & $0.31^{+0.07}_{-0.05}$ & 58$^{+16}_{-15}$ & Pipeline    & This work \\ [4pt] 
    N11 ELS 60 & O3\,V((f$^{\ast}$))& $45.7^{+2.3}_{-1.0}$ & $3.92^{+0.09}_{-0.05}$ & $5.57^{+0.10}_{-0.10}$ &$-6.3^{+0.1}_{-0.2}$ & (2738) & $0.32^{+0.05}_{-0.04}$ & 106 & GA (He) & MKE07\\ [2pt]
     $\cdots$ & $\cdots$       & $50.2^{+9.4}_{-4.7}$ & $4.12^{+0.33}_{-0.33}$ & $5.68^{+0.27}_{-0.16}$ & $-6.2^{+0.4}_{-2.1}$ & 3070 & $0.25^{+0.17}_{-0.10}$ & 0$^{+28}_{-0}$ & Pipeline    & This work \\ [4pt]
     VFTS 180 & O3\,If$^{\ast}$ & $40.5^{+0.2}_{-0.6}$ & $3.42^{+0.03}_{-0.02}$ & $5.85^{+0.02}_{-0.01}$ &$-5.0^{+0.0}_{-0.1}$ & 1927 & $0.65^{+0.03}_{-0.04}$ & 118$^{+22}_{-22}$ & GA (He+N) & RSK17\\ [2pt]
     $\cdots$ & $\cdots$       & $42.4^{+1.6}_{-2.0}$ & $3.50^{+0.14}_{-0.14}$ & $5.83^{-0.08}_{-0.09}$ & $-5.2^{+0.2}_{-0.3}$ & 2170 & $0.53^{+0.02}_{-0.06}$ & 77$^{+97}_{-20}$ & Pipeline    & This work \\ [4pt]
    Sk --67$^{\circ}$ 166 & O4\,If & $40.3^{+0.9}_{-0.8}$ & $3.65^{+0.00}_{-0.08}$ & $6.03^{+0.07}_{-0.07}$ &$-5.0^{+0.1}_{-0.0}$ & (1917) & $0.52^{+0.05}_{-0.04}$ & 97 & GA (He) & MKE07\\ [2pt]
     $\cdots$ & $\cdots$       & $37.7^{+1.6}_{-2.0}$ & $3.31^{+0.14}_{-0.19}$ & $5.81^{+0.09}_{-0.10}$ & $-5.2^{+0.2}_{-0.3}$ & 1775 & $0.40^{+0.14}_{-0.05}$ & 78$^{+97}_{-19}$ & Pipeline    & This work \\ [4pt]
     VFTS 244 & O5\,III(n)fpc & $41.1^{+0.4}_{-1.2}$ & $3.65^{+0.05}_{-0.08}$ & $5.58^{+0.01}_{-0.04}$ &$-5.6^{+0.0}_{-0.1}$ & 2123 & $0.30^{+0.02}_{-0.04}$ & 230$^{+14}_{-20}$ & GA (He) & RSK17\\ [2pt]
     $\cdots$ & $\cdots$       & $40.1^{+1.6}_{-2.0}$ & $3.50^{+0.14}_{-0.14}$ & $5.47^{-0.08}_{-0.10}$ & $-8.3^{+1.3}_{-0.1}$ & 2495:  & $0.18^{+0.05}_{-0.02}$ & 153$^{+24}_{-24}$  &Pipeline    & This work \\ [4pt]
    N11 ELS 32 & O7.5\,III(f) & $35.2^{+0.4}_{-0.7}$ & $3.45^{+0.06}_{-0.07}$ & $5.43^{+0.06}_{-0.06}$ &$-6.1^{+0.1}_{-0.1}$ & (1536) & $0.26^{+0.04}_{-0.04}$ & 96 & GA (He) & MKE07\\ [2pt]
     $\cdots$ & $\cdots$       & $35.4^{+3.5}_{-1.6}$ & $3.50^{+0.29}_{-0.29}$ & $5.45^{+0.15}_{-0.07}$ & $-6.2^{+0.4}_{-0.3}$ & 2115 & $0.15^{+0.07}_{-0.08}$ & 113$^{+20}_{-19}$ & Pipeline    & This work \\ [4pt]
     Sk --71$^{\circ}$ 41 & O9.5 Ib & 30.0               & 3.4                    &
     5.50                    & $-5.6$  & 1800 & $\cdots$ & 90 & by-eye & RHH18 \\ [2pt]
     $\cdots$ & $\cdots$ & $29.9^{+1.2}_{-1.2}$ & $3.31^{+0.14}_{-0.14}$ & $5.60^{+0.08}_{-0.08}$ & $-5.7^{+0.2}_{-0.2}$ & 1660 & $0.15^{+0.03}_{-0.00}$ & 55$^{+76}_{-13}$ & Pipeline & This work \\ [4pt]
    N11 ELS 33 & B0\,III-II(n) & $27.2^{+1.0}_{-0.9}$ & $3.21^{+0.09}_{-0.06}$ & $5.07^{+0.08}_{-0.08}$ &$-6.6^{+0.1}_{-0.3}$ & (1536) & $0.24^{+0.06}_{-0.02}$ & 256 & GA (He) & MKE07\\ [2pt]
     $\cdots$ & $\cdots$       & $29.9^{+2.0}_{-7.8}$ & $3.50^{+0.29}_{-0.67}$ & $5.16^{+0.12}_{-0.37}$ & $-6.6^{+0.3}_{-2.0}$ & 1260    & $0.37^{+0.18}_{-0.22}$ & 201$^{+24}_{-24}$ & Pipeline    & This work \\ [4pt]
     Sk --68$^{\circ}$ 41 & B0.5\,Ia & $24.5$ & $2.90$ & $5.56$ &$-6.05$ & 865 & $0.28$ & 150 & by-eye (Si, He) & MZM09\\ [2pt]
     $\cdots$ & $\cdots$       & $25.2^{+1.2}_{-1.2}$ & $3.12^{+0.14}_{-0.19}$ & $5.67^{+0.09}_{-0.09}$ & $-5.7^{+0.2}_{-0.2}$ & 1165 & $0.15^{+0.07}_{-0.00}$ &18$^{+12}_{-18}$  &Pipeline    & This work \\ [4pt]
     Sk --68$^{\circ}$ 26 & BC2\,Ia & $18.16^{+0.21}_{-0.20}$ & $2.19^{+0.02}_{-0.02}$ & $\cdots$ & $\cdots$ & $\cdots$ & $\cdots$ & $\cdots$ & by-eye (Si) & UKG17\\ [2pt]
     $\cdots$ & $\cdots$       & $19.0^{+0.8}_{-1.6}$ & $2.31^{+0.14}_{-0.33}$ & $5.64^{+0.09}_{-0.14}$ & $-6.7^{+0.4}_{-0.2}$ & 1245: & $0.20^{+0.23}_{-0.05}$ & 55$^{+77}_{-13}$ &Pipeline    & This work \\ [4pt]
     \hline
    \end{tabular}
    \end{center}
\tablebib{    {\bf MBK04} \citet{Massey+2004}; {\bf MKE06} \citet{Mokiem+2006-SMC}; {\bf MKE07} \citet{Mokiem+2007-LMC}; {\bf MZM09} \citet{Massey+2009}; {\bf RHH18} \citet{Ramachandran+2018b}; {\bf RPM12} \citet{Rivero-Gonzalez+2012}; {\bf RSK17} \citet{Ramirez-Agudelo+2017}; {\bf TLP04} \citet{Trundle+2004}; {\bf UKG17} \citet{Urbaneja+2017}. 
}
\end{table*}

 \section{Conclusions}\label{conclusions}

 We have presented a study of XShootU spectroscopy of 122 LMC and 103 SMC OB stars. In constast with previous studies that focus on a relatively narrow subset of spectral type and/or luminosity class, our analysis pipeline \citep{Bestenlehner2024} spans early O dwarfs to late B supergiants across both Magellanic Clouds, permitting
 \begin{itemize}
     \item Uniform spectral classification from the homogeneous XShootU dataset, permitting double-lined binaries and OeBe stars to be identified (see online material). With respect to literature spectral types we confirm a deficit of early O stars amongst the SMC ULLYSES sample, and identify a deficit of late B supergiants in the SMC sample (all 3 stars later than B3 are B8 supergiants), as shown in Fig.~\ref{XShootU_SpT}.
     \item Excluding OeBe and double lined binaries, we quantitatively analyse 97 LMC and 77 SMC OB stars. We establish that ULLYSES O stars in the LMC are more luminous (massive) than SMC counterparts (see Fig~\ref{HRD} and Fig~\ref{XShootU_mass}), with implications for evolutionary and stellar wind comparisons. High temperature O stars dominate the production of hydrogen and helium ionizing photons (Fig.~\ref{ion}). 
     \item Projected rotational velocities are modest, with only $\sim$10\% of XShootU OB stars possessing $\varv_{\rm e} \sin i \geq$ 200 km\,s$^{-1}$ (Fig.\ref{vsini}).
     \item We compare empirical wind velocities from ULLYSES to the temperature-wind velocity relation of \citet{Hawcroft+2023} indicating that the calibration provides an upper envelope to wind velocities (Fig.~\ref{v_inf}). The Milky Way $v_{\infty}/v_{\rm esc}$ relation from \citet{Lamers+1995} is a reasonable approximation to LMC OB stars -- albeit with a large observational scatter -- whereas SMC OB stars reveal a lower ratio, with $v_{\infty}/v_{\rm esc} \ll 1$ in some instances (Fig.~\ref{v_esc}). 
     \item We use empirical wind velocities to compare (reduced) wind momenta, $D_{\rm mom} = \dot{M} v_{\infty} (R_{\ast}/R_{\odot})^{0.5}$, of LMC and SMC OB stars with dense winds, $\log Q = \dot{M}(R_{\ast} v_{\infty})^{-3/2} \geq -13.2$, and find LMC luminous O stars exceed SMC counterparts by $\sim$0.27 dex (Fig.~\ref{mom}). Since H$\alpha$ is only a reliable indicator of mass-loss rates for stars with strong winds, we defer to detailed ULLYSES+XShootU studies \citep[e.g.][]{Backs+2024} for weak-wind stars.
     \item Overall our analysis pipeline results are in satisfactory agreement with bespoke studies of OB stars in the Magellanic Clouds (Table~\ref{lit}, Fig.~\ref{XShootU}). 
Therefore, we consider our pipeline to be well suited to the upcoming very large spectroscopic surveys such as 1001MC \citep{1001MC}.
 \end{itemize}


\section*{Data availability}\label{data_availability}

Tables A1-A2 (physical and wind properties of single and SB1 binaries) and Tables A3-A4 (updated spectral types of OB stars excluded from analysis) for LMC and SMC OB stars are only available in electronic form at the CDS via anonymous ftp to cdsarc.u-strasbg.fr (130.79.128.5) or via \url{http://cdsweb.u-strasbg.fr/cgi-bin/qcat?J/A+A/}.

Online material at \url{https://zenodo.org/XXX} includes blue visual spectral montages of LMC and SMC O dwarfs, giants, supergiants and B supergiants from XShootU, including late SMC B supergiants from BLOeM \citep{Shenar+2024} (Figs B.1--B.8). In addition, spectral fits and spectral energy distribution comparisons are provided for each star (model in red, observations in blue), in alphabetical order. Photometry (indicated as stars) is drawn from \citet{Vink+2023}. 

\begin{acknowledgements}
This study has been made possible courtesy of the Director’s Discretionary ULLYSES survey, which was implemented by a Space Telescope Science Institute (STScI) team led by Julia Roman-Duval, having been recommended by the Hubble UV Legacy Science Definition Working Group chaired by Sally Oey, convened in 2018 by the then STScI Director Ken Sembach. Based on observations made with ESO telescopes at the Paranal observatory under programme ID 106.211Z.001 and observations obtained with the NASA/ESA {\it Hubble Space Telescope}, retrieved
from the Mikulski Archive for Space Telescopes (MAST) at the
STScI. STScI is operated by the Association of Universities for Research in Astronomy, Inc. under NASA contract NAS 5-26555. Nidia Morrell kindly obtained and reduced the Magellan Clay MIKE spectroscopic datasets.

JMB and PAC acknowledge financial support from the Science and Technology Facilities Council via research grant ST/V000853/1 (P.I. Vik Dhillon). AACS is supported by the German \emph{Deut\-sche For\-schungs\-ge\-mein\-schaft, DFG\/} in the form of an Emmy Noether Research Group -- Project-ID 445674056 (SA4064/1-1, PI Sander). SB is supported by a grant from the Dutch Research School for Astronomy (NOVA). This research has made use of the SIMBAD database, operated at CDS, Strasbourg, France, FASTWIND stellar atmosphere code developed by Joachim Puls, and BONNSAI software developed by Fabian Schneider. We appreciate comments on a draft manuscript from Matheus Bernini-Peron, Rolf Kuiper, Lucimara Martins, Varsha Ramachandran and Aida Wofford, plus discussions regarding
masses of post-main sequence massive stars with Norbert Langer, Pablo Marchant and Fabian Schneider.
\end{acknowledgements}

\bibliographystyle{aa}
\bibliography{biblio}



\appendix

\section{Physical properties and spectral types of XShootU OB stars}

Table~\ref{table:targets} presents physical parameters of XShootU single and SB1 targets in the SMC and LMC, ordered by spectral type. Table~\ref{table:wind_mom} presents spectroscopic and evolutionary masses and ages together with inferred wind properties of SMC and LMC stars. Table~\ref{table:SMC} (SMC) and Table~\ref{table:LMC} (LMC) provide updated spectral types of OB stars excluded from analysis (binarity or OeBe), together with XShootU Wolf-Rayet stars, for completeness.

\onecolumn

\begin{small}
\begin{landscape}
\setlength\LTcapwidth{\linewidth}
\begin{longtable}{
l@{\hspace{1mm}} 
l@{\hspace{1mm}} 
l@{\hspace{1mm}} 
l@{\hspace{1mm}} 
c@{\hspace{1mm}} 
c@{\hspace{1mm}} 
c@{\hspace{1mm}} 
r@{\hspace{1mm}} 
c@{\hspace{1mm}} 
c@{\hspace{1mm}} 
c@{\hspace{1mm}} 
c@{\hspace{1mm}} 
c@{\hspace{1mm}} 
l@{\hspace{1mm}} 
l} 
\caption{Physical parameters of XshootU single+SB1 targets in the SMC and LMC, ordered in spectral type, plus measured radial velocities, $\varv_{\rm rad}$. Wind densities are parameterized via $Q = \dot{M}(R_{\ast} \varv_{\infty})^{-3/2}$ with respective units $M_{\odot}$\,yr$^{-1}$, $R_{\odot}$ and km\,s$^{-1}$. Results include ionizing photons predicted below 912\AA\ ($Q_{\rm H}$) and 504\AA\ ($Q_{\rm HeI}$) edges, but negligible emission is predicted below the ionized helium limit at 228\AA. Physical quantities shown in parentheses are not considered reliable.  Solely spectral type information is provided for ULLYSES archival VLT/Xshooter targets.} 
\label{table:targets} \\
\hline\hline
Target, Alias & Sp Type (lit) & Ref &	Sp Type (new) & $\varv_{\rm rad}$ & $T_{\rm eff}$ & $\log g$ & $R_{\ast}$ & $\log L$ & $Y$ & $\log$(N/H) & $\varv_{\rm e} \sin i$ & $\log Q_{\rm H}$ & $\log Q_{\rm He I}$ & Note\\ 
            &                 &      &                & km s$^{-1}$  & kK            &
cm s$^{-2}$ & $R_{\odot}$ & $L_{\odot}$  &   &+12  & km s$^{-1}$ & s$^{-1}$ & s$^{-1}$ \\
\hline
\endfirsthead
\multicolumn{4}{c}%
{\tablename\ \thetable\ -- \textit{Continued from previous page}} \\
\hline
Target, Alias & Sp Type (lit) & Ref &	Sp Type (new) & $\varv_{\rm rad}$ & $T_{\rm eff}$ & $\log g$ & $R_{\ast}$ & $\log L$ & $Y$ & $\log$(N/H) & $\varv_{\rm e} \sin i$ & $\log Q_{\rm H}$ & $\log Q_{\rm He I}$ & Note\\ 
            &                 &      &                & km s$^{-1}$  & kK            &
cm s$^{-2}$ & $R_{\odot}$ & $L_{\odot}$  &   &+12  & km s$^{-1}$ & s$^{-1}$ & s$^{-1}$ \\
\hline
\endhead
\hline \multicolumn{4}{r}{\textit{Continued on next page}} \\
\endfoot
\hline
\endlastfoot
\multicolumn{15}{c}{\bf SMC targets} \\ [2pt]
NGC346 MPG 355 = W3 & ON2 III(f*)    & WMH04 & ON2 III(f*)  &  177 & 59.6$^{+0.0}_{-12.1}$ &  4.12$^{+0.33}_{-0.38}$ &  11.0$^{+3.4}_{-2.5}$ &  6.14$^{ +0.05}_{- 0.30}$ &  0.25$^{+0.10}_{-0.09}$ &   8.4$^{ +0.2}_{-1.3}$ & 112$^{+20}_{- 19}$ & 50.1 &  49.7 & a\\ [2pt] 
AzV 14, Sk~9                                  & O5 V                  & MBK04  &  O3 V((f$^{*}$c))z + ?
& 149 &  44.7$^{+ 9.4}_{ -2.7}$ &  4.12$^{+ 0.38}_{-0.14}$ &  12.1$^{+5.2}_{-4.0}$ &  5.72$^{+ 0.29}_{- 0.11}$ &  0.18$^{+ 0.13}_{-0.02}$ &   7.2$^{+  0.9}_{-0.2}$ & 113$^{+ 20}_{- 19}$ &  49.5 &  48.8 & b\\  [2pt] 
AzV 435    & O3 V((f$^{*}$))   & MPP05 & O3 V((f$^{*}$))z + ? 
& 181 &  44.3$^{+15.2}_{ -3.5}$ &  3.69$^{+ 0.52}_{-0.14}$ &  18.4$^{+12.8}_{-9.7}$ &  6.07$^{+ 0.42}_{- 0.14}$ &  0.24$^{+ 0.07}_{-0.07}$ &   8.5$^{+  0.1}_{-1.0}$ & 113$^{+ 20}_{- 19}$ & 49.9 &  49.2 & b\\ [2pt] 
AzV 177 & O4 V((f))   & MPP05 & 	O3.5 Vn((fc))z & 109 &  44.7$^{+14.8}_{ -3.5}$ &  3.69$^{+ 0.57}_{-0.14}$ &   7.4$^{+5.0}_{-3.8}$ &  5.30$^{+ 0.41}_{- 0.14}$ &  0.30$^{+ 0.10}_{-0.14}$ &   7.2$^{+  1.0}_{-0.1}$ & 201$^{+ 24}_{- 24}$ & 49.1 &  48.5\\ [2pt]
AzV 377, 2dFS 1971                              & O5 V((f)) & EHI04     & O3.5 V((f))z & 172 &  44.7$^{+ 3.9}_{ -3.5}$ &  3.88$^{+ 0.57}_{-0.24}$ &   7.7$^{+1.6}_{-1.4}$ &  5.33$^{+ 0.13}_{- 0.12}$ &  0.41$^{+ 0.14}_{-0.09}$ &   7.2$^{+  1.0}_{-0.1}$ & 113$^{+ 19}_{- 19}$ & 49.1 &  48.4\\
\hline
\end{longtable}
\tablebib{ 
{\bf BTM13} \citet{Bouret+2013},
{\bf CGM86} \citet{Conti+1986},
{\bf CW11} \citet{CrowtherWalborn2011},
{\bf DEH19} \citet{Dufton+2019},
{\bf ECF04} \citet{Evans+2004-cmfgen},
{\bf EHI04} \citet{Evans+2004-2dF}, 
{\bf EKD15} \citet{Evans+2015},
{\bf ELH15} \citet{Evans+2015-2dF},
{\bf ELS06} \citet{Evans+2006},
{\bf ETH11} \citet{Evans+2011},
{\bf Fit88} \citet{Fitzpatrick1988},
{\bf Fit91} \citet{Fitzpatrick1991},
{\bf FBM09} \citet{Farina+2009},
{\bf GW87} \citet{GarmanyWalborn1987},
{\bf GCM87} \citet{Garmany+1987},
{\bf GKM18} \citet{Gvaramadze+2018},
{\bf Len97 } \citet{Lennon1997}, 
{\bf LOG13} \citet{Lamb+2013},
{\bf LOS16} \citet{Lamb+2016},
{\bf MBK04} \citet{Massey+2004},
{\bf MCP06} \citet{Mennickent+2006},
{\bf MLD95} \citet{Massey+1995},
{\bf MNM17} \citet{Massey+2017},
{\bf MPP05} \citet{Massey+2005},
{\bf MWD00} \citet{Massey+2000},
{\bf MZM09} \citet{Massey+2009},
{\bf NG04} \citet{NiemelaGamen2004},
{\bf OS98} \citet{OeySmedley1998}, 
{\bf PGM92} \citet{Parker+1992},
{\bf RHH18} \citet{Ramachandran+2018b},
{\bf RHO19} \citet{Ramachandran+2019},
{\bf RMP78} \citet{Rousseau+1978},
{\bf San69} \citet{Sanduleak1969},
{\bf SB97} \citet{SmithNeubigBruhweiler1997},
{\bf SSM22} \citet{Shenar+2022},
{\bf Wal77} \citet{Walborn1977},
{\bf Wal83} \citet{Walborn1983},
{\bf WFC02} \citet{Walborn+2002-FUSE},
{\bf WHE10} \citet{Walborn+2010},
{\bf WHL02} \citet{Walborn+2002},
{\bf WLH95} \citet{Walborn+1995},
{\bf WLH00} \citet{Walborn+2000},
{\bf WMH04} \citet{Walborn+2004},
{\bf WSS14} \citet{Walborn+2014},
}
\tablefoot{
\tablefoottext{a}{Temperature too hot. N\,{\sc v} reproduced, N\,{\sc iv} too weak and N\,{\sc iii} not matched}
\tablefoottext{b}{Strong He\,{\sc i} suggesting SB2?}
\tablefoottext{c}{Poor fit - clumped model required}
\tablefoottext{d}{Strong nebular lines, resulting in poor fit}
\tablefoottext{e}{$\log Q$ overestimated due to strong nebular H$\alpha$}
\tablefoottext{f}{Wings of H$\alpha$ in emission but not fitted. $v_{\rm e} \sin i$ underestimated.}
\tablefoottext{g}{Si\,{\sc iii} too strong.}
\tablefoottext{h}{Si\,{\sc iii} too strong, emission in wings of H$\alpha$, H$\beta$, H$\delta$.}
\tablefoottext{i}{ poor normalisation: lower $T_{\rm eff}$}
\tablefoottext{j}{N\,{\sc v} too strong, $T_{\rm eff}$ too high.}
\tablefoottext{k}{$T_{\rm eff}$ too low due to presence of He\,{\sc i} (crowded region, contamination?)}
\tablefoottext{l}{Data reduction issue with C\,{\sc iv}. Broad N\,{\sc iv} $\lambda$7116 complex}
\tablefoottext{m}{$T_{\rm eff}$ too low since N\,{\sc v} too weak, strong nebular contamination: He abundances too low, Clumped model needed. SB2 (Pollock et al. submitted)}
\tablefoottext{n}{$T_{\rm eff}$ too high, N\,{\sc v} too strong, N\,{\sc iii} too weak.}
\tablefoottext{o}{$T_{\rm eff}$ too high, He\,{\sc i} not reproduced.}
\tablefoottext{p}{C\,{\sc iv} too strong, N\,{\sc iii} not reproduced.}
\tablefoottext{q}{$T_{\rm eff}$ too low, N\,{\sc v} not reproduced, but He\,{\sc i-ii} ok.}
\tablefoottext{r}{$T_{\rm eff}$ too low and $\log Q$ to high due to nebular contamination}
\tablefoottext{s}{Poor normalisation, $T_{\rm eff}$ too high.}
\tablefoottext{t}{$T_{\rm eff}$ too high, N\,{\sc iv} too strong and N\,{\sc iii} too weak}
\tablefoottext{u}P{$T_{\rm eff}$ too low, very poor normalistion, He\,{\sc ii} not reproduced}
\tablefoottext{v}{H$\alpha$ not well matched: $\log Q$ underestimated.}
\tablefoottext{w}{Spectral energy distribution infers unphyical reddening $E(B-V) \leq 0$}
\tablefoottext{x}{H$\alpha$ emission wings not reproduced, oddly shaped H$\beta$. $T_{\rm eff}$ too low.}
}
\end{landscape}
\end{small}




\begin{small}
\begin{landscape}
\setlength\LTcapwidth{\linewidth}
\begin{longtable}{
l@{\hspace{2mm}} 
l@{\hspace{2mm}} 
c@{\hspace{2mm}} 
c@{\hspace{2mm}} 
c@{\hspace{2mm}} 
c@{\hspace{2mm}} 
c@{\hspace{2mm}} 
c@{\hspace{2mm}} 
c@{\hspace{2mm}} 
r@{\hspace{1mm}} 
r@{\hspace{-2mm}} 
r@{\hspace{2mm}} 
l@{\hspace{2mm}} 
l@{\hspace{1mm}} 
c@{\hspace{1mm}} 
c@{\hspace{1mm}} 
c@{\hspace{1mm}} 
c} 
\caption{Masses and wind properties of XshootU OB stars. Spectroscopic masses, $M_{\rm spec}$ (in $M_{\odot}$) are determined from surface gravities adjusted for centrifugal forces, $g_{c}$ (cm\,s$^{-2}$), while evolutionary masses, $M_{\rm evol}$ (in $M_{\odot}$), initial masses, $M_{\rm init}$ (in $M_{\odot}$) and ages $\tau$ (in Myr) are from {\sc bonnsai} ($\log L, T_{\rm eff}, v_{\rm e} \sin i$), applied to \citet{Brott+2011} and \citet{Kohler+2015} evolutionary models (values with colons exclude $v_{\rm e} \sin i$ constraint). For late B supergiants  masses and ages, indicated in [], are estimated from SMC evolutionary models of \citet{Schootemeijer+2019}. Wind properties are obtained from the wind density parameter $Q = \dot{M} (R_{\ast} \varv_{\infty})^{-3/2}$ plus measured wind velocities from C\,{\sc iv} $\lambda$1548, N\,{\sc v} $\lambda$1238 or Si\,{\sc iv} $\lambda$1393. Mass-loss rates (in $M_{\odot}$\,yr$^{-1}$) adopt a wind clumping factor of 10 ($f_{v}$ = 0.1), although are insensitive to H$\alpha$ for low wind densities $\log Q < -13.2$. Reduced wind momenta $\log D_{\rm mom} = \dot{M} \varv_{\infty} \sqrt{R/R_{\odot}}$ with units of g\,cm\,s$^{-2}$. Physical quantities shown in parentheses are not considered reliable.} 
\label{table:wind_mom} \\
\hline\hline
Target & Sp Type & $\log (L/L_{\odot}$) & $\log g_c$ & $M_{\rm spec}$ & $M_{\rm evol}$ & $M_{\rm init}$ & $\tau$ & $\log Q$ & \multicolumn{3}{c}{$\varv_{\infty}$ (km\,s$^{-1}$)} & Line & Ref & $\log \dot{M}$ & $\log D_{\rm mom}$ & 
$A_{\rm V}$ & Note \\ 
\hline
\endfirsthead
\multicolumn{4}{c}%
{\tablename\ \thetable\ -- \textit{Continued from previous page}} \\
\hline
Target & Sp Type & $\log (L/L_{\odot}$) & $\log g_c$ & $M_{\rm spec}$ & $M_{\rm evol}$ & $M_{\rm init}$ & $\tau$ & $\log Q$ & \multicolumn{3}{c}{$\varv_{\infty}$ (km\,s$^{-1}$)} & Line & Ref & $\log \dot{M}$ & $\log D_{\rm mom}$ & $A_{\rm V}$ & Note\\ 
\hline
\endhead
\hline \multicolumn{4}{r}{\textit{Continued on next page}} \\
\endfoot
\hline
\endlastfoot
\multicolumn{17}{c}{\bf SMC targets} \\ 
NGC346 MPG 355 & ON2 III(f*)   & 6.14$^{ +0.05}_{- 0.30}$ &  4.12$^{+0.33}_{-0.38}$  & 59.1$^{+ 109.5}_{- 30.7}$ & 55.8$^{+25.0}_{-16.4}$ & 63.4$^{+22.7}_{-18.4}$ & 0.0$^{+2.2}_{-0.0}$ & -12.6$^{ +0.4}_{ -0.6}$ & 3120 & $\pm$ & 30 & C\,{\sc iv} & a & --6.29$^{+0.43}_{-0.60}$ & 28.52$^{+0.42}_{-0.65}$ & 0.33 & 1\\[2pt] 
AzV 14 &  O3 V((f$^{*}$c))z + ?& 5.72$^{+ 0.29}_{- 0.11}$ & 4.12$^{+0.38}_{-0.14}$ & 70.7$^{+ 192.4}_{- 33.2}$ & 51.0$^{+6.7}_{-6.3}$ &51.8$^{+7.2}_{-5.8}$ & 2.1$^{+0.5}_{-1.5}$ & -13.4$^{+  0.4}_{ -1.6}$& 1000 & $\pm$ & 100 & N\,{\sc v} & c & --7.75$^{+0.41}_{-1.63}$  & 26.59$^{+0.48}_{-1.63}$ & 0.19 \\ [2pt]  
AzV 435 & O3 V((f$^{*}$))z + ? & 6.07$^{+ 0.42}_{- 0.14}$ & 3.70$^{+0.52}_{-0.14}$ & 61.7$^{+ 369.1}_{- 35.9}$  & 71.4$^{+33.5}_{-20.4}$ & 76.4$^{+32.4}_{-19.6}$ & 1.1$^{+0.7}_{-1.1}$ & -15.0$^{+  2.0}_{ -0.0}$& 1490 & $\pm$ & 100 & N\,{\sc v} & c & --8.84$^{+2.00}_{-0.11}$ & 25.76$^{+2.03}_{-0.16}$ & 1.52 & 1\\ [2pt] 
AzV 177 & O3.5 Vn((fc))z       & 5.30$^{+ 0.41}_{- 0.14}$ & 3.75$^{+0.57}_{-0.14}$ & 11.5$^{+  75.1}_{-  6.6}$  & 33.2$^{+11.4}_{-7.2}$ &33.6$^{+11.1}_{-7.1}$ & 1.6$^{+1.4}_{-1.6}$ & -13.6$^{+  0.9}_{ -1.4}$ & 2670 & $\pm$ & 10 & C\,{\sc iv} & a & --7.64$^{+0.96}_{-1.42}$ & 27.02$^{+1.03}_{-1.42}$ & 0.14 \\
\end{longtable}
\tablebib{
(a) SEI fits \citet{Hawcroft+2023}; (b) This study $v_{\infty} = v_{\rm black}$ \citep{Prinja+1990}; (c) This study $v_{\infty} \sim 0.85 v_{\rm edge}$ \citep{PrinjaCrowther1998, Crowther+2016}; (d) Eqn~\ref{Hawcroft} from \citet{Hawcroft+2023}
}
\tablefoot{
\tablefoottext{1}{LMC grid adopted}
\tablefoottext{2}{No HST UV spectroscopy}
}
\end{landscape}
\end{small}

\clearpage

\setlength\LTcapwidth{\linewidth}
\begin{longtable}{
l@{\hspace{1mm}}
l@{\hspace{1mm}}
r@{\hspace{3mm}}
l@{\hspace{1mm}}
l@{\hspace{1mm}}
r}
\caption{SMC XShootU stars excluded from analysis (binarity, OeBe star, or a Wolf-Rayet star), sorted by (new) spectral type.} 
\label{table:SMC} \\
\hline\hline
Target, Alias & Sp Type (lit) & Ref & Sp Type (new) & Note & Ref \\ 
\hline
AzV 255,  Sk~90$^{a}$                          & O8 V & MLD95 & O(n)e & Oe & This work \\ [1pt] 
NGC346 ELS 7 = MPG 324 = W6 & O4 V((f))   & WLH95 & O3 V((f$^{*}$c))z + ? &SB2? & This work\\ [1pt]  
Sk 183     &	O3 V((f$^{*}$))z    & RHO19 & O3 V((f$^{*}$))z + OB & SB2? & This work\\ [1pt] 
AzV 476 &	O4 IV--II((f))p+O9.5:& POH22 & O3.5 III((f$^{*}$c)) + ? & EB, SB2 &PGS13, POH22 \\ [1pt] 
2dFS 2553 & O6.5 III((f))e   & LOS16 & O4 III(f) + B &  SB2 & RSP24 \\ [1pt] 
NGC346 MPG 435 = W1  &	O4 If+O5-6 &	DEH19 & O4 If(n)p + ? & SB2 & Nie02, DEH19 \\
\hline
\end{longtable}
\tablebib{ 
{\bf CG82} \citet{CramptonGreasley1982},
{\bf DEH19} \citet{Dufton+2019},
{\bf EHI04} \citet{Evans+2004-2dF}, 
{\bf ELS06} \citet{Evans+2006},
{\bf FMG03} \citet{Foellmi+2003},
{\bf GCM87} \citet{Garmany+1987},
{\bf KMH14} \citet{Koenigsberger+2014},
{\bf LOS16} \citet{Lamb+2016},
{\bf LZW98} \citet{Lamers+1998},
{\bf MLD95} \citet{Massey+1995},
{\bf MPP05} \citet{Massey+2005},
{\bf MZM09} \citet{Massey+2009},
{\bf Nie02} \citet{Niemela2002}, %
{\bf PGS13} \citet{OGLE-III-SMC},
{\bf POH22} \citet{Pauli+2022},
{\bf RHB93} \citet{Reynolds+1993}
{\bf RHO19} \citet{Ramachandran+2019},
{\bf RSE12} \citet{Ritchie+2012},
{\bf RSP24} \citet{Ramachandran+2024}
{\bf SHT18} \citet{Shenar+2018},
{\bf Wal83} \citet{Walborn1983},
{\bf WLH95} \citet{Walborn+1995},
{\bf ZKW96} \citet{Zickgraf+1996}
}
\tablefoot{
\tablefoottext{a}{Initially selected for ULLYSES, but ultimately omitted from the UV spectroscopic programme.}
\tablefoottext{b}{Archival VLT/Xshooter (PI Kalari)}
}

\newpage

\setlength\LTcapwidth{\linewidth}
\begin{longtable}{
l@{\hspace{1mm}}
l@{\hspace{1mm}}
r@{\hspace{3mm}}
l@{\hspace{1mm}}
l@{\hspace{1mm}}
r}
\caption{LMC XShootU stars excluded from analysis (binarity, OeBe star, or a Wolf-Rayet star), sorted by (new) spectral type.}  
\label{table:LMC} \\
\hline\hline
Target, Alias & Sp Type (lit) & Ref & Sp Type (new) & Note & Ref \\ 
\hline
Sk --70$^{\circ}$ 60 & O4--5 V((f))pec & MZM09 & early O(f)pe & Oe & MZM09, This work\\ [1pt] 
LH 114--7             &	O2 III(f$^{*}$)+OB? & WHL02          & O2 III(f$^{*}$) & SB2? & This work \\  
PGMW 3070                & O6 V                  &	  PGM92 & O2: III: + O4: III: + B0: &SB2 & This work   \\ [1pt] 
VFTS 542, Mk~30, BAT99 113     & O2 If$^{*}$/WN5 + B0 & SSH19  &  O2 If$^{*}$/WN5 + ? & SB2 & SSH19 \\ [1pt] 
BI 13                    & O6.5 V             &    Mas02  & O3 V(f$^{*}$c) + OB & SB2? & This work \\ [1pt] 
Sk --67$^{\circ}$ 105                                 & O4f + O6         &	OL03 & O4 Ifc + O & SB2 & OL03 \\ [1pt] 
Sk --66$^{\circ}$ 100                              & O6 II(f)   &WLH95            &  O6.5 II(f) + ? & SB2? & This work\\ [1pt] 
$[\,$ST92$]\,$ 5--52, W61 3--14$^{a}$     &	O3V ((f*))+OB               & WHL02 & O7 V((f))z + ? & SB2? &This work\\ [1pt] 
BI 272 & O7\,II & CGM86 & O7:\,V+ B2: & SB2 & This work (MIKE)\\ [1pt] 
BI 189                                                 & O8 IV((f))e   &RHH18a         &	O8 Vz  & Oe & This work \\ [1pt] 
HV 5622     & B0 V                  &	    BOL09 & O8 V + B0.5: & EB, SB2?  & GSP11, This work \\ [1pt] 
LMC X-4$^{b}$                                             & O8 III   & NC02          &O8 III((f))p var & SB1 & HCC78 \\ [1pt] 
Sk --69$^{\circ}$ 83, HDE~269244          & O7.5 Iaf             &	    Fit88 & O8 Iaf + O7: & SB2? & This work  \\ [1pt] 
BI 184                                               &O8 (V)e   & RHH18b	& O9:V + early B & SB2? & This work \\ [1pt] 
Sk --70$^{\circ}$ 32 & O9.5 II:   &	Fit88 & O9 III + O9.5 III: & SB2? & This work \\ [1pt] 
Sk --67$^{\circ}$ 118  & O7 V & MLD95 & O9 III + B0: II: & SB2 & This work (MIKE) \\ [1pt] 
VFTS 66                                             & O9 V+B0.2 V & MSA20 & B0 V + ? & SB2 & MSA20  \\ [1pt] 
Sk --68$^{\circ}$ 23a & B1 III    &	   MLD95 & B0.7 II + B0.5: & SB2? & This work\\ [1pt] 
N206--FS 170                                     & B1 IV        &RHH18b         & B1 III: + B & SB2? &This work\\ [1pt] 
Sk --71$^{\circ}$ 35                                & B1 II       &	RHH18b           &  B1 Ia + O9: &SB2 & RSP24 \\ [1pt] 
Sk --67$^{\circ}$ 197  & B7 I & RMP78 & B8 Ia & Be? & This work \\ [1pt] 
LMCe055-1  & WN4/O4          &  MNM17 & $\cdots$        & EB & GSP11\\ [1pt]
Sk --67$^{\circ}$ 20, HD~32109, BAT99 7            & WN4b              &	    SSM96 &  $\cdots$ &$\cdots$ & $\cdots$\\ [1pt]
Sk --65$^{\circ}$ 55, BAT99 30    &	WN6h               & SSM96 &$\cdots$ & $\cdots$ & $\cdots$ \\ [1pt] 
Sk --71$^{\circ}$ 21, HD~36063, BAT99 32 &	WN6(h)    &	CS97 & WN6(h)+? & SB2 & SMS08, This work \\ [1pt] 
HDE~269927c, Sk --69$^{\circ}$ 249c, BAT99 120 & WN9h  & CS97 &  $\cdots$ &$\cdots$ & $\cdots$ \\ [1pt] 
Sk --68$^{\circ}$ 73, HDE~269445, R~99, BAT99 33 & WN9pec & Cro21 &$\cdots$ &  $\cdots$ & $\cdots$ \\  [1pt] 
Sk --67$^{\circ}$ 266, S 61, BAT99 133$^{c}$  & WN11h & CS07 & $\cdots$ &  $\cdots$ & $\cdots$ \\ [1pt]
Sk --68$^{\circ}$ 15, HD 32402, BAT99 11    & WC4                  &      SSM90   & $\cdots$ &  $\cdots$ & $\cdots$ \\ [1pt] 
Sk --69$^{\circ}$ 191, HD~37680, BAT99 61 & WC4 & SSM90   &$\cdots$  &$\cdots$ & $\cdots$ \\  
\hline
\end{longtable}
\tablebib{ 
{\bf BOL09} \citet{Blair+2009},
{\bf CGM86} \citet{Conti+1986},
{\bf Cro21} Crowther (2021, priv.\ com.),
{\bf CS97} \citet{CrowtherSmith1997},
{\bf Fit88} \citet{Fitzpatrick1988},
{\bf GCM87} \citet{Garmany+1987},
{\bf GSP11} \citet{OGLE-III-LMC},
{\bf HCC78} \citet{Hutchings+1978},
{\bf Mas02} \citet{Massey2002},
{\bf MLD95} \citet{Massey+1995},
{\bf MNM17} \citet{Massey+2017},
{\bf MSA20} \citet{Mahy+2020},
{\bf MZM09} \citet{Massey+2009},
{\bf NC02} \citet{NegueruelaCoe2002},
{\bf OL03} \citet{OstrovLapasset2003},
{\bf PGM92} \citet{Parker+1992},
{\bf RHH18a} \citet{Ramachandran+2018a},
{\bf RHH18b} \citet{Ramachandran+2018b},
{\bf RMP78} \citet{Rousseau+1978},
{\bf RSP24} \citet{Ramachandran+2024}
{\bf SMS08} \citet{Schnurr+2008},
{\bf SSH19} \citet{Shenar+2019},
{\bf SSM90} \citet{Smith+1990},
{\bf SSM96} \citet{Smith+1996},
{\bf WLH95} \citet{Walborn+1995},
{\bf WHL02} \citet{Walborn+2002},
}
\tablefoot{
\tablefoottext{a}{Initially selected for ULLYSES, but ultimately omitted from the UV spectroscopic programme.}
\tablefoottext{b}{Excluded from analysis owing to spectral variability between XShootU epochs}
\tablefoottext{c}{Archival VLT/Xshooter (PI Mehner)}
}

\clearpage

\section{Spectral montages of LMC and SMC OB stars}\label{montage}

\begin{figure}
\centering
  \includegraphics[width=16cm,bb=60 70 545 765]{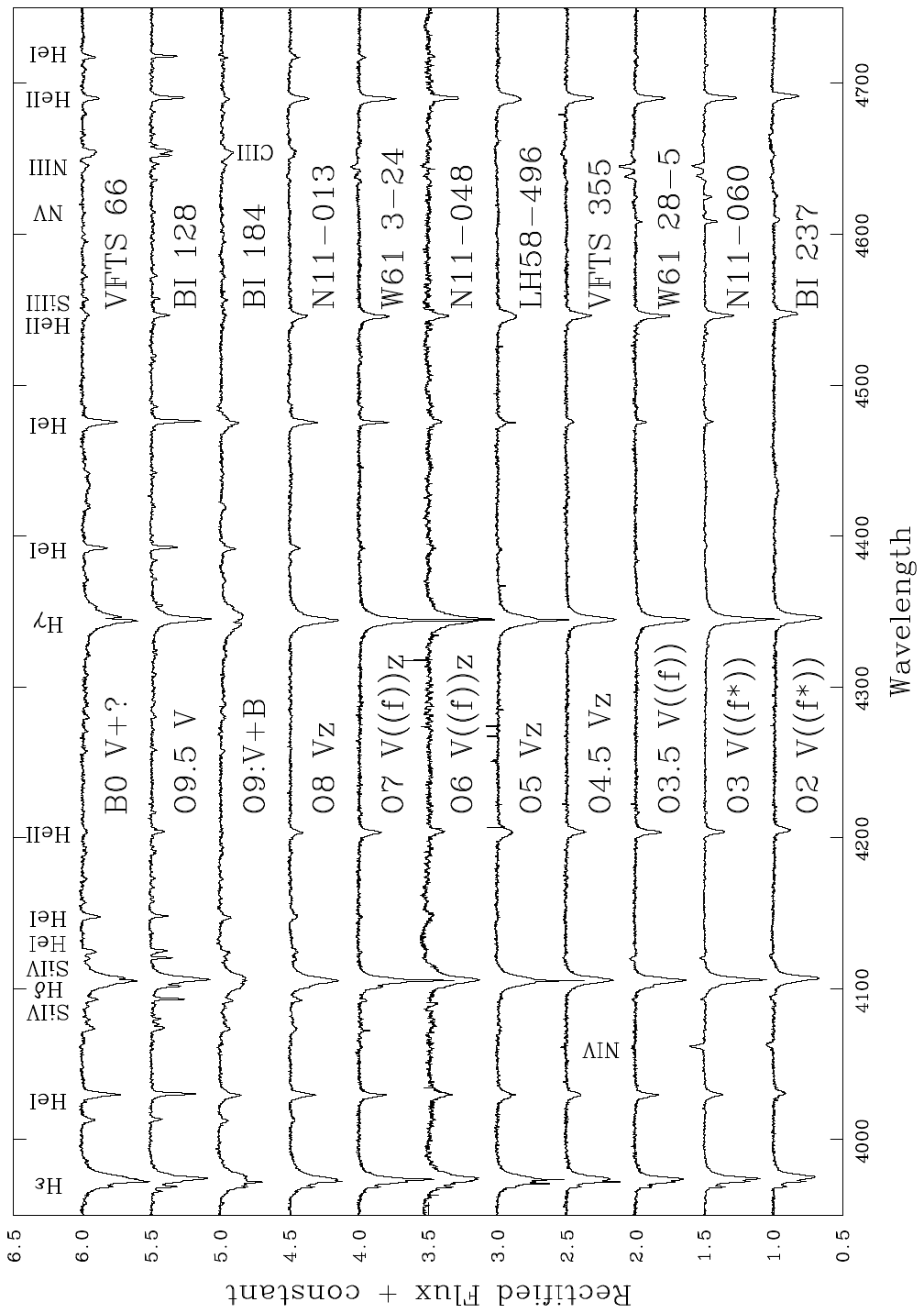}
  \caption{Spectral montage of LMC O2 to B0 dwarfs from XShootU}
  \label{LMC_Odwarf}
\end{figure}
\begin{figure}
\centering
  \includegraphics[width=16cm,bb=60 70 545 765]{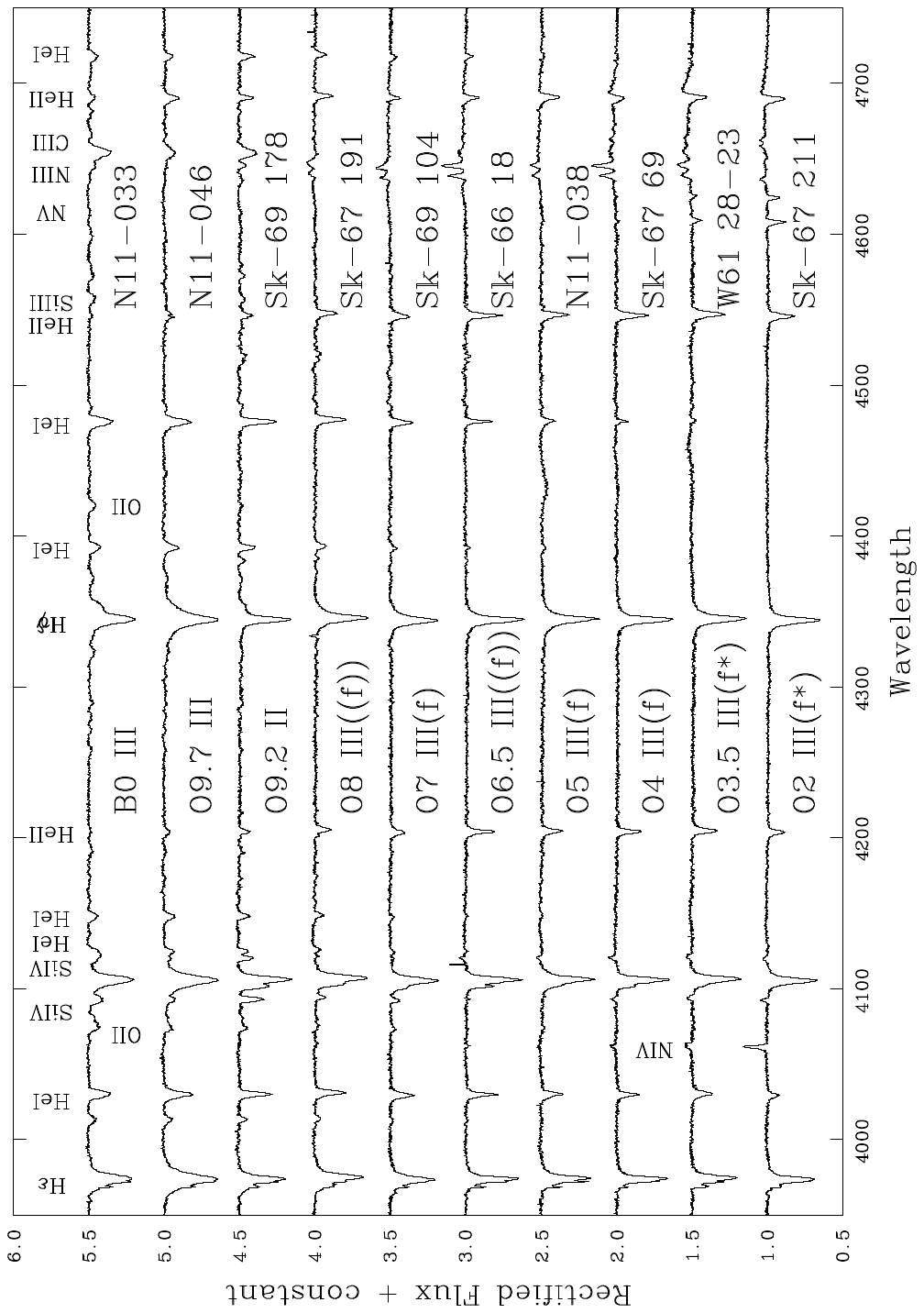}
  \caption{Spectral montage of LMC O2 to B0 giants from XShootU}
  \label{LMC_Ogiant}
\end{figure}
\begin{figure}
\centering
  \includegraphics[width=16cm,bb=60 70 545 765]{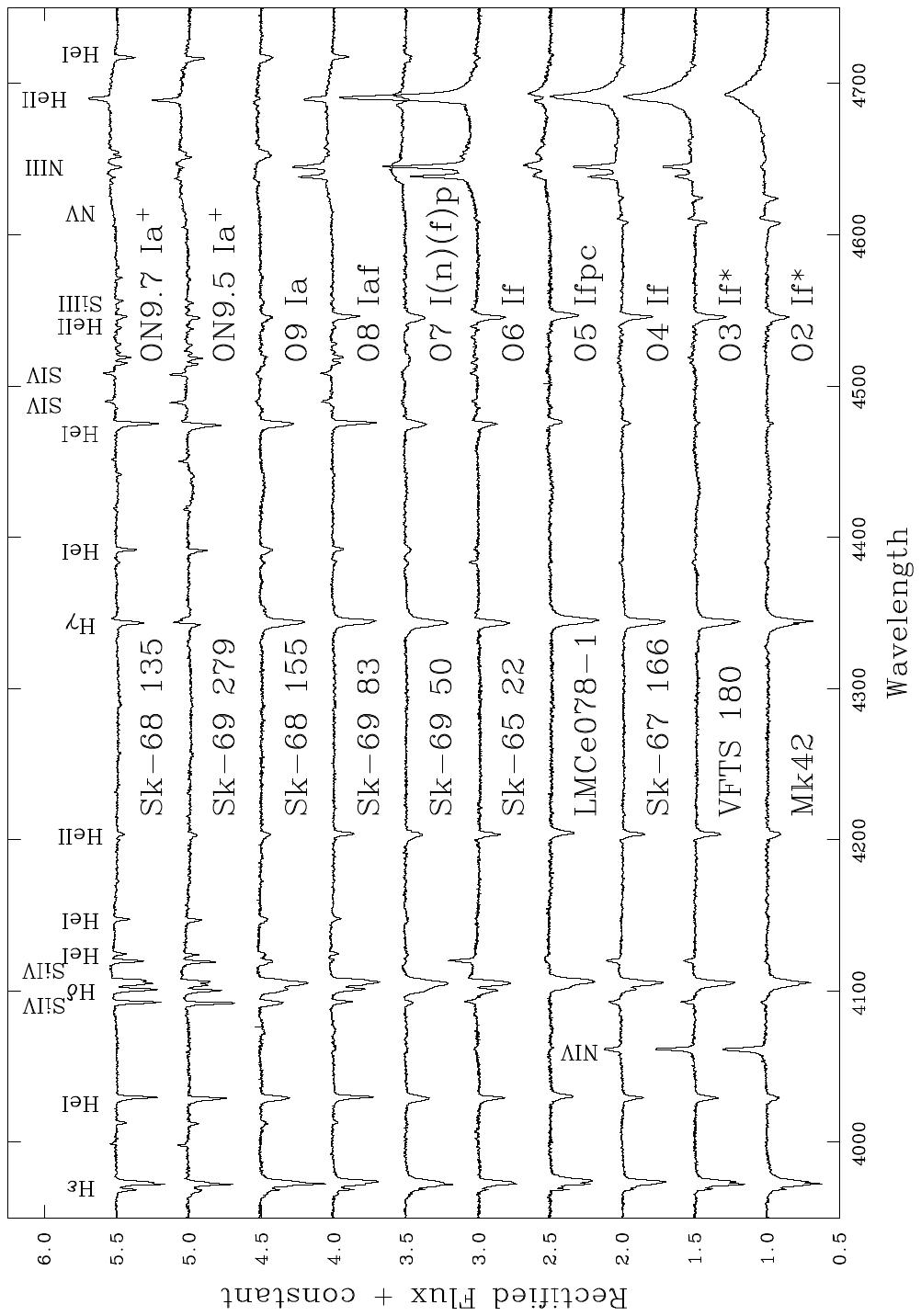}
  \caption{Spectral montage of LMC O2 to O9.7 supergiants from XShootU}
  \label{LMC_Osuper}
\end{figure}
\begin{figure}
\centering
  \includegraphics[width=16cm,bb=60 70 545 765]{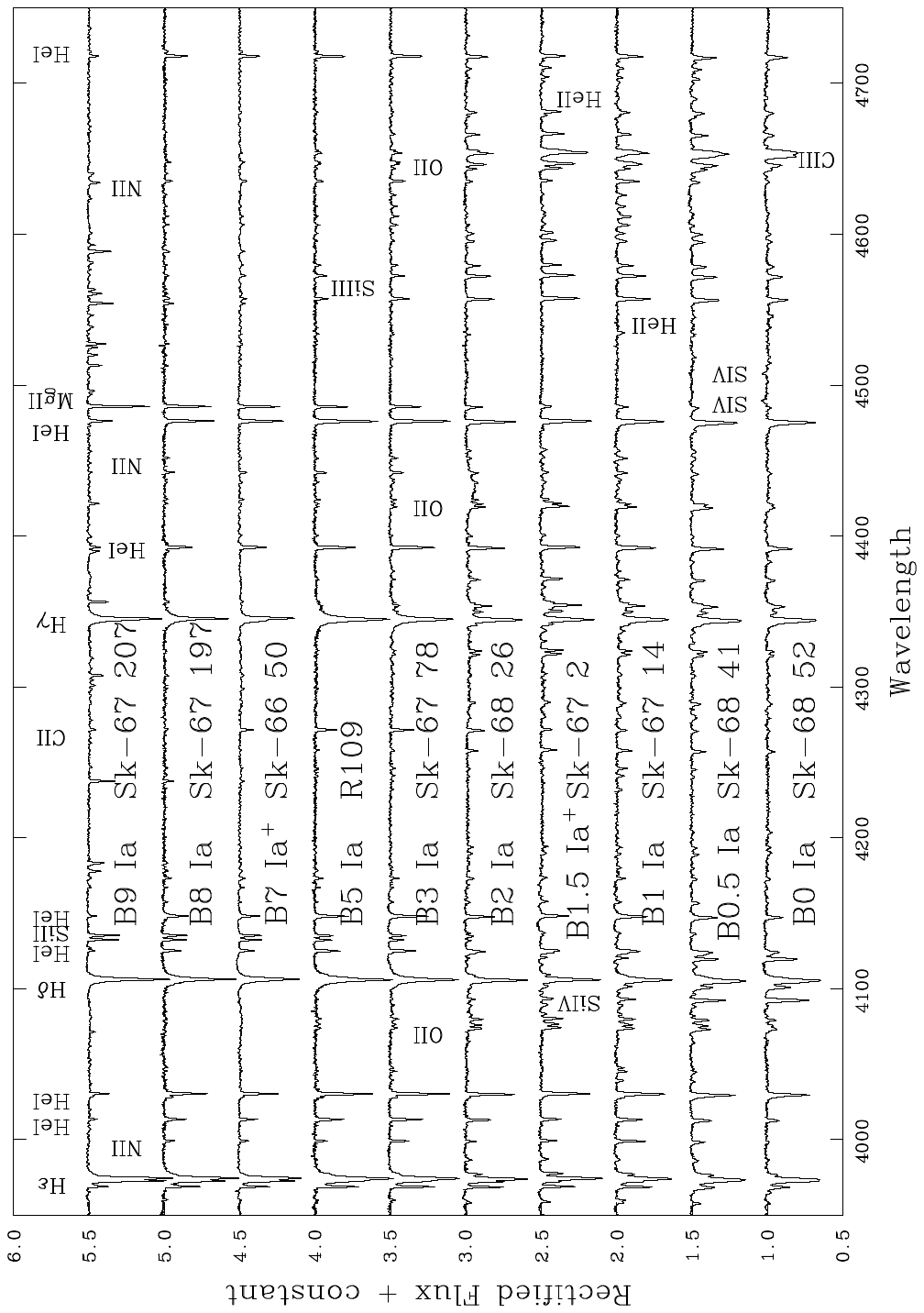}
  \caption{Spectral montage of LMC B0 to B9 supergiants from XShootU}
  \label{LMC_Bsuper}
\end{figure}
\begin{figure}
\centering
  \includegraphics[width=16cm,bb=60 70 545 765]{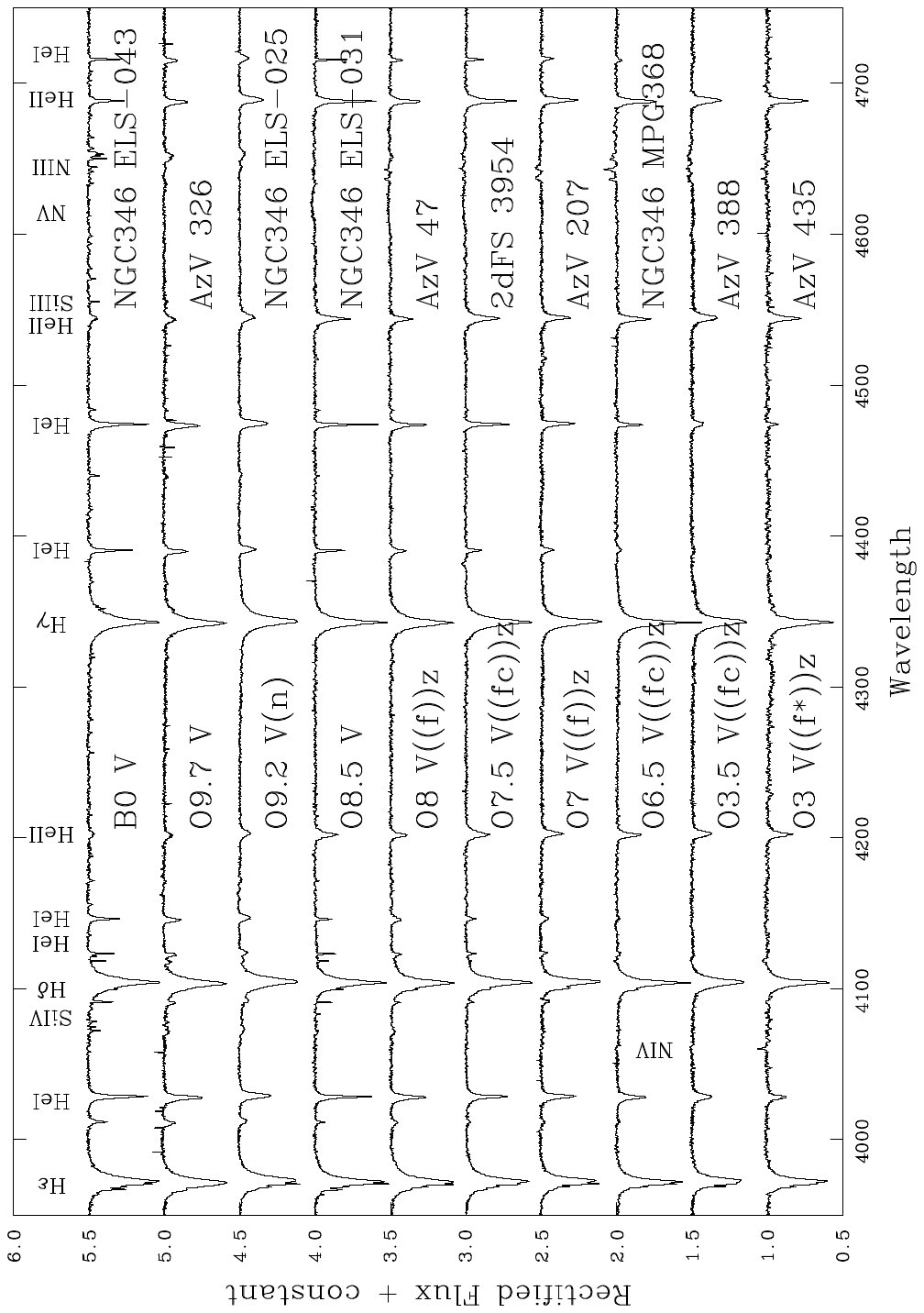}
  \caption{Spectral montage of SMC O2 to B0.2 dwarfs from XShootU}
  \label{SMC_Odwarf}
\end{figure}
\begin{figure}
\centering
  \includegraphics[width=16cm,bb=60 70 545 765]{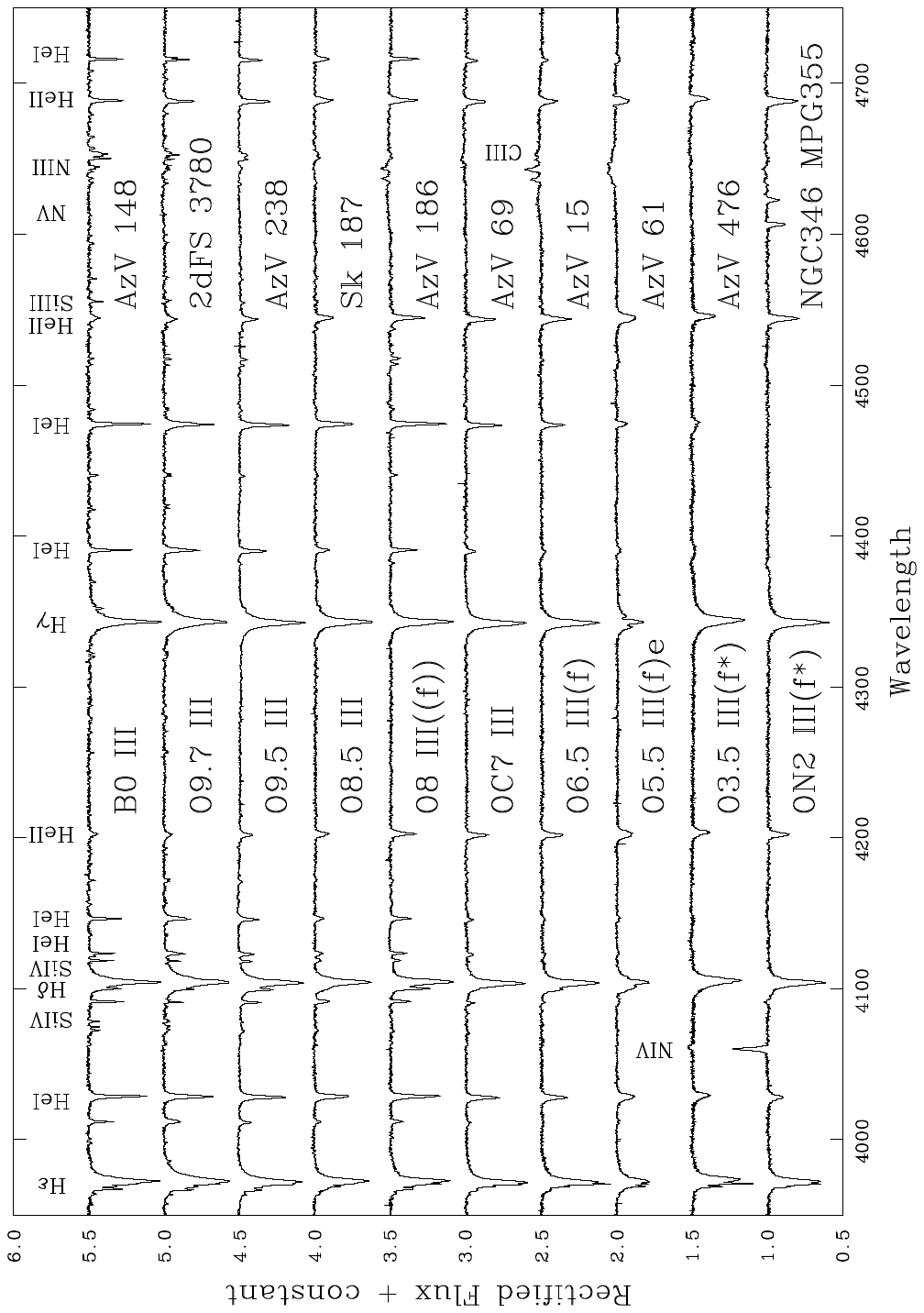}
  \caption{Spectral montage of SMC O2 to B0 giants from XShootU}
  \label{SMC_Ogiant}
\end{figure}
\begin{figure}
\centering
  \includegraphics[width=16cm,bb=60 70 545 765]{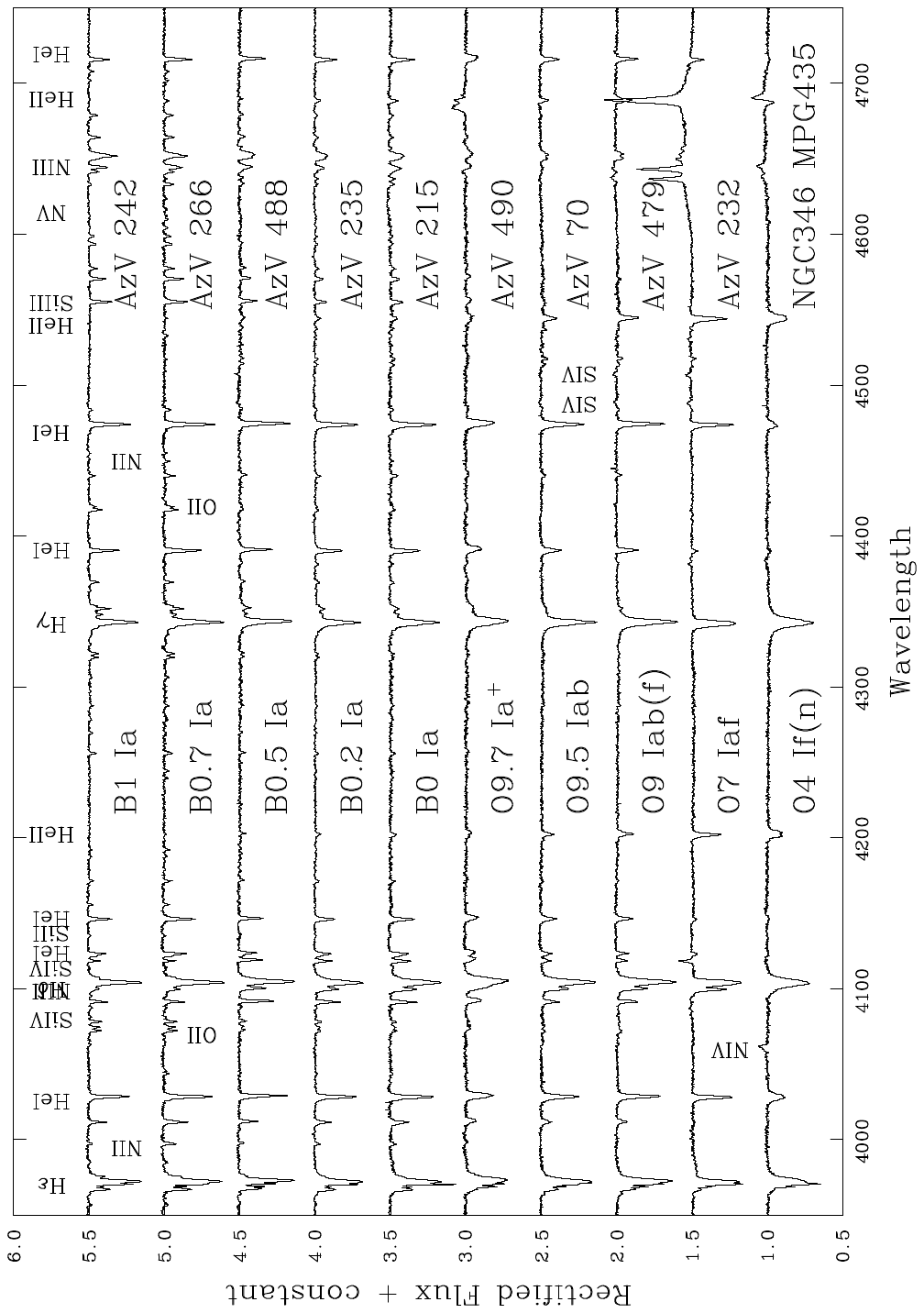}
  \caption{Spectral montage of SMC O4 to B1 supergiants from XShootU}
  \label{SMC_Osuper}
\end{figure}
\begin{figure}
\centering
  \includegraphics[width=16cm,bb=60 70 545 765]{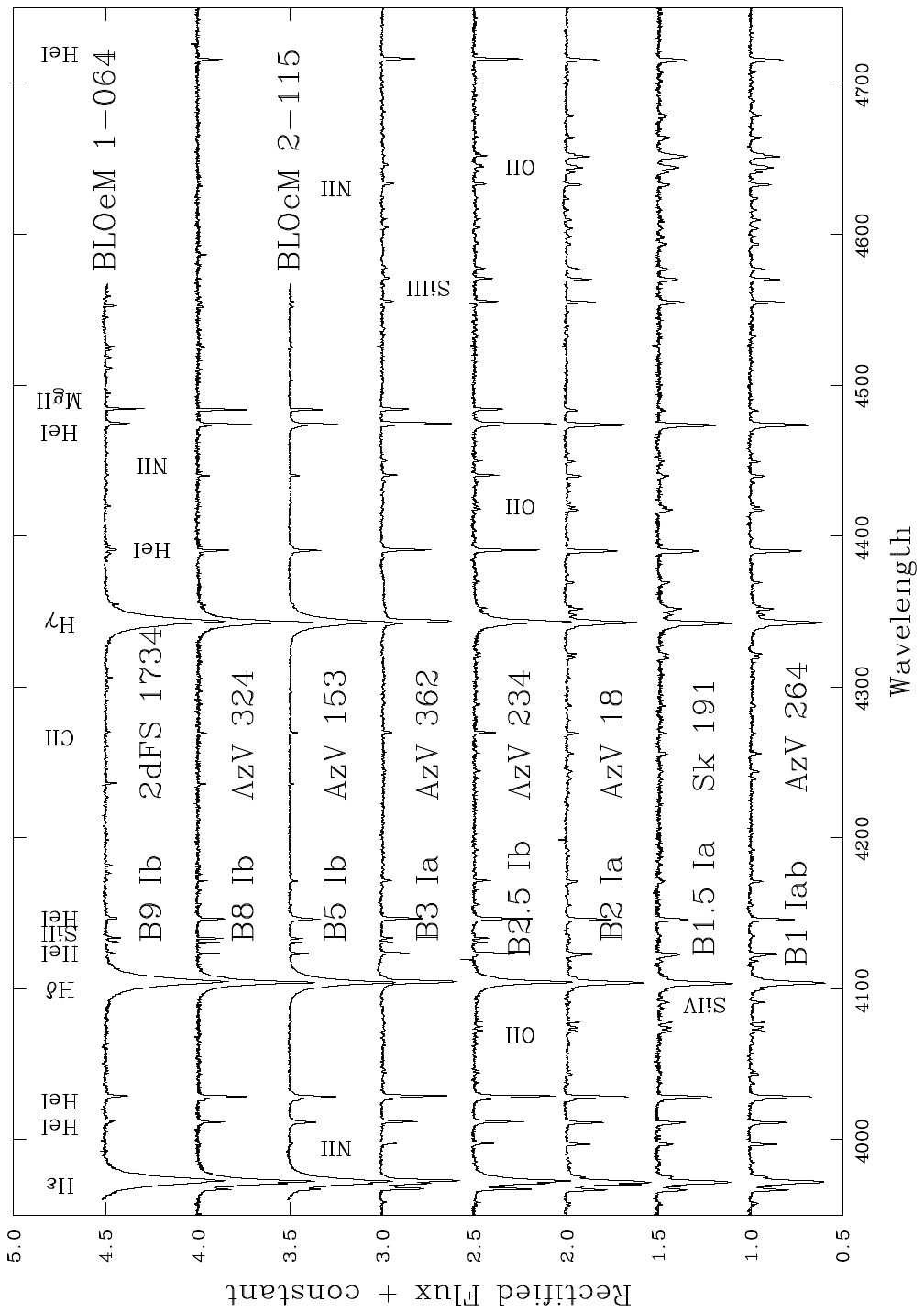}
  \caption{Spectral montage of SMC B1 to B8 supergiants from XShootU,
  plus SMC B5 (AzV 153) and B9 (2dFS 1734) supergiants from BLOeM \citep{Shenar+2024}.}
  \label{SMC_Bsuper}
\end{figure}

Figures~\ref{LMC_Odwarf}-\ref{SMC_Bsuper} present blue visual spectral montages of LMC and SMC O dwarfs, giants, supergiants and B supergiants from XShootU, including examples of  SMC late B supergiants from BLOeM \citep{Shenar+2024}.

\clearpage

\end{document}